%% file: main.tex
\documentclass[12pt]{article}
\usepackage{psfig}
\def\eqb{\addtocounter{equa}{1} \eqno{\enu}}
\def\enu{(\thesection.\theequa )}
\def\equaname#1{\relax
      \global\addtocounter{equa}{1}
      \xdef#1{(\thesection.\theequa )~ }}
\def\equaleb#1{\relax
      \global\addtocounter{equa}{0}
      \xdef#1{(\thesection.\theequa )~ }}
\def\eqn#1{ \equaname{#1}\eqno\enu}
\def\theequation{\thesection.\theequa}
\newcounter{subeq}
\def\beqas{
\def\theequation{\addtocounter{subeq}{1}
\thesection.\theequa\alph{subeq}}\beqa}
\def\equaaleb#1{\relax
      \global\addtocounter{subeq}{1}
      \xdef#1{(\thesection .\theequa \alph{subeq})~ }
      \global\addtocounter{subeq}{-1}                 }
\def\eeqas{\eeqa\setcounter{subeq}{0}\def\theequation{\thesection.\theequa}}
\def\req#1{(\ref{#1})}
\def\ncr{\nonumber\\}
\def\lb{\label}
\def\bi{\bibitem}
\def\rar{\rightarrow}

\def\bc{\begin{center}}
\def\ec{\end{center}}
\def\bt{\begin{tabbing}}
\def\et{\end{tabbing}}
\def\bdis{\begin{displaystyle}}
\def\edis{\end{displaystyle}}
\def\beq{\global\addtocounter{equa}{1}\begin{equation}}
\def\eeq{\end{equation}}
\def\beqa{\global\addtocounter{equa}{1}\begin{eqnarray}}
\def\eeqa{\end{eqnarray}}
\def\ben{\begin{enumerate}}
\def\een{\end{enumerate}}
\def\bfr{\begin{flushright}}
\def\efr{\end{flushright}}
\def\bra#1{\langle #1 |}
\def\ket#1{| #1 \rangle}
\def\emp{${\rm o}\hspace*{-0.2cm}/$}
\let\a=\alpha \let\b=\beta \let\g=\gamma \let\d=\delta
\let\e=\varepsilon   
  \let\l=\lambda \let\m=\mu
 \let\x=\xi \let\p=\pi \let\r=\rho \let\s=\sigma
\let\t=\tau 
\let\o=\omega 
 
 \let\Ph=\phi  \let\Ps=\Psi
\let\O=\Omega \let\S=\Sigma  \let\Th=\Theta
\let\L=\Lambda \let\G=\Gamma \let\D=\Delta

\def\0{\over } \def\1{\bf }     \def\2{{1\over2}} \def\4{{1\over4}}
\def\5{\bar }  \def\6{\partial } \def\7#1{{#1}\llap{/}}

\def\1#1{{\bf #1}}

\def\1#1{{\bf #1}}
\def\bm#1{{\mbox{\boldmath $ #1 $}}}

\def\et{\it et at.}
\def\hbsm{\hspace*{-0.35cm}}
\def\hbss{\hspace*{-0.15cm}}
\def\H{{\cal H}}
\def\tp{t_{\g\p}}

\def\nfi{Fig. 2.1 }

\def\nfii{Fig. 2.3 }
\def\nfiii{Fig. 2.4 }
\def\nfiv{Fig. 2.5 }
\def\nfv{Fig. 2.6 }
\def\nfvi{Fig. 2.7 }
\def\nfvii{Fig. 2.8 }
\def\nfviii{Fig. 2.9 }

\begin{document}

\begin{center}
{\Large \bf Photo Reactions on the Deuteron \\ in the $\D$-Resonance Region }\\

\vspace{1.0cm}

{\bf K. A. Bugaev$^{1,2}$,  
U. Oelfke$^{1,3}$ 
and P.U. Sauer$^1$}\\

\vspace{1.cm}

$^1$Institute for Theoretical Physics,
University of Hannover\\
30167 Hannover, Germany

\hfill \\

$^2$Bogolyubov Institute for Theoretical Physics\\
 Kiev, Ukraine\\

\hfill \\

$^3$ Theory Group, TRIUMF\\ 4004 Wesbrook Mall\\
Vancouver, B.C.\\
Canada V6T 2A3
\end{center}

\vspace{2.0cm}

\begin{abstract}
\noindent
Photo disintegration of the deuteron and photo pionproduction on
the deuteron are described in the region of the $\D$-resonance.
The two reactions are unitarily coupled.
A novel definition of electromagnetic exchange currents  
in the presence of pionproduction is given.
Numerical predictions are done for spin-independent and spin-dependent
observables of both reactions.
The sensitivity of the results with respect to the nucleon-$\D$ 
transition current and with respect to the irreducible 
nucleon-$\D$ potential is studied.
\end{abstract}

\vspace{4.cm}


\vspace{1.cm}

\newpage

\setcounter{section}{2}
\newcounter{equa}
\setcounter{equa}{0}

\begin{center}
{\Large\bf 1. Introduction}
\end{center}
\vspace{0.5truecm}

The two-nucleon system above pion threshold is the simplest many-nucleon
system which can be studied at intermediate-energy excitation. At
those energies the pion ($\pi$) and the $\Delta$-isobar
degrees of freedom become active and are clearly seen in elastic
two-nucleon scattering and in its unitarily
coupled inelastic channels with one pion, i.e., in the two-body
pion-deuteron (d) channel and in the three-body $\pi$NN channel
containing two free nucleons (N) and one pion. The hadronic
properties of the two-nucleon system above pion threshold are
often described in terms of hamiltonian force models \cite{poe,lee} with explicit
pion and $\Delta$-isobar degrees of freedom or in terms of
variants \cite{afn,fay,rin} of
it, in which the pion-nucleon $P_{33}$-resonance is accounted for by the
properties of a pion-nucleon potential. The theoretical description
of the two-nucleon system above pion threshold is technically
demanding due to its three-body aspect and
unfortunately not always successful, since the building blocks of the
employed force model are not tuned yet. An overview on the art of
describing the hadronic processes in the two-nucleon system
above pion threshold is given in Ref. \cite{gar}.
At intermediate energies the pion and the
$\Delta$-isobar can also be excited in electromagnetic (e.m.)
reactions, i.e., in inelastic electron scattering or
by  the absorption of real photons ($\gamma$). Bremsstrahlung in two-nucleon scattering
is the corresponding process of photon emission. This paper develops
a description for both photo reactions in the two-nucleon
system. The description is attempted to be consistent with the
description of hadronic reactions given previously \cite{poe}. The paper
applies the description to photo disintegration of the deuteron
and to photo pionproduction on the deuteron.

The simultaneous description of hadronic and e.m. reactions requires
the conceptual consistency between hadronic interaction operators and
e.m. current
operators. That consistency is the underlying theme of this paper.
It completes work started in Ref. \cite{uwethesis}
by two of the present authors.
In their pioneering work Arenh\"ovel and  Leidemann \cite{are1} studied
deuteron disintegration in e.m. reactions and beautifully proved
the importance of meson-exchange currents and isobar-excitation
for a successful description.
Wilhelm {\it et al.} \cite{are2, are3, are4} extended that work to
a simultaneous description of 
photo pionproduction on the deuteron.
Parallel theoretical developments can be found  in Ref. \cite{taohta1}. 
More
recent theoretical results on this subject can also be found in Ref. \cite{bbb}. 
The present paper will therefore be
unable to add much to our
understanding of those processes. However, we believe that a
thorough discussion of a consistent treatment of all pionic and e.m.
reactions from the deuteron is a worthwhile enterprise.

Sect. 2 describes the conceptual characteristics of the model used for
the hadronic interaction and for the e.m. current.
It discusses in some length the constraint of current conservation and
the redefinition of exchange currents in a coupled-channel theory which
considers the pion as an active degree of freedom.
It gives the e.m.  multi-channel transition
matrix for photon reactions in the two-nucleon
system. Sect. 3 describes
the actual parameterization of the force and of the current
in our practical calculations.
Sect. 4 applies the theoretical apparatus
to photo disintegration of the deuteron and to photo pionproduction on
the deuteron. It gives results for both types of reactions.
Sect. 5 contains our conclusions.

\vspace{0.4truecm}

\begin{center}
{\Large\bf 2. Theoretical Framework}\\
\vspace{0.5truecm}
{\large\bf 2.1 Force Model and Hadronic Reactions}
\end{center}
\vspace{0.3truecm}

Our approach for studying e.m.
reactions in the two-nucleon system employs the hadronic interaction of
Ref. \cite{poe}. That interaction describes  the two-nucleon system
below and above pion-threshold and its
coupled inelastic channels with at most one pion, i.e.,
the reactions NN$\rar$NN, NN$\leftrightarrow$$\p$d, NN $\rar$$\p$NN,
$\p$d$\rar$$\p$d and $\p$d$\rar$$\p$NN. The inclusion of
inelastic channels with at most
one pion limits \cite{poe} the validity of the force model
to c.m. energies below
500 MeV. The force model is defined in the framework of
nonrelativistic quantum mechanics.
This section briefly reviews its basic
ideas. Appendix A provides the details
needed for the objectives of this paper.

The Hilbert space for the description of hadronic and e.m. reactions
at intermediate energies has the purely
nucleonic sector $\H_N$ and two additional sectors $\H_\D$
and $\H_\p$ with non-nucleonic content.
In $\H_\D$ one of the nucleons is replaced by a $\D$-isobar.
The third sector $\H_\p$ contains one additional
pion besides nucleons. The projection operators on the three different
sectors $\H_N$,
$\H_\D$ and $\H_\p$ are labeled  $P_N$, $P_\D$ and $Q$, respectively,
with $P_N + P_\D + Q = 1$.

The hamiltonian $H$ consists of the kinetic energy  $H_0$
and of the interaction $H_1$, i.e., $H=H_0 +H_1$.
The eigenstates of the kinetic energy $H_0$ make up the different
Hilbert sectors. The interaction hamiltonian $H_1$ is given
in terms of instantaneous potentials and is decomposed according to
its action within and between the three different Hilbert sectors, i.e.,
 $$
H_1 = (P_N + P_\D ) H_1 (P_N + P_\D ) + P_\D H_1 Q + Q H_1 P_\D + Q H_1 Q.
\eqn\hf
 $$
The various contributions to $H_1$ are shown in Fig. 2.1 for a system of
baryon number two.
The potentials $P_a H_1 P_b$ with a, b = N, $\D$ act in the purely
baryonic sectors, i.e., $P_N H_1 P_N$ is the two-nucleon potential
of Fig. 2.1(a), $P_\D H_1 P_N$ the transition potential from two-nucleon to
nucleon-$\D$ states of Fig. 2.1(b) and $P_\D H_1 P_\D$ the potential between
nucleon-$\D$ states of Figs. 2.1(c) and 2.1(d).
The two-baryon states are coupled to the pionic states by the
one-baryon $\p$N$\D$ vertex $ Q H_1 P_\D $ of Fig. 2.1(e). The hermitian conjugate
parts $P_N H_1 P_\D$ and $ P_\D H_1 Q $ are not shown in Fig. 2.1. 
In the three-particle Hilbert sector $\H_\p$ the interaction
Q$H_1$Q consists of the two-nucleon potential in the presence of a
pion and of the pion-nucleon potential in nonresonant partial waves
according to Figs. 2.1(f) and 2.1(g) respectively.

A special feature of the considered force model is its mechanism
for pionproduction and pion absorption. Pionic states are not coupled
directly to purely nucleonic ones, i.e., $P_N H_1 Q
=Q H_1 P_N = 0$. Pion production and pion absorption are assumed to proceed
through the excitation and deexcitation of an $\D$-isobar in
two-nucleon processes. The $\D$-isobar is a bare particle
of spin and isospin $3\0 2$. It is
unobservable and gets dressed to the physical $P_{33}$-resonance of
pion-nucleon scattering by self-energy corrections due to coupling to
pion-nucleon states. This force model assumes the $P_{33}$ partial
wave to dominate pion-nucleon scattering in the region up to 300 MeV
pion lab energy. Its standard realization  assumes $P_{33}$ pion-nucleon scattering
to proceed exclusively through
the $\D$-isobar as shown in Fig. 2.2, i.e., a possible nonresonant potential
$Q H_1Q$ is taken to be zero in that partial wave. The $P_{33}$
pion-nucleon transition matrix
$t_{\p N}(\e_\D+i0) $ therefore takes the particular form \cite{poe,lee,drei}
$$
t_{\p N}(\e_\D +i0) \ket{{\bf k}_\D} = \nonumber \\
Q H_1 P_\D {1\0  \e_\D + i0 - { k_\D^2 \0 2 m_\D^0 }
-M_\D(\e_\D, k_\D ) + {i\0 2} \G_\D (\e_\D, k_\D )} P_\D H_1 Q
\ket{{\bf k}_\D}.  \eqb $$
In Eq. (2.2)  $m^0_\D = 1315.8$ MeV is the bare mass of the $\D$-isobar according to \cite{poe}, whereas
$M_\D (\e_\D ,k_\D) $ and $\G_\D (\e_\D, k_\D )$ are its
renormalized mass and width. The latter quantities depend
on the energy $\e_\D$ available for and on the total momentum ${\bf k}_\D$
of the pion-nucleon system; they become identical with the mass and
the width of the physical $P_{33}$-resonance for $\e_\D = 1232$ MeV at
resonance and for ${\bf k}_\D = 0$ in the pion-nucleon c.m. system.

The explicit parameterization of the hamiltonian (2.1) will be given in
Sect. 3.2. Its application to the two-nucleon system above pion threshold
leads to a complex multi-channel scattering problem. The employed
scattering theory is an extended version of the
Alt-Grassberger-Sandhas (AGS)  formulation \cite{ags},
extended to particle production in Ref. \cite{poe}. In Ref. \cite{poe}
methods for the practical solution of the resulting equations
are also given. Details of the
theoretical apparatus are repeated in Appendix A in order to make the present
paper self-contained. This section reviews only some further essentials:

The reaction channels differ by their particle content. There are
two-baryon channels without a pion, i.e., two-nucleon and nucleon-$\D$
channels, N and $\D$, respectively, denoted by the small Latin letters a, b, c ... in the following,
there are two-body channels in the three-particle Hilbert
sector $\H_\p$, in which two particles form a bound pair, denoted by small
Greek letters $\alpha$, $\beta$, $\gamma$ ..., the label for the
spectating particle,
and there are three-particle channels in $\H_\p$, in which all three
particles are free, denoted by the subscript 0.
In the two-nucleon system above pion threshold only three different
reaction channels are experimentally observable, i.e.,
\ben
\item the two-nucleon channel a = N with the asymptotic states
$\ket{\Ph_N ({\1 p}_N )}$
 of relative momentum  ${\1 p}_N$, $E_N ({\1 p}_N )$ being the corresponding
 energy,
\item the pion-deuteron channel $\alpha = \pi$ with asymptotic states
$\ket{\Ph_\p (\1 q_\p )} $
 of relative pion-deuteron momentum  $\1 q_\p $, $E_\p (\1 q_\p) $ being the
 corresponding energy and
\item the three-particle break-up channel 0 of two free nucleons and one free pion
with the asymptotic $\p$NN states $\ket{\Ph_0 (\1 p , \1 q )}$ of internal
momenta $\1 p$ and $\1 q$,  $E_0 (\1 p ,\1 q )$ being the corresponding
energy.
\een
When no confusion arises, we shall omit the arguments of the channel energy,
i.e., we shall use $E_N, E_\p$ and $E_0$.

The $S$-matrix elements between the different kinds of
initial and final states  $\ket{\Ph_i}$ and $\ket{\Ph_f}$ of
energy $E_i$ and $E_f$ with i, f = N, $\p$, 0 referring to the three distinct
channels are given in terms of the on-shell multi-channel transition matrix
$U(z)$, i.e.,
$$ \bra{\Ph_f} S \ket{\Ph_i} = \bra{\Ph_f} \Ph_i\rangle  - 2\p i \d
(E_f - E_i ) \bra{\Ph_f} U_{fi}(E_i + i0) \ket{\Ph_i}. \eqn\smat $$
The various channel components of the off-shell transition matrix
$U(z)$	follow from an appropriate decomposition of the full resolvent 
$$ G(z) = {1\0 z-H_0 - H_1}\eqb $$
according to Eq. (A.1) of Appendix A. They satisfy integral equations.
The multi-channel transition matrix $U(z)$ contains the complete information
about the dynamics generated by the hamiltonian (2.1). E.g., it
allows to construct the fully correlated scattering states
 \beqa
 \ket{\Ps^{(\pm )}_N ({\1 p}_N )}  &=&
\pm i0 G(E_N\pm i0 ) P_N \ket{\Ph_N ({\1 p}_N )},
 \nonumber
\\
 \ket{\Ps^{(\pm )}_\p (\1 q_\p )}
    &=& \pm i0 G(E_\p\pm i0 ) Q \ket{\Ph_\p (\1 q_\p )},
\\
 \ket{\Ps^{(\pm )}_0 (\1 p, \1 q)} &=& \pm i0 G(E_0\pm i0 ) Q
				 \ket{\Ph_0 (\1 p, \1 q)},   \nonumber
\label{wfcor}
\eeqa
in terms of the uncorrelated channel states $\ket{\Ph_N (\1 p_N)}$,
$\ket{\Ph_\p (\1 q_\p )}$
and $\ket{\Ph_0 (\1 p, \1 q)}$, i.e.,
\beqas
 P_b \ket{\Ps^{(\pm )}_N ({\1 p}_N)}&= & [\d_{bN} + g_{b0}^P (E_N
\pm i0) U_{bN}(E_N\pm i0)]
 P_N \ket{\Ph_N ({\1 p}_N)},
\\
 Q \ket{\Ps^{(\pm )}_N ({\1 p}_N)} &= &  g_\b^Q (E_N\pm i0) U_{\b N}(E_N\pm i0)
 P_N \ket{\Ph_N ({\1 p}_N)},
\eeqas
\beqas
 P_b \ket{\Ps^{(\pm )}_\p (\1 q_\p )}&= &   g_{b0}^P (E_\p\pm i0)
U_{b\p}(E_\p\pm i0)
 Q \ket{\Ph_\p (\1 q_\p )},
\\
 Q \ket{\Ps^{(\pm )}_\p (\1 q_\p)} & = &
 [\d_{\b\p}+g_\b^Q (E_\p\pm i0) U_{\b\p}(E_\p\pm i0)]
 Q \ket{\Ph_\p (\1 q_\p)},
\eeqas
\beqas
 P_b \ket{\Ps^{(\pm )}_0 (\1 q,\1 p )} & = &  g_{b0}^P (E_0\pm i0)
 U_{b0}(E_0\pm i0)
 Q \ket{\Ph_0 (\1 q,\1 p )},
\\
 Q \ket{\Ps^{(\pm )}_0 (\1 q,\1 p )} & = & g_{\b}^Q (E_0\pm i0) U_{\b 0}
(E_0\pm i0)
 Q \ket{\Ph_0 (\1 q,\1 p )},
\\
				     & = & [1+ g_{0}^Q (E_0\pm i0)
U_{00}(E_0\pm i0)]
 Q \ket{\Ph_0 (\1 q,\1 p )}. \label{wfcu}
\eeqas
The quantities $g^P_{b0}(z)$, $g^Q_\b(z)$ and $g^Q_0(z)$ are partial
resolvents defined in Eq. (A.2) of Appendix A. The form of
Eqs. (2.6) - (2.8) for the hadronic
scattering states is needed in Sect. 2.3 
to derive the corresponding transition amplitudes
for the e.m. processes in the two-nucleon system. That information
is unnecessary for the $S$-matrix (2.3) of hadronic reactions.
\newpage

\begin{center}
{\large\bf 2.2 The E.M. Current and the Constraint of Current\\
Conservation}
\end{center}
\vspace{0.3truecm}

The e.m. interaction of a system of nucleons whose dynamics is controlled
by the force model of Sect. 2.1 is described in this section. The
presentation is split into three parts: First, the general structure
of the current acting in
and between the different  Hilbert sectors $\H_N$, $\H_\D$ and $\H_\p$ is
explained and the corresponding e.m. hamiltonian $H_1^\gamma$
for the emission and
absorption of real photons is constructed.  Second, the constraint
of current conservation is discussed. 
Third, the interaction-dependent contributions to the current are 
constructed.
We consider Subsects. 2.2.2 and 2.2.3  
the conceptually most important ones of this paper.

\vspace{0.3truecm}
\begin{center}
{\large\bf 2.2.1 Structure of the E.M. Interaction}
\end{center}
\vspace{0.3truecm}

The photon field $A^\m(x)$ couples to the e.m. current operator
$j^\m (x) = $$(\rho (x), {\bf j}(x))$
of hadrons and forms the e.m. hamiltonian $H_1^\g$ by
$$ H_1^\g = \int d^3x A_\m ({\1 x} , 0) j^\m ({\1 x} , 0). \eqb $$
The photon field $A^\m (\1 x, 0)$ is employed in Coulomb gauge,
i.e., $A^\m (\1 x, 0) =(0, \1 A(\1 x,0))$ with
$\bm\nabla \cdot\1 A(\1 x,0) = 0$. Consequently,
the e.m. hamiltonian $H_1^\g$ couples the photon field only to
the spatial components $\1 j(\1 x, 0)$ of the current $j^\m(x)$.
The e.m. hamiltonian is  used in one-photon exchange
approximation, i.e., in lowest-order perturbation theory.
In this paper it will be applied to reactions with real photons only.
Thus, the operator of the e.m. field $A^\m ({\1 x} , 0)$ takes the form
$$
A^\m ({\1 x}, 0)={1\0 (2\p )^{3\0 2}} \int {d^3k \0 2\o} \sum_{\l =\pm 1}
  \e^\m ({\1 k},\l ) [ a({\1 k},\l ) e^{i{\1 k}{ \1 x}} + a^\dagger ({\1 k},\l
)
  e^{-i{\1 k\1x}}], \eqn\amu
$$
where $a(\1 k,\l )$ and $a^\dagger (\1 k,\l )$ are the annihilation and
creation operators for a photon of momentum $\1 k$, of energy $\o = |\1 k|$ and
of helicity $\l$ with $\e^\m (\1 k,\l )$ being the polarization vector.
The conventions of Ref. \cite{drell} for commutation rules and
normalization are adopted.

Though real photons in Coulomb gauge couple only to the spatial components
of the current, we nevertheless discuss the e.m. current in full --
in anticipation of the theoretical needs when describing general e.m.
reactions.
The current will be employed in the
Fourier-transformed form
$$
j^\m({\1 k}_\g) = \int d^3x e^{i {\1 k}_\g {\1 x}}
j^\m({\1 x}, 0). \eqb
$$
It couples
the various sectors of the Hilbert space considered, i.e.,
$$
j^\m({\1 k}_\g) = (P_N + P_\D + Q ) j^\m({\1 k}_\g) (P_N +P_\D +Q ), \eqb
$$
and it consists of one-baryon and at least
two-baryon pieces, $j^{[1]\m}({\1 k}_\g)$ and $j^{[2]\m}({\1 k}_\g)$, i.e.,
$$
j^\m({\1 k}_\g) = j^{[1]\m}({\1 k}_\g) + j^{[2]\m}({\1 k}_\g). \eqb
$$
It has to be consistent with the underlying hadronic interaction
and has at least to satisfy current conservation
$$
{\1 k}_\g \cdot\1 j ({\1 k}_\g ) = [H_0 + H_1, \r ({\1 k}_\g)]\eqn\conx
\equaleb\currentconservationoperator
$$
as one consistency condition. All calculations will be done in the
c.m. system; nevertheless, the current operator $j^{\m}({\1 k}_\g)$ also has to be
a Lorentz four-vector; boost properties
of the current are therefore additional consistency conditions, but
will not be considered here, as usual.
The one-nucleon current $ P_N j^{[1] \m}({\1 k}_\g) P_N$
is seen in elastic electron-nucleon scattering, the current
$ (P_{\D} + Q) j^{[1] \m}({\1 k}_\g) P_N $ in electro pionproduction on the nucleon.
The considered hadronic hamiltonian
of Subsect. 2.1 requires
interaction-dependent exchange currents of many-baryon nature. Only
currents of one-baryon and two-baryon characteristics are considered according
to Eq. (2.13).

\vspace{0.3truecm}
\begin{center}
{\large\bf 2.2.2 Channel Decomposition and Baryon-Number Decomposition}
\end{center}
\vspace{0.3truecm}

This subsection analyzes the condition \currentconservationoperator of current conservation.
The condition \currentconservationoperator has the form
\beqa
&& {\1 k}_\g \cdot \, (P_N + P_\Delta + Q) \,\left[ {\1 j}^{[1]}(\1 k_\g ) +
{\1 j}^{[2]}(\1 k_\g) \right] \, (P_N + P_\Delta + Q) = \ncr
&&(P_N + P_\Delta + Q) \, [H_0^{[1]} + H_1^{[1]} +
H_1^{[2]},{\r}^{[1]}(\1 k_\g ) + {\r}^{[2]}(\1 k_\g)] \,
(P_N + P_\Delta + Q),
\equaleb\currentconservationprojection
\eeqa
once the coupling between channels and the baryon-number
characteristics of the hamiltonian and of the current are made explicit.
The kinetic energy operator $H_0^{[1]}$ is of one-baryon nature
and that fact is notationally indicated in this subsection 
by the superscript one in square brackets in
contrast to the remainder of the paper. However, also the interaction
part of the hamiltonian has one-baryon pieces $H_1^{[1]}$ besides the
irreducible two-baryon ones $H_1^{[2]}$. In one-baryon pieces one
nucleon does not participate in any interaction: One-baryon pieces are
the $\pi$N$\Delta$
vertex of process (e) in \nfi  and the pion-nucleon potential of
process (g) in	Fig. 2.1.
Thus, we use the  hadronic hamiltonian
whose interaction  structure is given in Eq. (2.1) in the form
\beqa
H &=& P_N H_0^{[1]} P_N + P_\Delta H_0^{[1]} P_\Delta
      + Q H_0^{[1]} Q + \ncr
  &&  P_\Delta H_1^{[1]} Q + Q H_1^{[1]} P_\Delta + Q H_1^{[1]} Q +
      \ncr
  &&  (P_N + P_\Delta) H_1^{[2]} (P_N + P_\Delta) + Q H_1^{[2]} Q.
\equaleb\hamiltonianprojection
\eeqa
The condition of
current conservation splits up into eighteen separate constraints
differentiated by channel coupling and baryon number. Only twelve
among the eighteen constraints are independent, the other six
are related to the independent ones by the hermiticity of the
hamiltonian and of the current operator. As an example two constraints are spelt out
in detail for one particular channel coupling, the coupling of purely nucleonic
to pion-nucleon states, i.e.,
\beqa
\hspace*{-0.5cm} {\1 k}_\g \cdot Q \, {\1 j}^{[1]}({\1 k}_\g ) \, P_N &=&
Q H_0^{[1]} Q \, Q {\r}^{[1]}({\1 k}_\g) P_N -
Q {\r}^{[1]}({\1 k}_\g ) P_N \, P_N H_0^{[1]} P_N + \ncr
&&Q H_1^{[1]} (P_\Delta + Q) \, (P_\Delta + Q) {\r}^{[1]}({\1 k}_\g) P_N,
\equaleb\constraintqpnonebaryon
\eeqa
\beqa
{\1 k}_\g \cdot Q \, {\1 j}^{[2]}(\1 k_\g) P_N &=&
Q H_0^{[1]} Q \, Q {\r}^{[2]}({\1 k}_\g) P_N -
Q {\r}^{[2]}({\1 k}_\g ) P_N \, P_N H_0^{[1]} P_N  + \ncr
&&Q H_1^{[1]} Q \, Q {\r}^{[1]}({\1 k}_\g) P_N + \ncr
&&Q H_1^{[1]} (P_\Delta + Q) \, (P_\Delta + Q) {\r}^{[2]}({\1 k}_\g)
P_N + \ncr
&& Q H_1^{[2]} Q \, Q \left[{\r}^{[1]}(\1 k_\g )
+ {\r}^{[2]}(\1 k_\g)\right] P_N - \ncr
&& Q \left[{\r}^{[1]}(\1 k_\g ) + {\r}^{[2]}(\1 k_\g)\right]
(P_N + P_\Delta) \, (P_N + P_\Delta) H_1^{[2]} P_N.
\equaleb\constraintqpntwobaryon
\eeqa
The split into constraints of one-baryon and two-baryon nature
is not straightforward. E.g., the term
$ Q H_1^{[1]} Q \, Q {\r}^{[1]}({\1 k}_\g) P_N $
in Eqs. \constraintqpnonebaryon and \constraintqpntwobaryon has one-baryon {\it and}
two-baryon contributions depending on
whether the pion interacts with the same nucleon on which the
e.m. charge produced it or whether it is exchanged between the nucleons.
The constraints following from Eq. \currentconservationprojection\hbss, e.g., 
the special constraints \constraintqpnonebaryon and
\constraintqpntwobaryon\hbss, are still quite general for the
employed force model.

Compared with the coupled-channel current model of Ref. \cite{stru},
the employed current also has a coupling to pionic states, i.e.,
$Q j^\m({\1 k}_\g) (P_N + P_\D)$. In contrast to the employed
hadronic interaction  there is coupling from the nucleonic to the pionic
sector due to  $Q j^\m({\1 k}_\g) P_N $.
Since the hadronic interaction treats
the pion degree of freedom explicitly, there is  a competition between
the reducible two-baryon contribution due to the explicit e.m.
mechanism for the pionproduction
$(P_\D + Q) H_1^{[1]} Q g_0^Q(z) Q j^{[1] \m}({\1 k}_\g) P_N$
and the corresponding irreducible one-baryon and two-baryon currents
$ ( P_\Delta + Q) \left[ j^{[1] \m}({\1 k}_\g) + j^{[2] \m}({\1 k}_\g)
\right] P_N $. E.g., parts of traditional
exchange currents are resolved into simpler building blocks
due to the explicit propagation of the pion, in the same way
as parts of the three-nucleon force and of the two-nucleon
current become reducible due to the explicit propagation of the
$\D$-isobar. In the considered force model, the purely nucleonic currents
$ P_N  \left[ j^{[1] \m}({\1 k}_\g) + j^{[2] \m}({\1 k}_\g) \right] P_N $
remain unaltered and irreducible. That property gets changed in a force model
with an explicit $\pi$NN vertex $\left[3 - 6, 18, 19 \right]$, which therefore has different e.m.
properties. The competition between the e.m. mechanism of pion
production and of exchange currents of pion range is the further
theme of this section, a general conceptual problem to be respected
in actual parameterizations of the current.
\vspace{0.5truecm}

\newpage
\begin{center}
{\large\bf 2.2.3 Construction of Interaction-Dependent Contributions \\
to the Current}
\end{center}
\vspace{0.3truecm}

Appendix B gives the example of a coupled-channel current which satisfies
the constraints of Subsect. 2.2.2 exactly. However, that example is
purely phenomenological. In contrast, this subsection develops a current
with physics content.
The current is supposed to describe  e.m. processes with the inclusion
of pionproduction and absorption off nucleonic states.
Only the parts $(P_N + P_\D + Q )[ j^{[1]\m}({\1 k}_\g) + j^{[2]\m}({\1 k}_\g)] P_N$
of the current are therefore required.
The two-baryon part  $j^{[2]\m}({\1 k}_\g)$ is interaction-dependent.
However, also the one-baryon current $j^{[1]\m}({\1 k}_\g)$
contains an interaction-dependent part, $j^{[1]\m}_{exchange}({\1 k}_\g)$,
besides the interaction-free one, $j^{[1]\m}_{bare}({\1 k}_\g)$, i.e.,
\beq
j^{[1]\m}({\1 k}_\g) =	j^{[1]\m}_{bare}({\1 k}_\g) +
j^{[1]\m}_{exchange}({\1 k}_\g)\,.
\eeq
The one- and two-baryon interaction-dependent currents are derived
in the framework of the extended $S$-matrix method \cite{adam2,sauerbu}.
They will be defined up to first order in the potentials
$H_1$.
Thus, we have to expect that also the continuity equation for the current
can only be satisfied up to this order.

\vspace{1.cm}

\begin{center}
{\large\bf A. One-Baryon Interaction-Dependent Current }
\end{center}
\vspace{0.3truecm}

The one-baryon current for the use in Schr\"odinger theory is defined.
Its interac-\\tion-dependent part depends on the one-baryon potentials
$P_\D H_1^{[1]} Q$, $Q H_1^{[1]} P_\D$ and $Q H_1^{[1]} Q$ of Figs. 2.1(e)
and 1(g) employed for the description of pion-nucleon scattering.
Those potentials are derived from a field-theoretic Lagrangian ${\cal L}_I$
which couples the same channels through
$P_\D {\cal L}_I Q$, $Q {\cal L}_I P_\D$ and $Q {\cal L}_I Q$;
the potentials are defined to be instantaneous and to act between on-mass-shell
particles.
\nfii shows all one-baryon Feynman processes
up to first order in the interaction which contribute to the
effective current.
The effective field-theoretic current without interaction is denoted
by
   $J^{[1]\m}_{bare}( k_\g)$,
				     those with interaction by
   $J^{[1]\m}_{exchange}( k_\g)$.

The interaction-free current $J^{[1]\m}_{bare}( k_\g)$
contains the standard purely nucleonic contribution
    $j^\m_{NN}( k_\g)$,
the transition contribution
    $j^\m_{\D N}( k_\g)$
from the nucleon to the $\D$-isobar according to Ref. \cite{weberare} 
and the Born current
    $j^\m_{B}( k_\g)$
which connects nucleonic states with pion-nucleon states.
Fieldtheoretically, the Born current sums up five different
processes which \nfiii makes explicit;
however, the employed interaction model does not use an
explicit $\pi$NN-vertex as mechanism for pion-production
and pion-absorption, i.e.,
    $P_N H_1^{[1]} Q = Q H_1^{[1]} P_N = 0$;
thus, no part of the Born current is reducible into simpler building
 blocks, e.g., into a purely e.m. and  hadronic one.
The Born current will therefore be used as an irreducible
one-baryon current.
It is proportional to the $\p$NN coupling constant
or other meson nucleon-nucleon coupling constants,
but it
is considered to be interaction-free, a deliberately
chosen wording, which should not be misleading.
In the interaction-dependent processes (d) - (f) of \nfii  all
intermediate particles propagate covariantly and off-mass-shell.
The processes are considered of first order in the interaction.
If the pion-nucleon interaction in process (f) is described by
a field-theoretic four-point vertex, process (f) is manifestly
first-order in the interaction Lagrangian ${\cal L}_I$;
if it were defined by meson exchange, it were second order
in ${\cal L}_I$; even in the latter case we shall consider
process (f) as of first order in the interaction and lump
it together with processes (d) and (e).
The amplitudes for the effective field-theoretic current depend
on the four-momentum transfer $ k_\g$;
they are assumed to satisfy the continuity equation individually, i.e.,

\beqas
&& k_{\g \m}J^{[1]\m}_{bare}( k_\g)  = 0 \, ,\equaaleb\eodina\\
&& k_{\g \m}J^{[1]\m}_{exchange}( k_\g) = 0. \equaaleb\eodinb
\equaleb\eodint
\eeqas

As the one-baryon potential $H^{[1]}_1$, the
interaction-free current
   $j^{[1]\m}_{bare}({\1 k}_\g) $
is defined for the purpose of Schr\"odinger theory by the
corresponding field-theoretic processes
between on-mass-shell states, i.e.,  by the processes
(a) -- (c) of \nfii
with amplitude
   $J^{[1]\m}_{bare}( k_\g) $,

\beqas
&&\langle{\1 k}'_f|j^{[1]\m}_{bare}({\1 k}_\g)|{\1 k}_N \rangle
=
\langle{\1 k}'_f|J^{[1]\m}_{bare}({k'_f - k_N})|{\1 k}_N \rangle ,
\equaaleb\edvaa \\
&&{\1 k}_\g \cdot (P_N + P_\Delta + Q)j^{[1]\m}_{bare}({\1 k}_\g)P_N
=
(P_N + P_\Delta + Q)[H_0^{[1]},\rho^{[1]}_{bare}({\1 k}_\g)] P_N ,
 \quad	\quad  \quad \equaaleb\edvab
\eeqas
with ${\1 k}_\g = {\1 k'}_f - {\1 k}_N$.
In  Eq. \edvaa ${\1 k}'_f$ denotes the total on-mass-shell
momentum in the final state.
Thus,  ${\1 k}'_f$
is
${\1 k}'_N$, ${\1 k}'_\D$
or
${\1 k}'_\pi + {\1 k}'_N$,
respectively, depending on the final channel;
since initial and final particles are on their mass shells, i.e.,
$ k^0_\g =  k'^0_f -  k^0_N$, $k^0$ being the relativistic 
on-mass-shell energy,
the defined interaction-free current depends on the three-momentum
transfer
${\1 k}_\g$
only.
The noncovariant continuity equation \edvab follows from the
field-theoretic one  \eodina for
$J^{[1]\m}_{bare}( k_\g) $.

\nfiv lists all one-baryon processes up to first order in the
noncovariant potentials
$Q H^{[1]}_1 P_\D$,
$P_\D H^{[1]}_1 Q$
and
$Q H^{[1]}_1 Q$
which a description in terms of Schr\"odinger theory yields.
There are three different types of processes.
Their sum is required to exactly account for the one-baryon
Feynman processes described by \nfii, i.e.,
\beqa
&&\langle{\1 k}'_f|j^{[1]\m}_{bare}({\1 k}_\g) +
		   j^{[1]\m}_{exchange}({\1 k}_\g) +
		   J^{[1]\m}_{oms}({\1 k}_\g)
|{\1 k}_N \rangle  =	    \ncr
&&\langle{\1 k}'_f|J^{[1]\m}_{bare}({k'_f - k_N}) +
		   J^{[1]\m}_{exchange}({k'_f - k_N})
|{\1 k}_N \rangle \,\, .
\equaleb\enew
\eeqa
The above requirement defines the interaction-dependent part
$ j^{[1]\m}_{exchange}({\1 k}_\g)$
of the one-baryon current.
\nfv makes that definition explicit.
According to  Eq. \edvaa the interaction-free current
$j^{[1]\m}_{bare}({\1 k}_\g)$,
contained in the first three processes of \nfiv\hbss\,,
is chosen to account for the corresponding first three
Feynman processes of \nfii\hbss.
Processes (d) -- (f) in \nfiv are the second type of
processes;
they stand for the effective current
$J^{[1]\m}_{oms}({\1 k}_\g) $,
first order in the noncovariant potentials.
Though these processes of \nfiv look identical to the corresponding
Feynman processes (d) - (f) of \nfii\hbss, they cannot
account for them in full:
In contrast to the Feynman processes, the building blocks of $J^{[1]\m}_{oms}({\1 k}_\g)$, the
noncovariant potentials and the interaction-free current
$j^{[1]\m}_{bare}({\1 k}_\g)$,
are defined between on-mass-shell states; furthermore, the intermediate
states are of positive energy and propagate on-mass-shell,
but off-energy-shell according to the global propagators
of Schr\"odinger theory.
Since the effective current
$J^{[1]\m}_{oms}({\1 k}_\g)$ with on-mass-shell (oms) intermediate states
cannot account for the corresponding Feynman processes in full,
a correction
$j^{[1]\m}_{exchange}({\1 k}_\g) $
is required.
Thus, it is defined according to Eq. \enew  by
\beq
\langle{\1 k}'_f|
j^{[1]\m}_{exchange}({\1 k}_\g)
|{\1 k}_N \rangle
: =
\langle{\1 k}'_f|
J^{[1]\m}_{exchange}( k_f' - k_N) |{\1 k}_N \rangle  -
\langle{\1 k}'_f|
J^{[1]\m}_{oms}({\1 k}_\g) |{\1 k}_N \rangle  .
\eeq
It is represented also by the processes (a) -- (c)  in \nfiv\hbss;
they can graphically not be differentiated from the interaction-free
ones
$j^{[1]\m}_{bare}$,
since the external parameters are the same,
but their dynamic content is entirely different.
They are the third type of processes in \nfiv\hbss.
Since the divergence of the effective current
$J^{[1]\m}_{oms}({\1 k}_\g) $
can easily be shown to be
\beq
\langle{\1 k}'_f|
({k'_f - k_N})_\m
		   J^{[1]\m}_{oms}({\1 k}_\g)
|{\1 k}_N \rangle  =
\langle{\1 k}'_f|
[H_1^{[1]},\rho^{[1]}_{bare}({\1 k}_\g)]|{\1 k}_N \rangle ,
\eeq
the continuity equation for the complete one-baryon current
connecting nucleonic states with all other sectors of Hilbert space
is derived from the following sequence of steps:

\beqa
&&{\1 k}_\g \cdot
\langle{\1 k}'_f|
		 {\1 j}^{[1]}_{bare}({\1 k}_\g)  +
		 {\1 j}^{[1]}_{exchange}({\1 k}_\g)
|{\1 k}_N \rangle			 \ncr
&& = \langle{\1 k}'_f|
[H_0^{[1]},\rho^{[1]}_{bare}({\1 k}_\g)]
 -  ({k'_f - k_N})_\m  j^{[1]\m}_{exchange}({\1 k}_\g)
 +  ({k'_f - k_N})^0     \rho^{[1]}_{exchange}({\1 k}_\g)
|{\1 k}_N \rangle		  \ncr
&& = \langle{\1 k}'_f|
[H_0^{[1]},\rho^{[1]}_{bare}({\1 k}_\g)]
 +  ({k'_f - k_N})_\m  J^{[1]\m}_{oms}({\1 k}_\g)
 +  [H_0^{[1]},\rho^{[1]}_{exchange}({\1 k}_\g)]
|{\1 k}_N \rangle			  \ncr
&& = \langle{\1 k}'_f|
[H_0^{[1]} + H_1^{[1]},\rho^{[1]}_{bare}({\1 k}_\g) +
\rho^{[1]}_{exchange}({\1 k}_\g)]
 -  [H_1^{[1]},\rho^{[1]}_{exchange}({\1 k}_\g)]
|{\1 k}_N \rangle \quad
\equaleb\etri
\eeqa
The last term on the right-hand side of Eq. \etri is of second order
in the potential.
Thus, the continuity equation is satisfied up to first order in the
potential, i.e.,
\beqa
&&{\1 k}_\g \cdot (P_N + P_\Delta + Q)
     \left( {\1 j}^{[1]}_{bare}({\1 k}_\g)  +
	    {\1 j}^{[1]}_{exchange}({\1 k}_\g)	\right) P_N = \ncr
&& (P_N + P_\Delta + Q)
[H_0^{[1]} + H_1^{[1]},\rho^{[1]}_{bare}({\1 k}_\g) +
\rho^{[1]}_{exchange}({\1 k}_\g)] P_N +
    {\cal O}\left[(H_1^{[1]})^2\right]\,. \quad
\equaleb\etria
\eeqa
The definition of the exchange part
$ j^{[1]\m}_{exchange}({\1 k}_\g) $
of the one-baryon current and the proof of validity of the continuity
equation \etria for the one-baryon current rests on the current conservation
\eodint
for the amplitudes of the corresponding Feynman processes.
We called that condition an assumption;
it has always to be demonstrated that it holds for the form
of the field-theoretic interaction Lagrangian ${\cal L}_I$ and
of the field-theoretic currents
$j^{\m}_{NN}( k_\g) $,
$j^{\m}_{\D N}( k_\g) $ and
$j^{\m}_{B}( k_\g) $ employed.
Furthermore, the kinetic energy operator
$H_0^{[1]}$ used for the calculation of the relativistic Feynman amplitudes
and for the noncovariant Schr\"odinger  theory has to be the same, i.e.,
it has to be based on the relativistic on-mass-shell relation between energy
and three-momentum;
the noncovariant potentials are to be defined through the field-theoretic 
Lagrangian ${\cal L}_I$.

\vspace{0.5truecm}
\begin{center}
{\large\bf B. Two-Baryon Interaction-Dependent Current }
\end{center}
\vspace{0.3truecm}

We use the same strategy for the definition of the two-baryon
exchange current as for the definition of the one-baryon
exchange current.
We only choose a different presentation.

\nfvi shows all connected but reducible processes which arise for
the two-baryon current in the noncovariant framework
of Schr\"odinger theory and which are based on the one-baryon
current and the interaction $ H_1 $ up to first order.
The employed one-baryon current is the bare one, i.e.,
		 $ j^{[1]}_{bare}({\1 k}_\g) $,
since the one-baryon exchange part
		 $ j^{[1]}_{exchange}({\1 k}_\g) $
is itself already of first
order in the interaction.
The processes define the amplitudes
$(P_N + P_\Delta + Q) J^{[2]\m}_{oms}({\1 k}_\g) P_N$
which satisfy the condition
\beqa
&&
\langle {\1 k}'_{f_1} {\1 k}'_{f_2} |
(k'_{f_1} +  k'_{f_2} -  k_{N_1} - k_{N_2} )_\m
	   J^{[2]\m}_{oms}({\1 k_\g})
	  |{\1 k}_{N_1} {\1 k}_{N_2} \rangle   =   \ncr
&& \langle {\1 k}'_{f_1} {\1 k}'_{f_2} |
[H_1^{[1]} + H_1^{[2]},\rho^{[1]}_{bare}({\1 k}_\g) ]
	  |{\1 k}_{N_1} {\1 k}_{N_2} \rangle .
\eeqa
With respect to the commutator
$[H_1^{[1]} ,\rho^{[1]}_{bare}({\1 k}_\g) ]$,
only its part which is of two-baryon nature contributes.
We choose an ordering scheme in powers of the interaction
$ H_1 $.
However, one has to admit,
that this ordering scheme treats processes which physically belong together
differently. \nfvii demonstrates this fact:
Process (a) of \nfvii is of first order in $H_1$,
and therefore is included in the amplitudes
$ J^{[2]\m}_{oms}({\1 k_\g})$,
whereas process (b) is of second order in $H_1$ and is
therefore not considered there.
We remind that the hadronic part of the process (b) is taken out from
the instantaneous pion exchange of process (a) in the definition of the
interaction $ H_1 $ in order to avoid double counting of physical processes.

The processes of first order in the interaction $H_1$, depicted on \nfvi\hbss, 
cannot account for all corresponding processes
$ J^{[2]\m}_{exchange}( k_\g)$
which arise in a field-theoretic description of the same order.
The field-theoretic processes are assumed to satisfy
current conservation
$ k_{\g \m}J^{[2]\m}_{exchange}(\1 k_\g) = 0 $
and are employed for a standard definition of exchange currents, i.e.,
\beqa
&& \langle {\1 k}'_{f_1} {\1 k}'_{f_2} |
	   j^{[2]\m}_{exchange}({\1 k_\g})
	  |{\1 k}_{N_1} {\1 k}_{N_2} \rangle   =   \ncr
&& \langle {\1 k}'_{f_1} {\1 k}'_{f_2} |
	   J^{[2]\m}_{exchange}(k'_{f_1} +  k'_{f_2} -  k_{N_1} - k_{N_2} )
	 - J^{[2]\m}_{oms}({\1 k_\g})
	  |{\1 k}_{N_1} {\1 k}_{N_2} \rangle .
\eeqa
Though that definition of exchange currents is standard, the resulting
currents are not, since the underlying hadronic interaction is also not
standard due to its coupled-channel nature and due to the treatment of the pion as
an explicit degree of freedom.
We do not list many field-theoretic processes in diagrams and give
only one example in
\nfviii\hbss. It is argued traditionally, that the field-theoretic
processes (a) - (d) of \nfviii are not accounted for by the
Schr\"odinger processes (e) and (f) in full.
That difference defines the traditional two-body exchange current
$\langle {\1 k}'_{N_1} {\1 k}'_{\D_2} |
	 j^{[2]\m}_{traditional}({\1 k_\g})
	|{\1 k}_{N_1} {\1 k}_{N_2} \rangle  $
with $\D$-isobar exitation of the second nucleon.
However, the employed hamiltonian treats pionproduction and pion
absorption explicitly.
Thus, process (g) of \nfviii has to be taken out from the traditional
exchange current in an energy-independent way.
The appropriate definition of that
particular part of the exchange current is
\beqa
& & \langle {\1 k}'_{N_1} {\1 k}'_{\D_2} |
	j^{[2]\m}_{exchange}({\1 k_\g})
       |{\1 k}_{N_1} {\1 k}_{N_2} \rangle
	: =	     \ncr
& & \langle {\1 k}'_{N_1} {\1 k}'_{\D_2} |
	j^{[2]\m}_{traditional}({\1 k_\g}) - \ncr
& &
P_\D H_1^{[1]} Q    {Q\0
			k^0_{N1} + k^0_{N2}
 + i0 -  Q H_0^{[1]} Q }
	Q j^{[1]\m}_{bare}({\1 k_\g}) P_N
       |{\1 k}_{N_1} {\1 k}_{N_2} \rangle \, .
\equaleb\echet
\eeqa
In the definition \echet the one-baryon current
$ j^{[1]\m}_{bare}({\1 k_\g }) $
acts on nucleon one, baryon interchange has to be added explicitly.
This nonstandard definition of the interaction-dependent two-baryon currents
solves the problem of internal consistency
between the employed hadronic and e.m. mechanisms of pion
production and absorption and the irreducible\, e.m. pion-exchange current.
The definition of the\, e.m. exchange current is identical
in spirit to the definition of the two-nucleon potential
$P_N H_1 P_N$ in the interaction of Eq. (2.1) according to Ref. \cite{poe}:
Compared with a traditional potential
$V_{NN}$, $P_N H_1 P_N$ is not allowed to contain those
processes which the coupled-channel
description adds explicitly to the effective two-nucleon interaction.
Defining $P_N H_1 P_N$ in terms of a traditional
well-defined potential
$V_{NN}$,
those processes have to be taken out from
$V_{NN}$
in an energy-independent form.

The continuity equation for the two-baryon exchange current
takes the form
\beqa
&&\hspace*{-1.5cm}{\1 k}_\g \cdot (P_N + P_\Delta + Q)
	    {\1 j}^{[2]}_{exchange}({\1 k}_\g)	 P_N = \ncr
&&\hspace*{-1.5cm} (P_N + P_\Delta + Q)
[H_0^{[1]} + H_1^{[1]}+ H_1^{[2]} ,\rho^{[1]}_{bare}({\1 k}_\g) +
\rho^{[1]}_{exchange}({\1 k}_\g)] P_N + {\cal O}\left[(H_1  )^2\right] . 
\eeqa
Again as expected, it is satisfied up to order $(H_1)^2$ only.

The definition of the interaction-dependent one-baryon and
two-baryon currents is so general that e.m. and hadronic form factors 
can be added without problems.
On the other hand, we have to warn the reader in one respect:
The current model with its nonstandard parts defined in this subsection
could  not yet been used for the actual calculation of photo reactions in this
paper.
The current used here is of traditional form; its parameterization will be given
in Sect.\,\, 3.

\vspace{0.5truecm}
\begin{center}
{\large\bf 2.3 Photo Reactions in the Two-Nucleon System}
\end{center}
\vspace{0.3truecm}

This section describes processes with real photons in the two-nucleon
system whose dynamics is controlled by the force model of Subsect.\,\, 2.1.
Examples for the considered reactions are the photo absorption
processes on the deuteron, i.e.,
$\g$d$\rar$pn, $\g$d$\rar$$\p^0$d and $\g$d$\rar$$\p$NN, and
nucleon-nucleon bremsstrahlung without and with pionproduction, i.e.,
NN$\rar$NN$\g$, pn$\rar$d$\g$, NN$\rar \pi$d$\g$,
NN$\rar \pi$NN$\g$, $\pi$d$\rar$NN$\g$, $\pi$d$\rar \pi$d$\g$ and
$\pi$d$\rar \pi$NN$\g$. Only the first bremsstrahlung reaction is traditionally
measured.

Since the  energy balance is different in the absorption of a photon
with energy $\o_\g$ and in the emission of a photon with the same
energy $\o_\g$, both types of reactions have to be described
slightly differently. As for the hadronic reactions of Subsect.\,\, 2.1,
an e.m. multi-channel transition matrix $U^{\g A}(z)$ for the
photo absorption processes and the corresponding
transition matrix $U^{\g E}(z)$
for bremsstrahlung processes are introduced. They connect initial
and final hadronic channel states $\ket{\Ph_i}$ and $\ket{\Ph_f}$.
The channel states are those of Subsect.\,\, 2.1, i.e.,
the two-nucleon states $\ket{\Ph_N({\bf p}_N)}$,
the pion-deuteron states $\ket{\Ph_\pi({\bf q}_\pi)}$ and
the break-up states $\ket{\Ph_0 ({\bf p, q})}$ with two free nucleons
and one pion. The deuteron state
$\ket{\Ph_d({\bf k}_d)} = \ket{ d \, {\bf k}_d}$
with total momentum ${\bf k}_d$ and energy $E_d({\bf k}_d)$ has to be
added to that list of initial and final channel states; for that state
we make the formal identification $\ket{\Ps_d^{(\pm)}({\bf k}_d)} = $
$\pm i0 G(E_d \pm i0) \ket{\Ph_d({\bf k}_d)} $
$= \ket{\Ph_d({\bf k}_d)}$ for later ease in notation.
The $S$-matrix for photo absorption and photo emission is related
to the\, e.m.\, multi-channel transition matrices $U^{\g A}(z)$
and $U^{\g E}(z)$ by
\beqas
\bra{\Ph_f}S\ket{\Ph_i \g} &=& -2\p i \d (E_f - E_i - \o_\g )
 \bra{\Ph_f} U^{\g A}_{fi}(E_i + \o_\g +i0 )\ket{\Ph_i},
\equaleb\smatrixfig \equaaleb\smatrixfiga\\ 
			   &=& -2\p i \d (E_f - E_i - \o_\g )
 \bra{\Ps_f^{(-)}} H_1^{\g }\ket{\Ps_{i}^{(+)}},
\equaaleb\smatrixfigb
\eeqas

\beqas
\hspace{-.9cm}\bra{\Ph_f \g}S\ket{\Ph_i} &=& -2\p i \d (E_f + \o_\g  - E_i)
 \bra{\Ph_f} U^{\g E}_{fi}(E_i + i0)\ket{\Ph_i}\lb{ugs}, 
\equaleb\smatrixfgi \equaaleb\smatrixfgia\\
			   &=& -2\p i \d (E_f + \o_\g - E_i)
\bra{\Ps_f^{(-)}} H_1^{\g }\ket{\Ps_{i}^{(+)}}.\lb{hgs}
\equaaleb\smatrixfgib
\eeqas

The \,e.m. transition matrix elements are completely determined by
the hadronic and e.m. interactions of Eqs. (2.1) and (2.9). Eqs. \smatrixfigb
and \smatrixfgib give the on-shell elements in one-photon exchange
approximation. Their general off-shell elements have the same
structure as the on-shell elements of Eqs. \smatrixfig and \smatrixfgi which are
determined by the M{\emp}ller  operators for the step from
channel states to correlated wave functions according to
Eqs. (2.6) - (2.8). Thus, the on-shell e.m. multi-channel
transition matrices can be given in terms of the off-shall hadronic
multi-channel transition matrix $U(z)$ and the hamiltonian
$H_1^\g$ for the e.m. interaction. E.g., the e.m. multi-channel
transition matrix $U^{\g A}_{f d}$ for the absorption of
real photons by the deuteron has the following form

\beq
U^{\g A}_{Nd} (z)  =  \sum_{a=N, \D} (\d_{Na} + U_{Na} (z) g^P_{a0} (z) P_a
H_1^\g P_N ) + U_{N0} (z) g^Q_0 (z) Q H_1^\g P_N, \label{und}\\
\eeq
\beq
\hspace{-2.7cm}U^{\g A}_{\p d} (z)  =  \sum_{a=N, \D} U_{\p a}(z) g^P_{a0} (z) P_a
H_1^\g P_N  + U_{\p 0} (z) g^Q_0 (z) Q H_1^\g P_N,  \label{upd}\\
\eeq
\beq
\hspace{1.13cm}U^{\g A}_{0 d} (z)  =  \sum_{a=N, \D} U_{0 a}(z) g^P_{a0} (z) P_a
H_1^\g P_N  + (1+U_{00} (z) g^Q_0 (z)) Q H_1^\g P_N,  \label{uod}\\
\eeq

\noindent
i.e., for the processes
$\g$d$\rar$pn, $\g$d$\rar$$\p^0$d and $\g$d$\rar$$\p$NN in turn.
The derivation of Eqs. \req{und} - \req{uod} exploits the
special structure of the force
model according to Subsect.\,\, 2.1: The deuteron is a purely nucleonic state; it
does not have any  components in the Hilbert sectors
$\H_\D$ and $\H_\p$.


\def\H{{\cal H}}
\def\pp{{\1 p}'}
\def\pq{{\1 q'}}
\def\ho{H_1}
\def\hog{H_1^\g}
\def\ron{\r^{[1]}}
\def\jtw{{\1 j}^{[2]}}
\setcounter{section}{3}
\setcounter{equa}{0}

\vspace{0.7truecm}
\begin{center}
{\Large\bf 3. Parameterization of the Hamiltonian}
\end{center}
\vspace{0.4truecm}

We report our first calculations on photo reactions of the deuteron.
We extend early attempts \cite{uwethesis,sauerbu} of ours.
Our calculations are still incomplete and do not incorporate all
subtleties of those of Ref. $[9-12]$, but they have
special features compared to $[9-13]$ which should make them
interesting to others already in their present stage.

The parameterization of the kinetic energy operator $H_0$, of the
hadronic interaction hamiltonian $H_1$ and of the
e.m. interaction hamiltonian $H_1^\g$ are given in this section.

\vspace{0.6truecm}
\begin{center}
{\large\bf 3.1 The Kinetic Energy Operator $H_0$}
\end{center}
\vspace{0.3truecm}

The single-particle masses and momenta are denoted by $m_i$ and
${\bf k}_i$ with ${\rm i = N}, \D, \pi$ and $\g$.
The single-particle energies of the nucleon and the $\D$-isobar
$e_N (\1 k_N )$ and $e_\D (\1 k_\D )$ are assumed to be nonrelativistic, i.e.,
\beqas
e_N (\1 k_N ) &=& m_N + {\1 k_N^2\0 2 m_N}, \\
e_\D (\1 k_\D ) &=& m^0_\D + {\1 k_\D^2 \0 2 m^0_\D}.
\eeqas
The exact relativistic forms are used for the single-particle energies
of the pion and the photon $\o_\p (\1 k_\p )$ and $\o_\g (\1 k_\g )$, i.e.,
\beqas
\o_\p (\1 k_\p )&=& \sqrt{m_\p^2 + \1 k_\p^2}, \\
\o_\g (\1 k_\g )&=& |\1 k_\g|.
\eeqas
The chosen values of the masses are $m_N = 939$ MeV,
$m^0_\D = 1315.8$ MeV and
$m_\p = 139$ MeV.
The value for $m^0_\D$ was already chosen in the context of Eq. (2.2) such that
the interaction hamiltonian (2.1) describes pion-nucleon scattering
in the $P_{33}$ partial wave.

The calculations are carried out in the c.m. system, in which
the sum of all single-particle momenta ${\1 k}_i$ is zero.
Thus, Jacobi momenta are introduced for the
relative motion between particles. The kinetic energies of the
considered channels is reexpressed in terms of those momenta.

The kinetic energy of states in two-baryon channels is for two
nucleons characterized by
\beqas
P_N H_0 P_N \ket{ {\1 p}_N} &=& E_N (\1 p_N ) \ket{{\1 p}_N},\\
{\1 p}_N &=& {1\0 2} ({\1 k}_{N1} - {\1 k}_{N2} ),\\
{\1 K} &=& {\1 k}_{N1} + {\1 k}_{N2} = 0
\eeqas
with $E_N (\1 p_N ) = 2 e_N (\1 p_N )$, in case of nucleon-$\D$ states by
\beqas
P_\D H_0 P_\D \ket{ {\1 p}_\D} &=& E_\D (\1 p_\D ) \ket{{\1 p}_\D},\\
{\1 p}_\D &=& { m^0_\D {\1 k}_{N} - m_N {\1 k}_{\D}\0 m_N + m^0_\D}, \\
{\1 K} &=& {\1 k}_{N} + {\1 k}_\D = 0
\eeqas
with $E_\D (\1 p_\D ) = e_N (\1 p_\D ) + e_\D (\1 p_\D )$ . The basis states
$\ket{\1 p_N}$ and $\ket{\1 p_\D}$ are antisymmetrized plane-wave
states; $\ket{{\1 p}_N}$ coincides with the channel state
$\ket{\Ph_N({\bf p}_N)}$
of Subsect.\,\, 2.1.

In the three-particle Hilbert sector with two nucleons and one
pion, two different sets of Jacobi momenta are used. Either
the pion or one of the nucleons is chosen
to be the spectator. The corresponding momenta are distinguished
by the label of the spectator particle, i.e.,
\beqas
Q H_0 Q \ket{\1 p_\p \1 q_\p} &=& E_0 (\1 p_\p \1 q_\p )\ket{\1 p_\p \1 q_\p},\\
& & \ncr
\1 p_\p &=& {1\0 2} (\1 k_{N1} - \1 k_{N2} ),\\
& & \ncr
\1 q_\p &=& {\o_\p (\1 k_\p )(\1 k_{N1} +\1 k_{N2} ) -2m_N \1 k_\p \0
	    2 m_N + \o_\p (\1 k_\p )}, \\
& & \ncr
{\1 K} &=& {\1 k}_{N1} + {\1 k}_{N2} + {\1 k}_\p = 0,
\eeqas
with $E_0 (\1 p_\p \1 q_\p ) = 2 e_N (\1 p_\p ) + {\1 q_\p^2/4 m_N} + \o_\p (\1 q_\p )$,
and 
\beqas
Q H_0 Q \ket{\1 p_{N1} \1 q_{N1}} &=& E_0 (\1 p_{N1} \1 q_{N1} )
\ket{\1 p_{N1} \1 q_{N1}}, \\
& & \hfill \ncr
\1 p_{N1} &=& {\o_\p (\1 k_\p ) \1 k_{N2} - m_N \1 k_{\p}\0 m_N + \o_\p
(\1 k_\p )},
\\
& & \ncr
\1 q_{N1} &=& {m_N(\1 k_{N2} +\1 k_{\p} ) -(m_N + \o_\p (\1 k_\p ) )\1 k_{N1} \0
	    2 m_N + \o_\p (\1 k_\p )}, \\
& & \ncr
{\1 K} &=& {\1 k}_{N1} + {\1 k}_{N2} + {\1 k}_\pi = 0
\eeqas
with $E_0 (\1 p_{N1} \1 q_{N1} ) \approx e_N (\1 p_{N1} ) + e_N (\1 q_{N1} )
+ {\1 q_{N1}^2/2(\o_\p (\1 p_{N1})+ m_N}) + \o_\p (\1 p_{N1} )$. The basis
states $\ket{ \1 p_\pi \1 q_\pi} $ and
$\ket{ {\1 p}_{N1} {\1 q}_{N1}}$ are identical with the break-up
channel states $\ket{\Ph_0({\1 p} {\1 q})}$ of Subsect.\, 2.1.

The pion-deuteron channel state $\ket{\Ph_\pi({\1 q}_\pi)}$
of Subsect.\, 2.1 can be expanded in terms of the basis states
$\ket{{\1 p}_\pi {\1 q}_\pi}$; it has the energy
$E_{\p } (\1 q_{\p }) = 2 m_N +\e_d +{\1 q_{\p }^2 / 4 m_N}
+\o_{\p }(\1 q_{\p })$ with $\e_d$ denoting the binding energy
of the deuteron. The deuteron state $\ket{{\Ph}_d({\1 k}_d)} =
\ket{ d \, {\1 k}_d}$ of
Subsect.\, 2.3 is the tensor product of an internal wave function $\ket{d}$
expanded in terms of the basis states $\ket{{\1 p}_N}$ and of a two-nucleon c.m.
state $\ket{ {\1 k}_d}$; it has the energy
$E_d(\1 k_d) = 2 m_N +\e_d +{\1 k_d^2 / 4 m_N}$.

\vspace{0.7truecm}
\begin{center}
{\large\bf 3.2 Parameterization of the Hadronic Interaction $H_1$}
\end{center}
\vspace{0.3truecm}

The hadronic interaction (2.1) employed
is mainly taken from Ref. \cite{poe}.
The parameterization of the $\p$N$\D$-vertex in Fig. 2.1(e)
is unchanged compared with Ref. \cite{poe}.
The two-nucleon potential $P_N H_1 P_N$ of Fig. 2.1(a) is also employed
in the form of Ref. \cite{poe}, i.e., approximately phase-equivalent
with the Paris potential \cite{paris}.
The transition potential $P_\D H_1 P_N$ of Fig. 2.1(b) from two-nucleon to
nucleon-$\D$ states is
based on pion- and rho-exchange, i.e.,
\beq
P_\D H_1 P_N = V_{\D N} (\p ) + V_{\D N} (\r )
\eeq
with
\beqas
&&\bra{\1 p'_\D} V_{\D N} (\p ) \ket{\1 p_N}=\ncr
&&{(-)\0 (2\p )^3}
{f_{\p NN}\0 m_\p}\left({\L_{\p N}^2 -m_\p^2\0\L_{\p N}^2 + (\1 p'_\D -\1 p_N )^2}\right)
{f_{\p N\D}\0 m_\p}\left({\L_{\p \D}^2 -m_\p^2\0\L_{\p \D}^2
 + (\1 p'_\D -\1 p_N)^2}\right)\ncr
 &&\bm	\t (1)	\bm \t_{\D N} (2) {[\bm \s (1)\cdot (\1 p'_\D -\1 p_N )]
			       [\bm\s_{\D N} (2) \cdot (\1 p'_\D -\1 p_N )]
			      \0 m_\p^2 + (\1 p'_\D -\1 p_N )^2},\\
&&\bra{\1 p'_\D} V_{\D N} (\r ) \ket{\1 p_N}=\ncr
&&{(-)\0 (2\p )^3}
{f_{\r NN}\0 m_\r}\left({\L_{\r N}^2 -m_\r^2\0\L_{\r N}^2 + (\1 p'_\D -\1 p_N )^2}\right)
{f_{\r N\D}\0 m_\r}\left({\L_{\r \D}^2 -m_\r^2\0\L_{\r \D}^2 + (\1 p'_\D -\1 p_N
)^2}\right)\ncr
&&\bm \t (1)  \bm \t_{\D N} (2) {[\bm \s (1)\times (\1 p'_\D -\1 p_N )]
			       [\bm\s_{\D N} (2) \times (\1 p'_\D -\1 p_N )]
			      \0 m_\r^2 + (\1 p'_\D -\1 p_N )^2}
\eeqas
where ${\bm \s} ({\bm \s_{\D N}})$ and ${\bm \t} ({\bm \t_{\D N}})$ are the nucleonic (N$\D$-transition) spin and isospin operators,
respectively.

This form of the transition potential differs slightly from
the one used in Refs. \cite{bulla},
the $\d$-function contribution to the unregularized forms of the
potential in configuration space is not removed in contrast to
Refs. \cite{bulla}; thus, no term
proportional to $\bm \s (1)\cdot \bm\s_{\D N} (2)$ appears in
Eqs. (3.8); that change in the transition potential is according
to Ref. \cite{bulla} immaterial except for the inelasticity
in the $^3P_0$ partial wave of two-nucleon scattering.
The parameters
\beqa
{f_{\p NN}^2\0 4\p} &=&0.08, \quad\quad {f_{\p N\D}^2\0 4\p}=0.35,\ncr
{f_{\r NN}^2\0 4\p} &=&3.21, \quad\quad {f_{\r N\D}^2\0 4\p}=9.13,\\
\L_{\p N}= \L_{\p\D}\hspace{-.2cm}&=&\hspace{-.25cm}\L_{\r N} =\L_{\r \D} =650\hspace{.2cm} {\rm MeV},
\nonumber
\eeqa
in the transition potential yield a reasonable fit \cite{bulla} to
the total proton-proton cross sections up to
500 MeV in the c.m. system.
The nucleon-$\D$ potential $P_\D H_1 P_\D$ consists of a direct and
an exchange part according to the processes (d) and (c) of Fig. 2.1.
One time ordering of the pion-mediated	exchange part, however,
is generated dynamically by the explicit $\p$N$\D$-vertex. Therefore,
only the instantaneous
approximation for the pion contribution to the remaining time ordering
is contained in $P_\D H_1 P_\D$. As in Ref. \cite{poe}, the direct part of
$P_\D H_1 P_\D$ is altogether put to zero
for most calculations.
Only the calculations of Subsect. 4.3 will go beyond this choice.
Furthermore, all interactions in the Hilbert space with one pion
are chosen to be zero, i.e., $Q H_1 Q = 0$.

\vspace{0.6truecm}
\begin{center}
{\large\bf 3.3 Parameterization of the E.M. Interaction}
\end{center}
\vspace{0.3truecm}

This section gives the actual parameterization of the current $j^{\m}(\1 k_\g)$
in its 
Fourier-transformed form and of the\, e.m. hamiltonian $H_1^\g$ used in the calculations of this paper.
The description of photo reactions from the deuteron requires only
the part
$
(P_N + P_\D +Q) \, {\1 j}({\1 k}_\g) \, P_N
$
of the full\, e.m. current. 
The actual form of the\, e.m. current  will be given by its matrixelements
in momentum space. According to Eq. (2.13) one-body and two-body parts,
$\1 j^{[1]}({\1 k}_\g)$ and $\1 j^{[2]}({\1 k}_\g)$,
will be distinguished.
The currents $ P_\Delta {\1 j^{[1]} }({\1 k}_\g) P_N$ and 
$ Q {\1 j^{[1]} }({\1 k}_\g) P_N$
describe photo pionproduction on the single nucleon.
They are well tuned in Appendix C.
Compared to the Refs. \cite{are3, are4} they are chosen energy-independently and 
therefore do not spoil the unitarity of the theory.
However, for the photo reactions of the deuteron they are employed only in
simplified form as described in this section.
Of course, the parameters found in Appendix C, are then not optimal anymore. 
This is why we shall explore the sensitivity of our results on the strength  
of the nucleon-$\D$ transition current
$ P_\Delta \1 j^{[1] }({\1 k}_\g) P_N$.

\vspace{0.6truecm}
\begin{center}
{\large\bf 3.3.1 The E.M. One-Baryon Current}
\end{center}
\vspace{0.3truecm}

For the one-nucleon part of the  current $ P_N \1 j^{[1]}({\1 k}_\g) P_N$ 
of Fig. 2.5(a) the nonrelativistic convection and spin contributions
\beqa
\bra{\1 k'_N} P_N \1 j^{[1]}(\1 k_\g )P_N \ket{\1 k_N}
 &=& \d (\1 k'_N - \1 k_N -\1 k_\g ) e_P \ncr
&&\left[ e(\bm\t_0){\1 k'_N + \1 k_N\0 2m_N} +
{\m (\bm\t_0 )\0 2 m_N} i(\bm\s\times\1 k_\g )\right]
\eeqa
are used with $e (\bm\t_0 ) = {1\0 2} (1+\bm\t_0) $
and $\m (\bm\t_0 ) = \m_s +\bm\t_0\m_v,$\,\, $e_P$ being the positive elementary charge, $ \m_s =0.44$ and $\m_v=2.35 $,
$\m_s$ and $\m_v$ being the isoscalar and isovector
magnetic moments of the nucleon.
Relativistic corrections to this single-nucleon current, especially the ones
generated by the spin-orbit term, are known to be important for the process
of $\g d\rar pn$ at low energies in forward direction \cite{weberare,friar,jaus}. However, the influence of these
additional currents on the observables is expected to be small at energies
in the regime of the $\D$-resonance \cite{weberare} and therefore are neglected
in our present parameterization.

In the single-baryon transition current
$P_\D \1 j^{[1]}({\1 k}_\g) P_N$
of Fig. 2.5(b) we keep only its dominant part due to magnetic dipole
transition. The smaller contribution arising from the electric quadrupole
transition is neglected, i.e.,
\beqa
\bra{\1 k'_\D} P_\D \1 j^{[1]}(\1 k_\g )P_N \ket{\1 k_N}
&=&
 \d (\1 k'_\D - \1 k_N -\1 k_\g ) e_P\ncr
&&(\bm \t_{\D N})_0
{G^{\D N}_{M1} \0 2 \m_{\D N}}i \left( \bm \s_{\D N}\times {m_N \1 k'_\D - m^0_\D \1 k_N\0 m_N +m^0_\D} \right).
\eeqa
Appendix C suggests  $G_{M1}^{\D N} = 3.65 $ as  optimal value;
however, we shall employ a range of values in order to explore sensitivities.
In contrast to the form of Appendix C,
the current (3.11) is not fully transverse, i.e.,
$\1 k_\g \cdot P_\D \1 j^{[1]}(\1 k_\g )P_N \neq 0. $
Here $\m_{\D N}$ denotes the reduced mass of the nucleon and
of the $\D$-isobar. 

The one-baryon part of the e.m. current $Q\1 j^{[1]}({\1 k}_\g) P_N$ in Fig. 2.5(c) describes
the Born process of photo pionproduction from the nucleon
as displayed in Fig. 2.4.
It includes $\rho$- and  $\o$-exchanges. 
Its parameterization is defined explicitly in Appendix C.
However, only its contribution to the
$P_{33}$-partial wave is kept in the present calculations,
it is kept without any rescattering there. Thus, it does not contribute
to photo disintegration in our calculations.
Retained and omitted contributions arising from the Born process are displayed
in Fig. 3.1.

\vspace{0.6truecm}
\begin{center}
{\large\bf 3.3.2 The E.M. Two-Baryon Current}
\end{center}
\vspace{0.3truecm}

The two-nucleon current $P_N \1 j^{[2]}({\1 k}_\g) P_N$
is approximated by its traditional pion contribution which we assume to be dominant.
The corresponding current
contains the pair and the pion-in-flight parts which are shown
in Fig. 3.2(a) and 3.2(b). They are used in their nonrelativistic forms~\cite{stru}:
\beqas
&&\bra{\1 k'_{N1}\1 k'_{N2}} P_N \1 j^{[2]}_{\p, pair} (\1 k_\g )
P_N\ket{\1 k_{N1} \1 k_{N2}} = i {1\0 (2\p )^3} e_P\ncr
&&\d (\1 k'_{N1} + \1 k'_{N2} - \1 k_{N1} - \1 k_{N2} -\1 k_\g )
\left({f_{\p NN}\0 m_\p }\right)^2 ( \bm\t_1\times\bm\t_2 )_0 \ncr
&&\left(
	{\L^2_{\p N} - m^2_{\p}  \0
	 \L^2_{\p N} + (\1 k'_{N1} -\1 k_{N1} )^2 }
	{\bm\s_2 [\bm\s_1\cdot (\1 k'_{N1} -\1 k_{N1} )]\0
	(\1 k'_{N1} -\1 k_{N1} )^2 +m_\p^2} \,\, - \right. \ncr
&&\left.	{\L^2_{\p N} - m^2_{\p}  \0
	 \L^2_{\p N} + (\1 k'_{N2} -\1 k_{N2} )^2 }
 {\bm\s_1 [\bm\s_2\cdot (\1 k'_{N2} -\1 k_{N2} )]\0
      (\1 k'_{N2} -\1 k_{N2} )^2 +m_\p^2} \right) \, ,    \\
&&\bra{\1 k'_{N1}\1 k'_{N2}} P_N \1 j^{[2]}_{\p, pion} (\1 k_\g )
P_N\ket{\1 k_{N1} \1 k_{N2}} = i {1\0 (2\p )^3} e_P\ncr
&&\d (\1 k'_{N1} + \1 k'_{N2} - \1 k_{N1} - \1 k_{N2} -\1 k_\g )
\left({f_{\p NN}\0 m_\p }\right)^2 (\bm \t_1\times\bm\t_2 )_0 \ncr
&&(\1 k'_{N1} - \1 k_{N1} - \1 k'_{N2} +\1 k_{N2} )
\left[
	{\L^2_{\p N} - m^2_{\p}  \0
	 \L^2_{\p N} + (\1 k'_{N1} -\1 k_{N1} )^2 } \cdot
	{\L^2_{\p N} - m^2_{\p}  \0
	 \L^2_{\p N} + (\1 k'_{N2} -\1 k_{N2} )^2 }  \right]^{1\0 2}  \ncr
&&{ [\bm\s_1\cdot (\1 k'_{N1} -\1 k_{N1} )]\0
      (\1 k'_{N1} -\1 k_{N1} )^2 +m_\p^2}
 { [\bm\s_2\cdot (\1 k'_{N2} -\1 k_{N2} )]\0
(\1 k'_{N2} -\1 k_{N2} )^2 +m_\p^2} \, ,  \qquad
\eeqas
with $\L_{\p N}= 200\hspace*{.2cm} {\rm MeV}$.
All other two-baryon currents are neglected. 

\medskip

The currents of Subsect. 3.3.1 are  bare ones in the definition
of Sect. 2.2. None of the nontraditional interaction-dependent currents
defined in Sect. 2.2 are included in the present calculations.

\vspace{0.6truecm}
\begin{center}
{\Large\bf 4. Results}
\end{center}
\vspace{0.3truecm}

Photo disintegration of the deuteron and photo pionproduction
on the deuteron are unitarily coupled reactions. A simultaneously
consistent description of both processes is required in order to be conceptually
satisfactory; such a description is given with the interactions of Sect.\, 3.

\vspace{.5cm}

\begin{center}
{\large\bf 4.1 Photo Disintegration of the Deuteron}
\end{center}

\vspace{.3cm}

We show some characteristic results of our calculations in Figs. 4.1 -- 4.3.
We give them for the kinematic region in which resonance 
production becomes important in the unitarily coupled pionic channel.
We present the
total cross-section as function of the photon lab energy
as well as the unpolarised  differential cross-section $d \sigma / d \Omega$ and
the photon asymmetry for the photon lab energies 260 MeV and 300 MeV.

The dependence of our results on the magnetic dipole strength
$G_{M1}^{\D N}$ in the nucleon-$\D$ transition current is studied. 
Results for three values are displayed. We note that the value  
$G_{M1}^{\D N} = 3.65$ suggested in Appendix C
by the single-nucleon data
yields a rather poor
agreement with experimental results. 
Of course, the strength parameter $G_{M1}^{\D N}$ has to be increased 
compared with 3.65 when other mechanisms for $\D$-isobar excitation as done in the present calculation, i.e.,
the electric quadrupole resonance production and the rescattering of the pion produced in a Born process are left out,
as explained in Subsect. 3.3.1.
The value of $G_{M1}^{\D N}$, preferable for this reaction,
is essentially suggested by Fig. 4.1, and it is
slightly above 5.16.  

\vspace{.6cm}

\begin{center}
{\large\bf 4.2 Photo Pionproduction on the Deuteron }
\end{center}

\vspace{.3cm}

We show some characteristic results of our calculations in Figs. 4.4 -- 4.6
and give them for the kinematic region in which resonance pionproduction
becomes important. 
We present 
the unpolarized differential cross-section $ d \s / d \O$
and the deuteron vector polarization $t_{11}$ for the 
photon lab energies 260 MeV and 300 MeV.

The dependence of our results on the magnetic dipole strength
$G^{\D N}_{M1} $
in  the nucleon-$\D$ transition current is studied.
Results for three values are displayed.
We note that -- in contrast to photo disintegration --
the value suggested in Appendix C seems to yield a reasonable account of the
few existing experimental data.
Of course, the strength parameter $G^{\D N}_{M1} $
has to be increased slightly when other mechanisms for pion
production  as done in the present calculation, i.e., the
electric quadrupole resonance production and the
rescattering of the pion produced in a Born process are left out as explained in Subsect.\, 3.3.1.
With respect to the small-angle results of the Fig. 4.6
the calculation does not appear fully converged with respect to the 
inclusion of two-baryon angular momentum.
The vector polarization $t_{11}$ is not sensitive
with respect to the choice of $G^{\D N}_{M1}$.


\vspace{0.6cm}

\begin{center}
{\large\bf 4.3 The Role of Irreducible Nucleon-$\D$ Potential }
\end{center}

\vspace{0.3cm}

Ref.  \cite{penahar91} noted the dependence of some observables 
for the two-nucleon system above pion threshhold on the choice of the
irreducible nucleon-$\D$ potentials of Figs. 1(c) and 1(d).
Sensitivity occured when the observables were described without such 
a potential
as done for the results of Subsects. 4.2 and 4.3, and,
alternatively, with two distinct nucleon-$\D$ potentials derived
either
from a one-boson exchange model or from a nonrelativistic quark model \cite{quarkm}.
Ref. \cite{penahar91} noted that dependence for photo reactions of the deuteron, too. 
We study it as well and show characteristic results in
Figs. 4.7 -- 4.11  for the kinematic region in which
$\D$-resonance contributions are important. 

We present the total
photo disintegration cross section as function of the photon lab energy,
the unpolarized differential cross section 
$d \s / d \O$ and the photon asymmetry $\S^\g$ 
at 260 MeV and 300 MeV photon lab energy for photo disintegration.
We also  present
the unpolarized differential cross section $d \s / d \O$
and the vector polarization $t_{11}$ for photo pionproduction
at 260 MeV and 300 MeV photon lab energy. 
Though no clear preference for one particular nucleon-$\D$ potential arises, 
we conclude: The quark-model based potential is smoother and therefore 
yields physically smoother corrections without apparently 
unphysical structures.

\vspace{0.6truecm}
\begin{center}
{\Large\bf 5. Conclusions}
\end{center}
\vspace{0.3truecm}

We are used to the fact that interaction-dependent
meson-exchange currents arise in a theoretical description of 
e.m. processes, when meson degrees of freedom are frozen.
When, however, mesons become active, the traditional definition 
of interaction-dependent currents has to be revised.
E.g., such a revision becomes necessary for\, e.m. pionproduction.
The paper gives a redefinition of interaction-dependent
currents when theoretically describing pionproduction.
We think this is the conceptual
value of the paper; its value remains, even if the quantitative significance
of the novel interaction-dependent currents could not be checked yet in
actual calculations.

The practical calculations performed employ traditional currents with well-known exchange contributions.
In contrast to Ref. \cite{are3, are4}
the currents are energy-independent thus, hermitian, in order to preserve
the unitarity of the theoretical description.
Photo disintegration of the deuteron and photo pionproduction on the deuteron
are calculated simultaneously in a consistent frame work.
The sensitivity of results on the parameters of the nucleon-$\D$ transition current
and on the form of the irreducible nucleon-$\D$ potential got well established. 
We hope that the sensitivity will allow at a later, more refined, stage
of theoretical description to learn more about the important parts of the e.m.
current and on the nucleon-$\D$ potential from photo reactions on the
deuteron.



%
%
\def\eqb{\addtocounter{equa}{1} \eqno{\enu}}
\def\enu{(\Alph{section}.\theequa )~ }
\def\equaname#1{\relax
      \global\addtocounter{equa}{1}
      \xdef#1{(\thesection.\theequa )~ }}
%
%
\def\equaleb#1{\relax
      \global\addtocounter{equa}{0}
      \xdef#1{(\Alph{section}.\theequa )~ }}
\def\eqn#1{ \equaname{#1}\eqno\enu}
\def\theequation{\Alph{section}.\theequa}
\setcounter{subeq}{0}
\def\beqas{
\def\theequation{\addtocounter{subeq}{1}
\Alph{section}.\theequa\alph{subeq}}\beqa}
\def\eeqas{\eeqa\setcounter{subeq}{0}\def\theequation{\Alph{section}.\theequa}}
\def\equaaleb#1{\relax
      \global\addtocounter{subeq}{1}
      \xdef#1{(\Alph{section}.\theequa \alph{subeq})~ }
      \global\addtocounter{subeq}{-1}                 }

\def\vm#1{|\1 #1 |}
\def\H{{\cal H}}
\def\tp{t_{\g\p}}
\def\gpa{g^P_{a0} (z)}
\def\gpb{g^P_{b0} (z)}
\def\gqp{g^Q_\p (z)}
\def\gqo{g^Q_0 (z)}
\def\pp{{\1 p}'}
\def\pq{{\1 q'}}

\setcounter{section}{1}
\setcounter{equa}{0}

\newpage
\begin{center}
{\Large\bf Appendix A: Hadronic Scattering Theory}
\end{center}
\vskip 0.3truecm

The $S$-matrix needed for describing the hadronic reactions in the
two-nucleon system above pion threshold are determined by the on-shell
elements of the multi-channel transition matrix $U(z)$. Its various components
$U_{fi}(z)$ are defined by appropriate decompositions of the full hadronic
resolvent $G(z)$ of Eq. (2.4), i.e.,
\beqa
P_b G(z) P_a &=&  \d_{ba}\gpa + \gpb U_{ba} (z) \gpa, \ncr
Q G(z) P_a &=& g^Q_\b (z) U_{\b a} (z) \gpa, \ncr
Q G(z) P_a &=& \gqo U_{0a} (z) \gpa, \ncr
P_b G(z) Q &=& \gpb U_{b\b} (z) g^Q_\b (z), \\
Q G(z) Q &=& \d_{\b\a} g^Q_\a (z) + g^Q_\b (z) U_{\b\a} (z)
 g^Q_\a (z), \ncr
Q G(z) Q &=& \gqo U_{0\a} (z) g^Q_\a (z), \ncr
Q G(z) Q &=& \gqo + \gqo U_{00} (z) \gqo. \nonumber\lb{dec}
\eeqa
In Eq. \req{dec}
\beqa
\gpa &=&{P_a\0 z - P_a H_0 P_a}, \quad	  \ncr
\gqo &=&{Q\0 z - Q H_0 Q}, \\
g^Q_\a (z) &=&{Q\0 z - Q H_0 Q - v_\a^Q} \quad
\nonumber
\eeqa
represent partial resolvents, $v_{\alpha}^Q$ being the potential
between the pair of particles $\alpha$ in the Hilbert sector
${\cal H}_\pi$ as illustrated in Figs. 2.1(f) and 2.1(g) and as
notationally
defined in Eq. (A.4). The resulting
relations between the  $S$-matrix elements and the on-shell multi-channel
transition matrix $U(z)$ are
\beqa
\bra{\Ph_N ( \pp_N)} S \ket{\Ph_N ({\1 p}_N)}&=& \d(\pp_N - {\1 p}_N)
-2\p i \d (E_N (\pp_N ) - E_N ({\1 p}_N)) \ncr &&\bra{\Ph_N (\pp_N )}
U_{NN} (E_N ({\1 p}_N )+ i0) \ket{\Ph_N ({\1 p}_N)},\ncr
\bra{\Ph_\p (\pq_\p )} S \ket{\Ph_N ({\1 p}_N)} &=&
-2\p i \d (E_\p (\pq_\p ) - E_N ({\1 p}_N)) \ncr &&\bra{\Ph_\p ({\pq}_\p )}
U_{\p N} (E_N ({\1 p}_N )+ i0) \ket{\Ph_N ({\1 p}_N)},\ncr
\bra{\Ph_0 (\pq, \pp)} S \ket{\Ph_N ({\1 p}_N)}&=&
-2\p i \d (E_0 (\pq ,\pp) - E_N ({\1 p}_N)) \ncr &&\bra{\Ph_0 (\pq ,\pp )}
U_{0N} (E_N ({\1 p}_N )+ i0) \ket{\Ph_N ({\1 p}_N)}, \ncr
\bra{\Ph_N ( \pp_N)} S \ket{\Ph_\p ({\1 q}_\p)}&=&
-2\p i \d (E_N (\pp_N ) - E_\p ({\1 q}_\p)) \ncr &&\bra{\Ph_N (\pp_N )}
U_{N\p} (E_\p ({\1 q}_\p )+ i0) \ket{\Ph_\p ({\1 q}_\p)},\\
\bra{\Ph_\p (\pq_\p )} S \ket{\Ph_\p ({\1 q}_\p )}&=& \d(\pq_\p -{\1 q}_\p )
-2\p i \d (E_\p (\pq_\p ) - E_\p ({\1 q}_\p ))\ncr && \bra{\Ph_\p (\pq_\p )}
U_{\p \p} (E_\p ({\1 q}_\p) + i0) \ket{\Ph_\p ({\1 q}_\p)},\ncr
\bra{\Ph_0 (\pq, \pp)} S \ket{\Ph_\p ({\1 q}_\p)}&=&
-2\p i \d (E_0 (\pq ,\pp) - E_\p ({\1 q}_\p)) \ncr &&\bra{\Ph_0 (\pq ,\pp )}
U_{0\p} (E_\p ({\1 q}_\p )+ i0) \ket{\Ph_\p ({\1 q}_\p)}.\ncr
\nonumber
\eeqa
All on-shell and off-shell elements of the multi-channel
transition matrix $U(z)$ are determined by the hamiltonian (2.1) through
integral equations. Ref. \cite{poe} chooses the integral equation
for its two-baryon components $U_{b a}(z)$ as fundamental equation and relates
all other components of the multi-channel transition matrix to
those. Redefining the various interaction terms in the hamiltonian
(2.1) for compactness, i.e.,
\beqas
 H_1 &=& (P_N + P_\D )H_1 (P_N + P_\D ) \; + \; P_\D H_1 Q
 \; + \; Q H_1 P_\D \; + \; Q H_1 Q,\\
 H_1 &=& \quad\quad\sum_{a,b=N,\D} v_{ab}^P
\quad\quad\quad \quad +  \sum_{a=N,\D} v^{PQ}_a +
\sum_{a=N,\D} v^{QP}_a
+\sum_{\a } v^Q_\a,
\eeqas
the integral equation for the two-baryon components $U_{b a}(z)$
takes the form
\beq
U_{ba}(z) =\sum_{c=N,\D} [v_{bc}^P + v_b^{PQ} g^Q (z) v_c^{QP} ]
		     [\d_{ca} + g^P_{c0} (z) U_{ca} (z) ] .
\lb{driveq}
\eeq
The remaining components are  related to the baryonic ones by
quadrature, i.e., by
\beqa
U_{\b a} (z) &=& u_{\b 0} (z) \gqo (z) \sum_b v_b^{QP} [\d_{ba} + g_{b0}^P (z)
U_{ba} (z) ],\ncr
U_{0a} (z) &=& [1 + u_{00} (z) \gqo ] \sum_b v^{QP}_b [\d_{ba} +\gpb U_{ba}
(z)],\ncr
U_{b\a} (z) &=&\sum_a [\d_{ba} + U_{ba} (z) \gpa ] v_a^{PQ} \gqo u_{0\a}(z),\\
U_{\b\a} (z) &=& u_{\b\a} (z) + u_{\b 0} (z) \gqo \ncr && \{\sum_{ba} v_b^{PQ}
[\d_{ba} \gpa + \gpb U_{ba} (z) \gpa ]v_a^{PQ} \} \gqo u_{0\a} (z),\ncr
U_{0\a} (z) &=& u_{0\a} (z) +
[1+u_{00} (z)\gqo ] \ncr &&\{\sum_{ba} [\d_{ba}\gpa + \gpb U_{ba} (z) \gpa ]
v_a^{PQ}\}\gqo u_{0\a} (z),\ncr
U_{00} (z) &=& u_{00} (z) +
[1+u_{00} (z)\gqo ] \ncr &&\{\sum_{ba} [\d_{ba}\gpa + \gpb U_{ba} (z) \gpa ]
v_a^{PQ}\}\gqo [1 +u_{00} (z)].\nonumber   \lb{threq}
\eeqa
The relations (A.5) and (A.6) also require the solution of the three-particle
scattering problem
in the pionic sector ${\cal H}_\pi$ of the Hilbert space without pion
production and absorption. That solution is given by the full
propagator
\beq
g^Q (z) ={1\0 z -Q H_0 Q - \sum_\a v_\a^Q}
\eeq
and its corresponding multi-channel three-body transition matrix
$u(z)$ determined by the usual form of the AGS-equations.
Once all components of the multi-channel transition matrix
$U(z)$ are determined, the scattering wave functions
$\ket{\Ps_N^{(\pm )}({\bf p}_N)}$,
$ \ket{\Ps_\pi^{(\pm)}({\bf q}_\pi)}$
and $\ket{\Ps_0^{(\pm)}({\bf p, q})}$ of Eq. (2.5) can be given
according to Eqs. (2.6) - (2.8).

\def\H{{\cal H}}
\def\tp{t_{\g\p}}
\def\gpa{g^P_{a0} (z)}
\def\gpb{g^P_{b0} (z)}
\def\gqp{g^Q_\p (z)}
\def\gqo{g^Q_0 (z)}
\def\pp{{\1 p}'}
\def\pq{{\1 q'}}
\def\ho{H_1}
\def\hog{H_1^\g}
\def\ron{\r^{[1]}}
\def\jon{ j^{[1]\m} }
\def\jtw{{\1 j}^{[2]}}
\def\jtm{ j^{[2]\m}}
\setcounter{section}{2}
\setcounter{equa}{0}

\newpage
\begin{center}
{\Large\bf Appendix B: Example for a Coupled-Channel \\
\hfill \\
		       Interaction-Dependent Current}
\end{center}
\vspace{0.3truecm}

This appendix proves that a coupled-channel e.m. current for the
description of disintegration of and pionproduction on the deuteron
, i.e., $(P_N + P_\Delta + Q) \,\bigl[ {\1 j}^{[1]}(\1 k_\g ) +$ \\
${\1 j}^{[2]}(\1 k_\g) \bigr] \, P_N $ 
can be given which satisfies the condition \currentconservationprojection
of Subsection 2.2.2 exactly. The example is purely phenomenological and
therefore is not and should not be used in practical calculations.
However, the example proves that the coupled-channel current has to have
interaction-dependent pieces already in its one-baryon part, and its
two-baryon part receives additional nontraditional contributions.

We make some simplifying assumptions: There be no interactions in the Hilbert
sector with a pion, i.e.,  $ Q H^{[1]}_1 Q = 0 $ and $ Q H^{[2]}_1 Q = 0 $; 
Siegert's hypothesis hold exactly, i.e., $ \r^{[2]} (\1 k_\g )  = 0 $, 
and the remaining one-baryon charge density has the vanishing components
$ P_\D \r^{[1]} (\1 k_\g ) P_N = 0$ and
$ Q \r^{[1]} (\1 k_\g ) P_\D = 0$.
Under these assumptions the constraints \currentconservationprojection
for the required parts of the current are
\beqas
{\1 k}_\g \cdot P_N \, {\1 j}^{[1]}({\1 k}_\g ) \, P_N &=&
   P_N H_0^{[1]} P_N \, P_N {\r}^{[1]}({\1 k}_\g) P_N -
   P_N {\r}^{[1]}({\1 k}_\g ) P_N \, P_N H_0^{[1]} P_N, \hspace*{.9cm} 
\equaaleb\constroa \\
{\1 k}_\g \cdot P_\Delta \, {\1 j}^{[1]}({\1 k}_\g ) \, P_N &=&
   P_\Delta H_1^{[1]} Q \, Q {\r}^{[1]}({\1 k}_\g) P_N, 
\equaleb\constro \equaaleb\constrob \\
{\1 k}_\g \cdot Q \, {\1 j}^{[1]}({\1 k}_\g ) \, P_N &=&
   Q H_0^{[1]} Q \, Q {\r}^{[1]}({\1 k}_\g) P_N -
   Q {\r}^{[1]}({\1 k}_\g ) P_N \, P_N H_0^{[1]} P_N , 
\equaaleb\constroc
\eeqas

\beqas
{\1 k}_\g \cdot P_N \, {\1 j}^{[2]}(\1 k_\g) P_N &=&
   P_N H_1^{[2]} P_N P_N {\r}^{[1]}(\1 k_\g) P_N - 
   P_N {\r}^{[1]}(\1 k_\g ) P_N P_N H_1^{[2]} P_N, \hspace*{.9cm} 
\equaleb\constrd \equaaleb\constrda \\
{\1 k}_\g \cdot P_\Delta \, {\1 j}^{[2]}(\1 k_\g) P_N &=&
   P_\Delta H_1^{[1]} Q \, Q {\r}^{[1]}({\1 k}_\g) P_N	+ 
   P_\Delta H_1^{[2]} P_N P_N  {\r}^{[1]}({\1 k}_\g) P_N - \ncr
   &&P_\Delta {\r}^{[1]}(\1 k_\g) P_\Delta P_\Delta) H_1^{[2]} P_N,
\equaaleb\constrdb \\
{\1 k}_\g \cdot Q \, {\1 j}^{[2]}(\1 k_\g) P_N &=&
   Q {\r}^{[1]}(\1 k_\g)  P_N  P_N  H_1^{[2]} P_N\,.
\equaaleb\constrdc
\eeqas
The  constraints \constro and \constrd are not entirely closed; they require
knowledge on the one-baryon charge
$P_\Delta {\r}^{[1]}(\1 k_\g) P_\Delta $ for the condition \constrdb\hbsm\,\,.
The fact that $ P_\Delta H_1^{[1]} Q \, Q {\r}^{[1]}({\1 k}_\g) P_N $
has one-baryon and two-baryon parts and therefore arises in Eqs. \constrob
and \constrdb has already been discussed
in Subsect. 2.2.2.

The traditional nucleonic and Born currents of one-baryon nature,
$P_N j^{[1]\m}_{bare}({\1 k}_\g) P_N$ and
  $Q j^{[1]\m}_{bare}({\1 k}_\g) P_N$,
satisfy Eqs. \constroa and \constroc respectively; there is no need for
corresponding interaction-dependent contributions, i.e., 
for $P_N j^{[1]\m}_{exchange}({\1 k}_\g) P_N = 0$ and
  \mbox{$Q j^{[1]\m}_{exchange}({\1 k}_\g) P_N = 0$.}
The traditional transition current
$P_\D j^{[1]\m}_{bare}({\1 k}_\g) P_N$
is purely spatial and transverse, satisfying Eq. \constrob with vanishing 
right-hand side. The transition current requires an interaction-dependent
contribution; a possible form satisfying Eq. \constrob is
\beq
P_\D \1 j^{[1]}_{exchange}(\1 k_\g )P_N \ket{\1 k_N} =
-  P_\D H_1^{[1]} Q { 1 \0 {  k_N^0  + i 0 -
Q H_0^{[1]} Q}} Q \1 j^{[1]}_{bare}(\1 k_\g) P_N \ket{\1 k_N}; 
\equaleb\delnexchangecurrent \hspace*{.9cm}
\eeq
it depends on the interaction $P_\Delta H_1^{[1]} Q$ and is, as the
corresponding bare parts, also purely spatial.
The interaction-dependent current 
$P_\D \1 j^{[1]}_{exchange}(\1 k_\g) P_N $
is due to channel-coupling.
The effective and reducible contribution to the transition current 
$P_\D H_1^{[1]} Q ( z  - Q H_0^{[1]}Q )^{-1} Q \1 j^{[1]\m}(\1 k_\g) P_N $
has to be taken out from its irreducible part in an averaged and energy-independent form.

The traditional two-nucleon exchange current 
$P_N {\1 j}^{[2]}_{traditional}(\1 k_\g) P_N $
satisfies Eq. \constrda\hbsm. The traditional transition  current of two-baryon
nature $P_\D {\1 j}^{[2]}_{traditional}(\1 k_\g) P_N $
solves Eq. \constrdb\hbsm, with a right-hand side containing terms two and three 
only, taking
\beq
\bra{\1 k'_\D}P_\D \ron(\1 k_\g )  P_\D \ket{\1 k_\D} = 
\d (\1 k'_\D - \1 k_N -\1 k_\g )e_P \frac{1}{2} (1 + \t_{\D3} ) .
\eeq
Thus, a nonstandard additional contribution is required, e.g.,
\beqa
&&P_\D \1 j^{[2]}_{nonstandard} (\1 k_\g )P_N
\ket{\1 k_{N1} \1 k_{N2}} = \ncr
&&  - \;
P_\D  H_1^{[1]} Q \,
{ 1 \0 {  k^0_{N1} + k^0_{N2} + i 0 - Q H_0^{[1]} Q} } \,
Q \1 j^{[1]}_{bare} (\1 k_\g ) P_N \ket{\1 k_{N1} \1 k_{N2}} ;
\eeqa
it is of two-baryon nature and satisfies Eq. \constrdb with only the first term
on the right-hand side. Thus, 
\beq
P_\D \1 j^{[2]} (\1 k_\g )P_N = P_\D \left[ \1 j^{[2]}_{traditional}(\1 k_\g ) 
+  \1 j^{[2]}_{nonstandard}(\1 k_\g ) \right] P_N.
\eeq
The two-baryon current from nucleonic states to states with one pion is
entirely nonstandard, i.e., 
\beqa
&& \bra{\1 k'_\p \1 k'_{N1}\1 k'_{N2}} Q \1 j^{[2]} (\1 k_\g )P_N
  \ncr
&&  - \;
\bra{\1 k'_\p \1 k'_{N1} \1 k'_{N2}} Q \1 j^{[1]}_{bare} (\1 k_\g )P_N \,
{ 1 \0 {  k'^0_\p +  k'^0_{N1} + k'^0_{N2} -
P_N H_0^{[1]} P_N}} \,
P_N H_1^{[2]} P_N .
\eeqa
The interaction-dependent nonstandard irreducible two-baryon currents arise,
since corresponding reducible ones are always generated and therefore
have to be taken out from the corresponding traditional processes in
an averaged energy-independent form; the traditional part of 
$Q \1 j^{[2]} (\1 k_\g )P_N $ is zero.


\setcounter{section}{3}
\setcounter{equa}{0}
\def\F{{\cal F}}
\def\M{{\cal M}}
\def\N{{\cal N}}
\def\MM{{{\cal M}_{\p N \g N}}}
\def\MF{M_{1+}^{({3 \0 2})}}
\def\EF{E_{1+}^{({3 \0 2})}}
\def\MFB{M_{1+}^{({3 \0 2})B}}

\def\MFRS{M_{1+}^{({3 \0 2})RS}}
\def\EFB{E_{1+}^{({3 \0 2})B}}
\def\EFRS{E_{1+}^{({3 \0 2})RS}}
\def\MFBO{M_{1+}^{({3 \0 2})Born}}
\def\EFBO{E_{1+}^{({3 \0 2})Born}}
\def\gmd{G_{M1}^{\D N}}
\def\gme{G_{E2}^{\D N}}

\input C1.tex

\newpage

\bc
{\bf Acknowledgements}
\ec

\vspace{0.2cm}

The authors acknowledge helpful discussions with
\ F.\ Fernandez, \ H.\ Garcilazo, \ A.\ Valcarce and
P. Wilhelm.
They thank K. Chmielewski and  A. Kolezhuk
for their help in using some software.
The work benefitted from the DAAD grant, Contract No. 314/Al-e-dr,
which allowed to upkeep the collaboration with the Nuclear Theory Group of the University of Salamanca;
the warm hospitality of the Salamanca Nuclear Theory Group 
is greatly acknowledged.
The work was supported in part by the 
Deutsche
Forschungsgemeinschaft (DFG) under the Contract No. Sa 247/14-1.

\newpage
\input paptref.tex

\newpage
\input F1.tex

\newpage

\input peps.tex

\end{document}

%% file: C1.tex
\newpage

\setcounter{section}{3}
\setcounter{equa}{0}
\def\F{{\cal F}}
\def\M{{\cal M}}
\def\N{{\cal N}}
\def\MM{{{\cal M}_{\p N \g N}}}
\def\MF{M_{1+}^{({3 \0 2})}}
\def\EF{E_{1+}^{({3 \0 2})}}
\def\MFB{M_{1+}^{({3 \0 2})B}}

\def\MFRS{M_{1+}^{({3 \0 2})RS}}
\def\EFB{E_{1+}^{({3 \0 2})B}}
\def\EFRS{E_{1+}^{({3 \0 2})RS}}
\def\MFBO{M_{1+}^{({3 \0 2})Born}}
\def\EFBO{E_{1+}^{({3 \0 2})Born}}
\def\MFBOBARE{M_{1+ bare}^{({3 \0 2})Born}}
\def\EFBOBARE{E_{1+ bare}^{({3 \0 2})Born}}
\def\gmd{G_{M1}^{\D N}}
\def\gme{G_{E2}^{\D N}}

\begin{center}
{\Large\bf Appendix C: 
Photo Pionproduction on\\ \hfill \\ the Single Nucleon }
\end{center}

\vspace{ 0.3cm}

A general description of photo pionproduction on the single nucleon is given.  The process
is used in order to fit phenomenological parameters in the parts
$P_{\Delta}{\mathbf j}^{[1]}({\mathbf k}_{\gamma})P_{N}$ and 
$Q{\mathbf j}^{[1]}({\mathbf k}_{\gamma})P_{N}$ of the single-baryon current.

\vspace{0.5cm}

\begin{center}
\noindent{\large\bf C1.
$S$-Matrix and E.M. Multipole Amplitudes of \\ Photo Pionproduction}
\end{center}

\vspace{0.3cm} 

A detailed description of the photo pionproduction on the 
single nucleon can be found in
the review of Berends et al. \cite{ber}.  Our summary of formulae is
based on the original work of Refs. [28 -- 33].

We adopt the following kinematics: In the initial state a photon of
momentum ${\1 k_\g}$ and polarization vector $\e^\m (\1 k_\g)$ is
absorbed by a nucleon of momentum ${\1 k_N}$. The
momenta of the produced pion and the final nucleon are labeled by
${\1 k_\p}$ and ${\1 k}'_N$.  The $S$-Matrix connecting the initial
photon-nucleon and the final pion-nucleon states is related to the
invariant scattering amplitude $\M_{\p N \g N}$ by 
\beq 
\bra{\p N} S
\ket{\g N} = (-) i (2 \p)^4 \d^4 (k_\p + k'_N - k_\g - k_N ) {\MM \0
\N} \, .  
\eeq
The normalization constant
\beq
\N = (2\p )^6 \sqrt{2 E_\g ({\1 k_\g})} \sqrt{2 \o_\p ({\1 k_\p})} \sqrt{{E_N ({\1 k_N}
) \0
m_N}} \sqrt {{E_N({\1 k}'_N) \0 m_N}}\,, \\
\eeq
with $E_\g ({\1 k_\g})$, $\o_\p ({\1 k_\p})$, $E_N ({\1 k_N})$ and $E_N
({\1 k}'_N)$ being the energies of
the corresponding particles, refers to the state normalization of Ref.
\cite{drell}.
We work in the c.m. system, i.e., ${\1 k_{N}} = - {\1 k_{\g}}$ and ${\1
k}'_{N} = - {\1 k_{\p}}$.
Energy conservation connects the momenta ${\mathbf k}_\g$ and ${\mathbf
k}_\pi$, i.e., 
\beq
W= \vm {k_\g} + \sqrt {m_N^2 + \1 k_\g^2} = \sqrt{m_\p^2 + \1 k^2_\p}
+
\sqrt {m_N^2 + \1 k^2_\p}\,;
\eeq
$W$ is the energy available for the process; 
covariantly, it is the invariant mass defined by 
 $W^2 = (k_N+k_\g)^2$.
We use the Coulomb gauge for the photon field, i.e., $\e^{\m} ({\1
k_{\g}}) =
 (0, {\bm \e}
({\1 k_{\g}}))$ with ${\bm \e} ({\1 k_{\g}}) \cdot {\1 k_{\g}} = 0$.

According to Lorentz symmetry and gauge invariance $\MM$ is
built up from  contributions of four independent operators acting in the
nucleonic spin space; their relative weight is controlled by four 
scalar functions $\F_i (W, \cos \Th)$. 
Thus, the general form of the scattering amplitude is
\beqa
\MM (W, \cos \Th) &=& 4 \p {W \0 m_N} \chi'_s \biggl[i {\bm \s} \cdot {\bm \e}
 \F_1(W, \cos \Th )
 + {{\bm \s} \cdot
\1 k_\p {\bm \s} \cdot(\1 k_\g \times {\bm \e}) \0\vm {k_\p}\vm
{k_\g}}\F_2(W, \cos \Th ) \ncr &+& 
i {{\bm\s}\cdot \1 k_\g  \1 k_\p \cdot {\bm\e}\0  
\vm {k_\g}\vm {k_\p}} \F_3 (W, \cos \Th )
+ i {{\bm\s}\cdot \1 k_\p \1 k_\p \cdot{\bm\e}\0 \1 k^2_\p}
 \F_4 (W, \cos \Th )
\biggr] \chi_s \ncr
\eeqa
with $\chi_s$ and $\chi'_s$ being the Pauli spinors of the nucleon in
the initial and final states and with ${\bm\e}$ abbreviating  
${\bm \e} ({\1 k_{\g}})$.
The scattering amplitude is a function of the
available energy $W$ and of the scattering angle $\Theta$ in the
c.m. system. 

The four independent spin operators depend on the scattering angle $\Theta$;
the four scalar functions ${\cal F}_i(W,\cos\Theta)$ also depend on
it. The latter dependence is usually expanded \cite{chew, gold} in terms of 
the
Legendre polynomials $P_l (\cos \Th)$ and of the electric and magnetic
 multipole amplitudes
$E_{l\pm}$ and $M_{l\pm}$, i.e.,
\beqas
\F_1 (W, \cos \Th) &=& \sum_{l=0}^{\infty}
\Bigl[ \left[ l M_{l+} (W) + E_{l+} (W) \right] P'_{l+1} (\cos
\Th)  \ncr &&
+ \left[(l+1) M_{l-} (W) + E_{l-} (W) \right] P'_{l-1} (\cos \Th) \Bigr] \, , \\
\F_2 (W, \cos \Th) &=& \sum^{\infty}_{l=1} \Bigl[ \left[ (l+1) M_{l+} (W)
 + lM_{l-} (W) \right] P'_l(\cos
\Th) \Bigr] \, , \\
\equaleb\ffunctions
\F_3 (W, \cos \Th) &=& \sum_{l=1}^{\infty} \Bigl[ \left[ -M_{l+} (W) + E_{l+} (W) \right]
 P''_{l+1} (\cos\Th) \ncr
&&+ \left[ M_{l-} (W) + E_{l-} (W) \right] P''_{l-1} (\cos \Th) \Bigr] \,,  \\
\F_4 (W, \cos \Th) &=& \sum_{l=1}^{\infty} \Bigl[ \left[ M_{l+} (W) - E_{l+} (W)\right.
 \ncr &&- \left.M_{l-} (W)  -
E_{l-} (W) \right] P''_l (\cos \Th) \Bigr] \,.
\eeqas
In Eq. \ffunctions $P_l' (P_l'')$ denotes the first (second) derivative of $P_l (\cos\Th)$
 with respect
to $\cos \Th$. 
Eq. (C.5)  can be inverted, i.e.,
\beqas
\hspace*{-.6cm}E_{l+} (W) &=& {1 \0 {2(l+1)}} \int^1_{-1} dx \Bigl[\F_1 (W,x) P_l (x) - \F_2 (W,x)
P_{l+1} (x)
-\F_3 (W,x) \times \ncr 
&&{l \0 {2l+1}} (P_{l+1} (x) - P_{l-1} (x)) - \F_4 (W,x)
{{l+1} \0 {2l+3}} (P_{l+2} (x) - P_l (x)) \Bigr], \hspace*{1.0cm}\\
\hspace*{-.6cm}E_{l-} (W) &=& {1 \0 {2l}} \int^1_{-1} dx \Bigl[ \F_1 (W,x) P_l (x) - \F_2 (W,x)
P_{l-1} (x)
+ \F_3 (W,x) \times \ncr 
&& {{l+1} \0 {2l+1}} (P_{l+1} (x) - P_{l-1} (x) ) -
 \F_4 (W,x) {l \0
{2l-1}} (P_{l+2} (x) - P_l (x)) \Bigr], \hspace*{1.0cm}\\
\hspace*{-.6cm}M_{l+}(W) &=& {1 \0{2(l+1)}} \int^1_{-1} dx \Bigl[ \F_1 (W,x) P_l(x) - \F_2 (W,x)
P_{l+1} (x) \ncr 
&&+ \F_3 (W,x) {1 \0 {2l+1}} (P_{l+1} (x) - P_{l-1} (x))\Bigr],\hspace*{1.0cm}\\
\hspace*{-.6cm}M_{l-}(W) &=& {1 \0 {2l}} \int^1_{-1} dx \Bigl[-\F_1 (W,x) P_l (x) + \F_2 (W,x)
P_{l-1} (x) \ncr 
&&-\F_3 (W,x) {1 \0 {2l+1}} (P_{l+1} (x) - P_{l-1} (x)) \Bigr].
\eeqas

In addition, the invariant scattering amplitude $\MM$ 
is built up from different isospin contributions distinct by the charge of the
produced pion; the various parts show up in the
scalar functions $\F_i(W,\cos\Th)$, i.e.,
\beq
\F_i(W,\cos\Th)  = \F_i^{(+)}(W,\cos\Th) \d_{\a , 0} 
+ \F_i^{(-)}(W,\cos\Th) {1 \0 2} [\bm\t_\a , \bm\t_0] +
\F_i^{(0)}(W,\cos\Th) \bm\t_\a \,.\\
\eeq
The subscript $\a$ refers to a specific isospin channel of the process, i.e.
$\a = (-) Q_\p$ where $Q_\p$ is the charge of the pion, $\bm\t_\a$
being the usual nucleonic Pauli matrices.
 Instead
of the amplitudes $\F_i^{(+)}$ and $\F_i^{(-)}$ the linear combinations
\beqas
\F_i^{({1 \0 2})}(W,\cos\Th) =& \F_i^{(+)}(W,\cos\Th) + 2 \F_i^{(-)}(W,\cos\Th)\\
\F_i^{({3 \0 2})}(W,\cos\Th) =& \F_i^{(+)} (W,\cos\Th) - \F_i^{(-)} (W,\cos\Th)
\eeqas
referring to  
definite total isospin of the final pion-nucleon state, i.e.,
${1\over2}$ and ${3\over2}$, are introduced.
The isospin contributions to the e.m. multipole amplitudes
$M_{l\pm}$ and $E_{l\pm}$ will be explicitly indicated
by the superscripts $\b = 0, {1 \0 2}$, or ${3 \0 2}$ of the scalar functions
$\F_i^{(\b)}(W,\cos\Th)$
and the multipoles
$M_{l\pm}^{(\b )}$ and $E_{l\pm}^{(\b )}$.
\\

\vspace{0.5cm}

\begin{center}
{\large\bf C2. Single-Baryon Current to Be Fitted}
\end{center}
\vskip 0.3truecm

The parts of the single-baryon current required for 
photo pionproduction are the transition current
\beqa
&&\bra {{\1 k}_\D } P_\D {\mathbf j}^{[1]}({\mathbf k}_\gamma) P_N 
\ket {{\1 k}_N )}
= 
\d ({\1 k}_\D - {\1 k}_N - {\1 k}_\g ) {e_p \0 {2 m_N}} (\bm\t_{\D N})_0
 \ncr &&
\Bigl[ 
i G_{M1}^{\D N} [{\bm \s}_{\D N} \times {\1 k}_\g] +
G_{E2}^{\D N} [ {\bm \s}_{\D N} {\bm\s} \cdot {\1 k}_\g - {i \0 2}
 ({\bm \s}_{\D N}
\times {\1 k}_\g)] 
\Bigr]
\equaleb\dcur
\eeqa
and the Born current $Q{\mathbf j}^{[1]}({\mathbf k}_\gamma) P_N$. 
The transition current has magnetic dipole and electric quadrupole
contributions with the strength parameters $G_{M1}^{\Delta N}$ and
$G_{E2}^{\Delta N}$, respectively. In Eq. \dcur the current is written
in the pion-nucleon c.m. system, i.e., for ${\mathbf k}_\Delta=0$; the
calculation of this appendix is done in that system. The Born current
$Q{\mathbf j}^{[1]}({\mathbf k}_\gamma) P_N$
is taken over from Ref. \cite{ols}; it is on its pion-nucleon side
augmented by the phenomenological form factor 
\beq
F_{B} (\1 k^2) = {(\L_{B})^2 \0 {(\L_{B})^2 + \1 {k^2} }} \,\, . \\
\label{borncut}
\eeq
As for the corresponding form factor in the $\pi$N$\Delta$-vertex $QH_1
P_\Delta$, the momentum ${\mathbf k}$ is chosen to be the relative
pion-nucleon momentum; it only becomes the true pion momentum in the
pion-nucleon c.m. system.

Given the structure of the single-baryon current, the resonant
multipole 
amplitudes have three contributions \cite{yang,ohta}, 
\beqas
M_{1+}^{({3 \0 2})}(W)=& M_{1+}^{({3 \0 2})Res}(W) +M_{1+}^{({3 \0
2})Born}(W) +M_{1+}^{({3 \0 2})Inter}(W)\,,\\
E_{1+}^{({3 \0 2})}(W)=& E_{1+}^{({3 \0 2})Res}(W) +E_{1+}^{({3 \0
2})Born}(W) +E_{1+}^{({3 \0 2})Inter}(W)\,,
\equaleb\offmsh
\eeqas
the pure resonant, Born and interference (or often called rescattering) contributions being
labeled with the superscript $Res$, $Born$ and $Inter$,
respectively. The three processes are displayed in Fig. C.1. They
have the following analytic forms
\beqas
M_{1+}^{({3 \0 2})Res}(W) &=& 
\,{1 \0 6 W} \,\, \left[ {|{\1 k}_\p| \0 {\sqrt {4 \p}} }
                   { f_\D ({\1 k}_\p^2) \0 {m_\p} }
g^\D (W) G_{M1}^{\D N} {e_p \0 { \sqrt {4 \p}}} \vm{k_\g} 
\right] \,,\\
E_{1+}^{({3 \0 2})Res} (W) &=&
{-1 \0 12 W}  \left[ {|{\1 k}_\p| \0 {\sqrt {4 \p}} }
                   { f_\D ({\1 k}_\p^2) \0 {m_\p} }
g^\D (W) G_{E2}^{\D N} {e_p \0 { \sqrt {4 \p}}} \vm{k_\g}
\right] \,, 
\equaleb\dmultips
\eeqas
\vspace{-0.6cm}
\beqas
\MFBO (W) &=&  F_{B} (\1 k_\p^2) \MFBOBARE (W) \,, \\
\EFBO (W) &=&  F_{B} (\1 k_\p^2) \EFBOBARE (W) \, 
\equaleb\bornmultips
\eeqas
and
\beqas
\hspace*{-0.7cm}
M_{1+}^{({3 \0 2})Inter}(W) =& g^\D (W) \left[ - {i \0 2} \G_\D (W,0) F_{B} (\1 k_\p^2) +
 I_P (W)\right] \MFBOBARE (W)\,,\,\,\\
\hspace*{-0.7cm}
E_{1+}^{({3 \0 2})Inter}(W) =& g^\D (W) \left[ - {i \0 2} \G_\D (W,0) F_{B} (\1 k_\p^2) +
 I_P (W)\right] \EFBOBARE (W)\,,\,\,
\equaleb\rsmultips
\eeqas
with
\beqas
\quad f_\D (\1 k^2 ) &=& f_{\p N \D}{\L_{\p\D}^2 -m_\p^2
\0\L_{\p\
\D}^2 +\1 k^2}\,,\\
\qquad g^\D (W) &=& {1 \0 { W +i0 - M_\D (W,0) + {i \0 2 } \G_\D
(W,0)}}\,
\eeqas
according to Eq. (2.2).
The last term in the brackets of the Eq. \rsmultips is the principle value integral
\beq
I_P(W) = {{f_\D ({\1 k_\p^2})} \0 {m_\p}} \vm {k_\p} {4\p \0 3} {1 \0
{(2 \p)^3}\
} {\cal P}
\int^{\infty}_0 {{d p} \0 {\o_\p (p)}} {{f_\D (p^2)} \0 {m_\p}} 
{{p^3 F_{B} (p^2)} \0 {W - E_N (p) - \o_\p (p)}} \, . \\
\eeq
The Born contribution of Eq. \bornmultips is too lengthy to be given explicitly; it is
taken over from Ref. \cite{ols}; only the additionally introduced
form factor $F_{B} ({\1 k_\p^2})$ has to be remembered. 

By construction, the resonant multipole amplitudes satisfy Watson's
theorem \cite{wat}, i.e.,
\beqas 
M_{1+}^{({3 \0 2})}(W) &=& \left|M_{1+}^{({3 \0 2})}(W)\right|
e^{i\delta_{P_{33}}(W)}, \\ 
E_{1+}^{({3 \0 2})}(W) &=& \left|E_{1+}^{({3 \0 2})}(W)\right|
e^{i\delta_{P_{33}}(W)}.
\label{watson} 
\eeqas
In the last equations  $\delta_{P_{33}}(W)$ is the pion-nucleon phase
shift in the $P_{33}$ partial wave.
This result gets obvious by using the equations above for the total 
multipoles, i.e.,
\beqas
\hspace*{-0.9cm}\MF(W) &=& g^\D (W) \biggl[
{1 \0 6 W}  {|{\1 k}_\p| \0 {\sqrt {4 \p}} }
                   { f_\D ({\1 k}_\p^2) \0 {m_\p} }
G_{M1}^{\D N} {e_p \0 { \sqrt {4 \p}}} \vm{k_\g}
\ncr
&& + 
\left( \left[W - M_\D (W,0)\right] F_{B} (\vm {k_\p}) + I_p (W)
\right)\MFBOBARE (W)\biggr] \,, \\
\hspace*{-0.9cm}\EF(W) &=& g^\D (W) \biggl[
{-1 \0 12 W}  {|{\1 k}_\p| \0 {\sqrt {4 \p}} }
                   { f_\D ({\1 k}_\p^2) \0 {m_\p} }
G_{E2}^{\D N} {e_p \0 { \sqrt {4 \p}}} \vm{k_\g} 
\ncr
&& + 
\left( \left[ W - M_\D (W,0) \right] F_{B} (\vm {k_\p}) + I_p (W)
\right)\EFBOBARE (W)\biggr] \,.
\equaleb\tmultips
\eeqas

\vskip 0.5truecm

\begin{center}
{\large\bf
C3. Results for Multipole Amplitudes}
\end{center}
\vskip 0.3truecm

Sample results for multipole amplitudes in the kinematic regime up to
$1.4$ GeV c.m. energy $W$ are given in Figs. C.2 - C.8.  Figs. C.5 and C.6
display the resonant magnetic dipole $M_{1+}^{({3 \0 2})}$, Figs. C.7 and C.8
the electric quadrupole $E_{1+}^{({3 \0 2})}$ and Figs. C.2 - C.4  some
nonresonant multipoles. 

Figs. C.5 - C.8 show results of our fit. We consider the fit
satisfactory. The optimized parameters for the single-baryon current
$P_\Delta \1{j}^{[1]}({\mathbf k}_\gamma) P_N$ are $G_{M1}^{\Delta
N}=3.65$ as magnetic dipole strength and $G_{E2}^{\Delta N}=0.10$ as
electric quadrupole strength. The cut-off $\Lambda_{B}$ of
Eq. (\ref{borncut}) for the Born current turns out to be better chosen  
differently for 
different multipoles, i.e., $\L_{B}^{M1}=245.5$~MeV and
$\L_{B}^{E2}=379.0$~MeV; no common satisfactory value could be
found. In the Born current the $\rho$-meson does not contribute to
either resonant multipole on symmetry grounds; in contrast, the
$\omega$-meson does contribute but does so noticeably only for the
magnetic dipole $M_{1+}^{({3 \0 2})}$.

Figs. C.2 - C.4 show sample results for nonresonant multipole
amplitudes.  The results do not use a cut-off form factor, i.e., 
$F_{B} (k^2) = 1$.
No distinct fit is carried out for each partial wave; thus, the
results are predictions.  The multipole
$E_{0+}^{\left({1\over2}\right)}$ is an example in which the
$\rho$-meson contributes markedly, $M_{1-}^{({3 \0 2})}$ one in which
the $\omega$-meson does so.

%% file: paptref.tex

\def\np#1{{\it Nucl. Phys.} {\bf #1}}
\def\prl#1{{\it Phys. Rev. Lett.} {\bf #1}}
\def\jp#1{{\it J. of Phys.} {\bf #1}}
\def\pr#1{{\it Phys. Rev.} {\bf #1}}
\def\pl#1{{\it Phys. Lett.} {\bf #1}}
\def\prp#1{{\it Phys. Rep.} {\bf #1}}
\def\ppnp#1{{\it Prog. Part. Nucl. Phys.} {\bf #1}}
\def\fbs#1{{\it Few Body Systems} {\bf #1}}

%% file: F1.tex

\newpage
\begin{picture}(0,0)%
\includegraphics{fn1.pstex}%
\end{picture}%
\setlength{\unitlength}{0.002500in}%
\begingroup\makeatletter\ifx\SetFigFont\undefined
\def\x#1#2#3#4#5#6#7\relax{\def\x{#1#2#3#4#5#6}}%
\expandafter\x\fmtname xxxxxx\relax \def\y{splain}%
\ifx\x\y   
\gdef\SetFigFont#1#2#3{%
  \ifnum #1<17\tiny\else \ifnum #1<20\small\else
  \ifnum #1<24\normalsize\else \ifnum #1<29\large\else
  \ifnum #1<34\Large\else \ifnum #1<41\LARGE\else
     \huge\fi\fi\fi\fi\fi\fi
  \csname #3\endcsname}%
\else
\gdef\SetFigFont#1#2#3{\begingroup
  \count@#1\relax \ifnum 25<\count@\count@25\fi
  \def\x{\endgroup\@setsize\SetFigFont{#2pt}}%
  \expandafter\x
    \csname \romannumeral\the\count@ pt\expandafter\endcsname
    \csname @\romannumeral\the\count@ pt\endcsname
  \csname #3\endcsname}%
\fi
\fi\endgroup
\begin{picture}(490,495)(110,290)
\end{picture}

\vspace*{1.0cm}
\noindent
\begin{minipage}[t]{16.0cm}
{\bf Fig. 2.1.} Graphical definition of the interaction hamiltonian
$H_1$ in a Hilbert space of
baryon number two. The solid vertical line denotes a nucleon,
the thick one a $\D$-isobar,
the dotted line a pion.
Horizontal dashed lines represent the action of an instantaneous
hermitian potential. Only process (a) contributes to the
isospin-singlet partial waves.
\end{minipage}

\clearpage
\newpage
\begin{figure}
\mbox{\psfig{figure=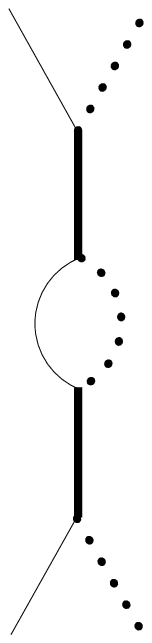,height=12cm,width=12cm}}

{\bf Fig. 2.2.} Schematic representation of  $\p$N scattering in 
the $P_{33}$ partial wave.
Only one characteristic process of an infinite order is shown.

\end{figure}

\newpage
\begin{picture}(0,0)%
\includegraphics{fn3.pstex}%
\end{picture}%
\setlength{\unitlength}{0.002500in}%
\begingroup\makeatletter\ifx\SetFigFont\undefined
\def\x#1#2#3#4#5#6#7\relax{\def\x{#1#2#3#4#5#6}}%
\expandafter\x\fmtname xxxxxx\relax \def\y{splain}%
\ifx\x\y   
\gdef\SetFigFont#1#2#3{%
  \ifnum #1<17\tiny\else \ifnum #1<20\small\else
  \ifnum #1<24\normalsize\else \ifnum #1<29\large\else
  \ifnum #1<34\Large\else \ifnum #1<41\LARGE\else
     \huge\fi\fi\fi\fi\fi\fi
  \csname #3\endcsname}%
\else
\gdef\SetFigFont#1#2#3{\begingroup
  \count@#1\relax \ifnum 25<\count@\count@25\fi
  \def\x{\endgroup\@setsize\SetFigFont{#2pt}}%
  \expandafter\x
    \csname \romannumeral\the\count@ pt\expandafter\endcsname
    \csname @\romannumeral\the\count@ pt\endcsname
  \csname #3\endcsname}%
\fi
\fi\endgroup
\begin{picture}(490,495)(110,290)
\end{picture}

\vspace*{3.7cm}
\noindent
\begin{minipage}[t]{16.0cm}
{\bf Fig. 2.3.} Feynman processes for the one-baryon current up to first order
in the field-theoretic Lagrangian ${\cal L}_I$.
The first three processes are interaction-free;
their amplitudes are denoted by $J^{[1]\m}_{bare}( k_\g)$;
they are the purely nucleonic current
    $j^\m_{NN}( k_\g)$,
the transition one
    $j^\m_{\D N}( k_\g)$  from the nucleon to the $\D$-isobar and
the Born one	$j^\m_{B}( k_\g)$
which produces a pion on the nucleon;
they are assumed to be conserved.
The processes (d) to (f) are interaction-dependent;
they are of first order in the interaction;
their amplitudes are denoted by $J^{[1]\m}_{exchange}( k_\g)$.
\end{minipage}

\newpage
\noindent
\begin{picture}(0,0)%
\includegraphics{fn4.pstex}%
\end{picture}%
\setlength{\unitlength}{0.002500in}%
\begingroup\makeatletter\ifx\SetFigFont\undefined
\def\x#1#2#3#4#5#6#7\relax{\def\x{#1#2#3#4#5#6}}%
\expandafter\x\fmtname xxxxxx\relax \def\y{splain}%
\ifx\x\y   
\gdef\SetFigFont#1#2#3{%
  \ifnum #1<17\tiny\else \ifnum #1<20\small\else
  \ifnum #1<24\normalsize\else \ifnum #1<29\large\else
  \ifnum #1<34\Large\else \ifnum #1<41\LARGE\else
     \huge\fi\fi\fi\fi\fi\fi
  \csname #3\endcsname}%
\else
\gdef\SetFigFont#1#2#3{\begingroup
  \count@#1\relax \ifnum 25<\count@\count@25\fi
  \def\x{\endgroup\@setsize\SetFigFont{#2pt}}%
  \expandafter\x
    \csname \romannumeral\the\count@ pt\expandafter\endcsname
    \csname @\romannumeral\the\count@ pt\endcsname
  \csname #3\endcsname}%
\fi
\fi\endgroup
\begin{picture}(490,495)(110,290)
\end{picture}

\vspace*{8.0cm}
\noindent
\begin{minipage}[t]{15.0cm}
{\bf Fig. 2.4.} Field-theoretic processes contained in the Born current
  $j^\m_{B}( k_\g)$.
The first four processes are solely derived from interactions with the pion.
The fifth process involves rho- and omega-meson exchange;
the vector mesons are diagrammatically indicated by double lines.
Since the employed hadronic interaction
does not use the pion-nucleon vertex as mechanism for the pion
production and absorption, i.e., 
$Q H_1^{[1]} P_N = 0$,
 the current is not reducible.
\end{minipage}

\newpage
\noindent
\begin{picture}(0,0)%
\includegraphics{fn5.pstex}%
\end{picture}%
\setlength{\unitlength}{0.002500in}%
\begingroup\makeatletter\ifx\SetFigFont\undefined
\def\x#1#2#3#4#5#6#7\relax{\def\x{#1#2#3#4#5#6}}%
\expandafter\x\fmtname xxxxxx\relax \def\y{splain}%
\ifx\x\y   
\gdef\SetFigFont#1#2#3{%
  \ifnum #1<17\tiny\else \ifnum #1<20\small\else
  \ifnum #1<24\normalsize\else \ifnum #1<29\large\else
  \ifnum #1<34\Large\else \ifnum #1<41\LARGE\else
     \huge\fi\fi\fi\fi\fi\fi
  \csname #3\endcsname}%
\else
\gdef\SetFigFont#1#2#3{\begingroup
  \count@#1\relax \ifnum 25<\count@\count@25\fi
  \def\x{\endgroup\@setsize\SetFigFont{#2pt}}%
  \expandafter\x
    \csname \romannumeral\the\count@ pt\expandafter\endcsname
    \csname @\romannumeral\the\count@ pt\endcsname
  \csname #3\endcsname}%
\fi
\fi\endgroup
\begin{picture}(490,495)(110,290)
\end{picture}

\vspace*{7.8cm}
\begin{minipage}[t]{15.0cm}
{\bf Fig. 2.5.} Processes of Schr\"odinger theory for the one-baryon current
up to first order 
in the interaction $H_1^{[1]}$.
The arrows indicate that the corresponding particles in intermediate states
are on their respective mass shells and propagate according to
the global propagators of Schr\"odinger theory.
\end{minipage}

\newpage
\begin{picture}(0,0)%
\includegraphics{fn6.pstex}%
\end{picture}%
\setlength{\unitlength}{0.002500in}%
\begingroup\makeatletter\ifx\SetFigFont\undefined
\def\x#1#2#3#4#5#6#7\relax{\def\x{#1#2#3#4#5#6}}%
\expandafter\x\fmtname xxxxxx\relax \def\y{splain}%
\ifx\x\y   
\gdef\SetFigFont#1#2#3{%
  \ifnum #1<17\tiny\else \ifnum #1<20\small\else
  \ifnum #1<24\normalsize\else \ifnum #1<29\large\else
  \ifnum #1<34\Large\else \ifnum #1<41\LARGE\else
     \huge\fi\fi\fi\fi\fi\fi
  \csname #3\endcsname}%
\else
\gdef\SetFigFont#1#2#3{\begingroup
  \count@#1\relax \ifnum 25<\count@\count@25\fi
  \def\x{\endgroup\@setsize\SetFigFont{#2pt}}%
  \expandafter\x
    \csname \romannumeral\the\count@ pt\expandafter\endcsname
    \csname @\romannumeral\the\count@ pt\endcsname
  \csname #3\endcsname}%
\fi
\fi\endgroup
\begin{picture}(490,495)(110,290)
\end{picture}

\vspace*{10.5cm}
\noindent
\begin{minipage}[t]{15.0cm}
{\bf Fig. 2.6.} Graphical definition of the multichannel one-baryon
interaction-dependent current
$(P_\D + Q ) j^{[1]\m}_{exchange}({\1 k}_\g) P_N $.
The meaning of the arrows for particles in intermediate states is given in Fig. 2.5.
\end{minipage}

\newpage
\begin{picture}(0,0)%
\includegraphics{fn7.pstex}%
\end{picture}%
\setlength{\unitlength}{0.002500in}%
\begingroup\makeatletter\ifx\SetFigFont\undefined
\def\x#1#2#3#4#5#6#7\relax{\def\x{#1#2#3#4#5#6}}%
\expandafter\x\fmtname xxxxxx\relax \def\y{splain}%
\ifx\x\y   
\gdef\SetFigFont#1#2#3{%
  \ifnum #1<17\tiny\else \ifnum #1<20\small\else
  \ifnum #1<24\normalsize\else \ifnum #1<29\large\else
  \ifnum #1<34\Large\else \ifnum #1<41\LARGE\else
     \huge\fi\fi\fi\fi\fi\fi
  \csname #3\endcsname}%
\else
\gdef\SetFigFont#1#2#3{\begingroup
  \count@#1\relax \ifnum 25<\count@\count@25\fi
  \def\x{\endgroup\@setsize\SetFigFont{#2pt}}%
  \expandafter\x
    \csname \romannumeral\the\count@ pt\expandafter\endcsname
    \csname @\romannumeral\the\count@ pt\endcsname
  \csname #3\endcsname}%
\fi
\fi\endgroup
\begin{picture}(490,495)(110,290)
\end{picture}

\vspace*{12.0cm}
\noindent
\begin{minipage}[t]{15.0cm}
{\bf Fig. 2.7.} Two-baryon current $(P_N + P_\D + Q) J^{[2]\m}_{oms}({\1 k}_\g) P_N$
in the noncovariant framework of Schr\"odinger theory. They are based on
the bare one-baryon current $j^{[1]\m}_{bare}({\1 k}_\g)$
and on the interaction $H_1^{[1]} + H_1^{[2]}$ up to first order.
The meaning of the arrows for particles in intermediate states is given in Fig.
2.5.
All connected, but reducible
processes are shown in which the current acts first;
exchange processes are not shown; furthermore, the processes in which
the interaction $H_1$ acts first are not shown for consistency
of presentation. In the three rows the
contributions $P_N J^{[2]\m}_{oms}({\1 k}_\g) P_N$,
$P_\D J^{[2]\m}_{oms}({\1 k}_\g) P_N$ and
$Q J^{[2]\m}_{oms}({\1 k}_\g) P_N$ are given
consecutively.
\end{minipage}

\newpage
\clearpage
\begin{figure}
\mbox{\psfig{figure=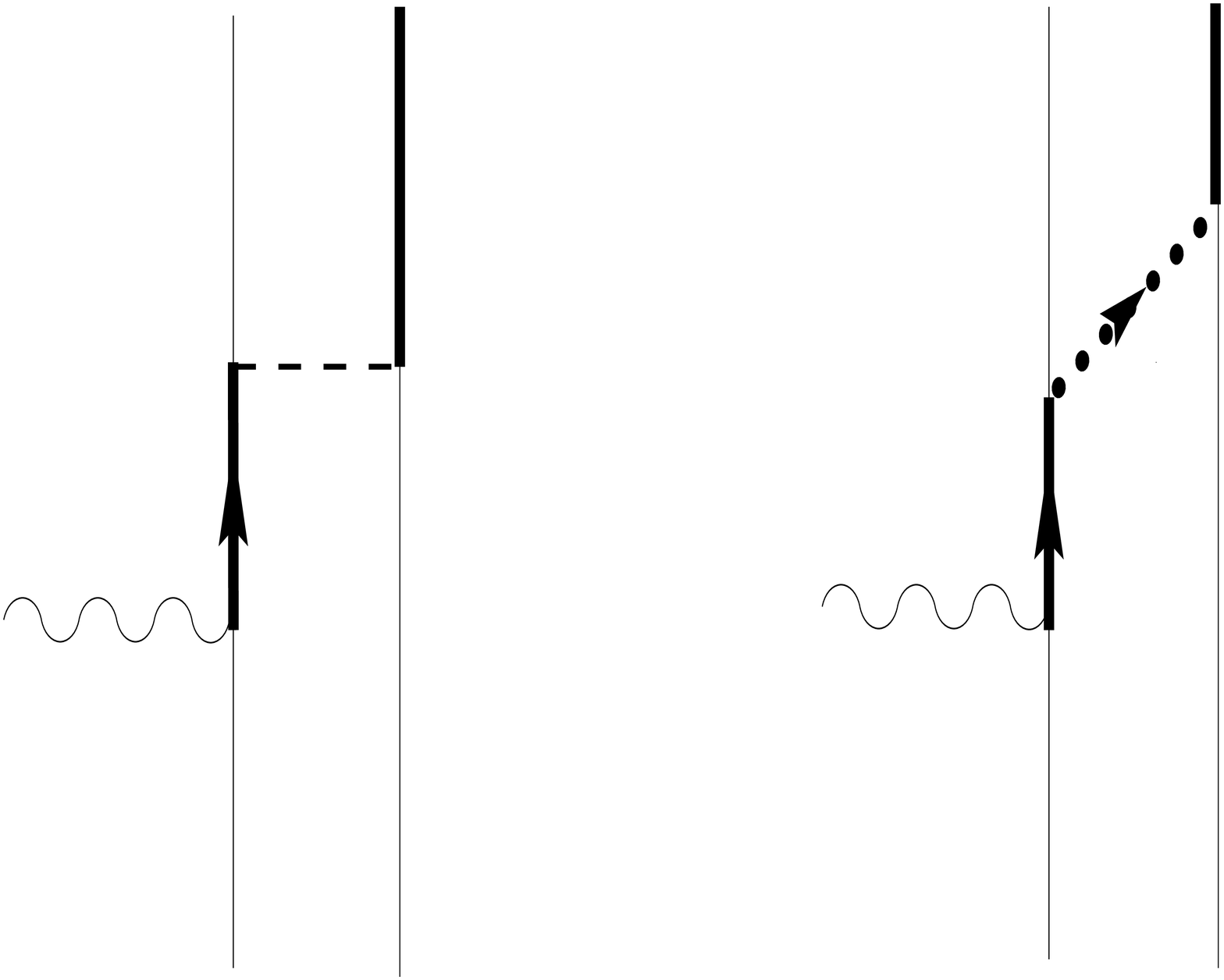,height=12cm,width=12cm}}

{\bf Fig. 2.8.} Contributions to the two-baryon current 
$P_\D J^{[2]\m}_{oms}({\1 k}_\g)P_N$
in the noncovariant framework of Schr\"odinger theory.
In orders of the employed interaction $H_1$, the left process is of first
in $H_1^{[2]}$,
the right one of second order $H_1^{[1]}$.
The meaning of the arrows for particles in intermediate states is given in Fig.
2.5.
\end{figure}

\newpage
\noindent
\begin{picture}(0,0)%
\includegraphics{fn9.pstex}%
\end{picture}%
\setlength{\unitlength}{0.002500in}%
\begingroup\makeatletter\ifx\SetFigFont\undefined
\def\x#1#2#3#4#5#6#7\relax{\def\x{#1#2#3#4#5#6}}%
\expandafter\x\fmtname xxxxxx\relax \def\y{splain}%
\ifx\x\y   
\gdef\SetFigFont#1#2#3{%
  \ifnum #1<17\tiny\else \ifnum #1<20\small\else
  \ifnum #1<24\normalsize\else \ifnum #1<29\large\else
  \ifnum #1<34\Large\else \ifnum #1<41\LARGE\else
     \huge\fi\fi\fi\fi\fi\fi
  \csname #3\endcsname}%
\else
\gdef\SetFigFont#1#2#3{\begingroup
  \count@#1\relax \ifnum 25<\count@\count@25\fi
  \def\x{\endgroup\@setsize\SetFigFont{#2pt}}%
  \expandafter\x
    \csname \romannumeral\the\count@ pt\expandafter\endcsname
    \csname @\romannumeral\the\count@ pt\endcsname
  \csname #3\endcsname}%
\fi
\fi\endgroup
\begin{picture}(490,495)(110,290)
\end{picture}

\vspace*{9.0cm}
\noindent
\begin{minipage}[t]{15.0cm}
{\bf Fig. 2.9.} Graphical definition of the two-baryon
exchange current of pion range
\mbox{$\langle {\1 k}'_{N_1} {\1 k}'_{\D_2} |
	 j^{[2]\m}_{exchange}({\1 k_\g})
	|{\1 k}_{N_1} {\1 k}_{N_2} \rangle $}.
The meaning of the arrows for particles in intermediate states is given in Fig.
2.5.
The four field-theoretic processes (a) - (d) of opposite time ordering
in which the $\pi$N$\D$-vertex occurs prior to the $\pi$NN-vertex
are required for the definition, but not shown.
\end{minipage}

\newpage

\begin{figure}
\mbox{\psfig{figure=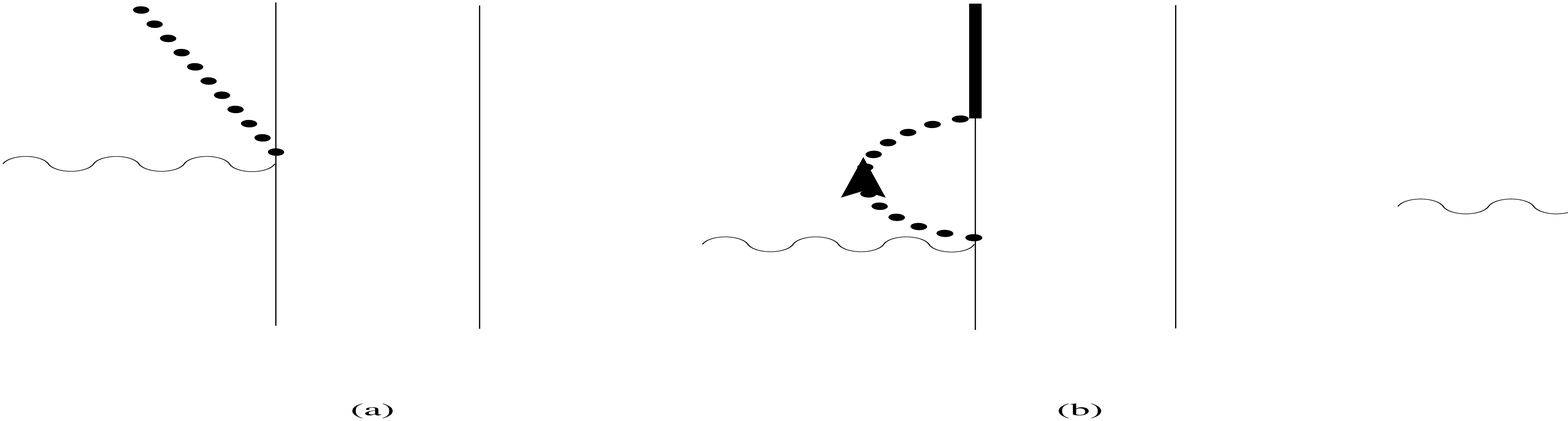,height=12cm,width=12cm}}

{\bf Fig. 3.1.}
Irreducible and reducible contributions arising from the Born current
in the two-baryon system. Contribution (a), however only in the $P_{33}$
partial wave, is retained in the present calculations for pionproduction
without any further rescattering.
The reducible processes (b) and (c) can contribute to
disintegration and to pionproduction due to further
final-state interaction;
neither of these two processes is retained in the present calculations.
\end{figure}

\newpage
\clearpage

\noindent
\begin{picture}(0,0)%
\includegraphics{fn11.pstex}%
\end{picture}%
\setlength{\unitlength}{0.002500in}%
\begingroup\makeatletter\ifx\SetFigFont\undefined
\def\x#1#2#3#4#5#6#7\relax{\def\x{#1#2#3#4#5#6}}%
\expandafter\x\fmtname xxxxxx\relax \def\y{splain}%
\ifx\x\y   
\gdef\SetFigFont#1#2#3{%
  \ifnum #1<17\tiny\else \ifnum #1<20\small\else
  \ifnum #1<24\normalsize\else \ifnum #1<29\large\else
  \ifnum #1<34\Large\else \ifnum #1<41\LARGE\else
     \huge\fi\fi\fi\fi\fi\fi
  \csname #3\endcsname}%
\else
\gdef\SetFigFont#1#2#3{\begingroup
  \count@#1\relax \ifnum 25<\count@\count@25\fi
  \def\x{\endgroup\@setsize\SetFigFont{#2pt}}%
  \expandafter\x
    \csname \romannumeral\the\count@ pt\expandafter\endcsname
    \csname @\romannumeral\the\count@ pt\endcsname
  \csname #3\endcsname}%
\fi
\fi\endgroup
\begin{picture}(490,495)(110,290)
\end{picture}

\vspace*{6.0cm}
\noindent
\begin{minipage}[t]{15.0cm}
{\bf Fig. 3.2.} Traditional meson-exchange currents. 
On the left side the contact or pair process is shown, on the right side
the meson-in-flight one.
\end{minipage}

%% file: peps.tex
\newpage

\begin{figure}
\mbox{\psfig{figure=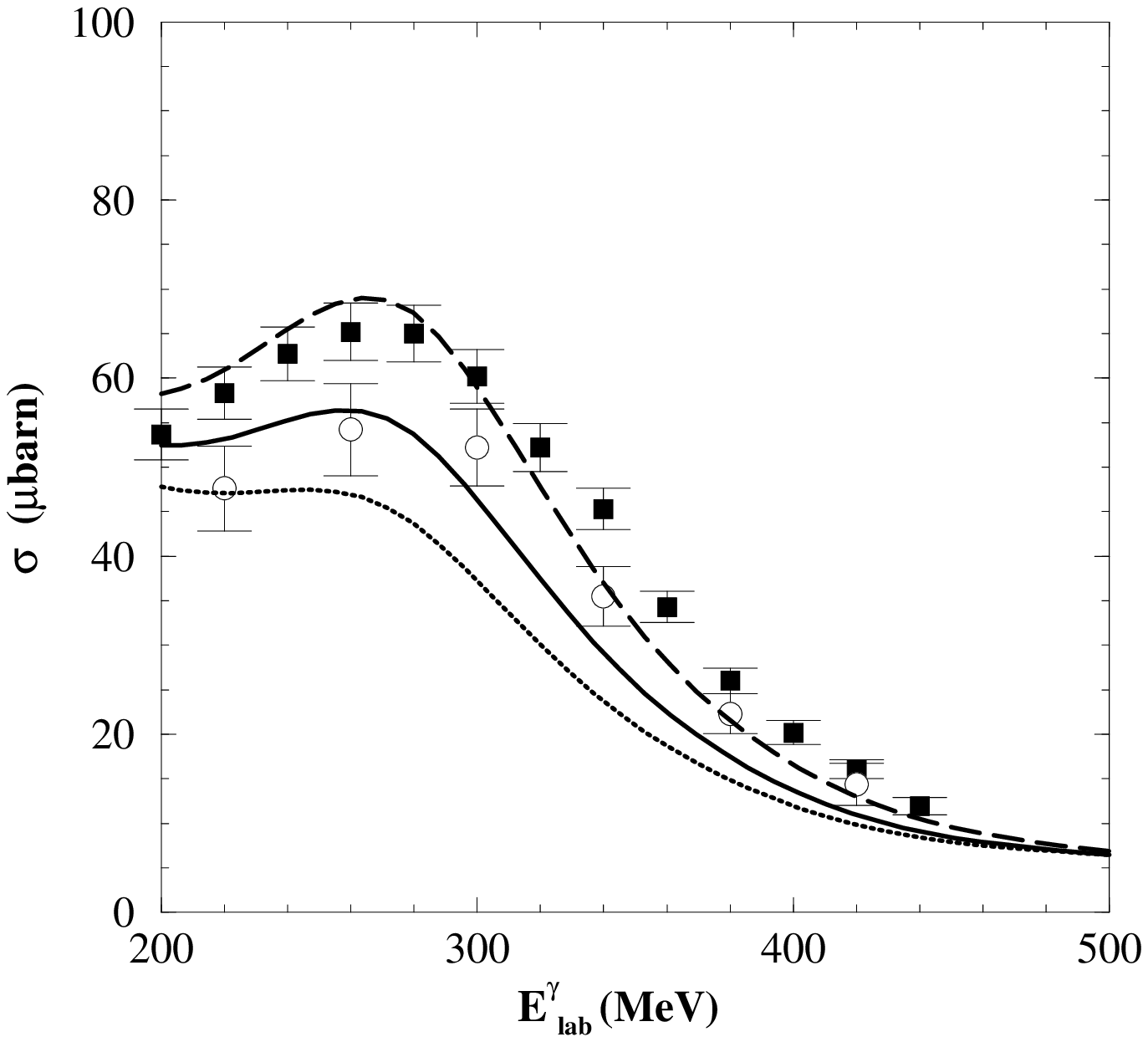,height=12cm,width=12cm}}

{\bf Fig. 4.1.}  Total cross-section of
photo disintegration of the deuteron 
as a function of the photon lab energy.
The sensitivity of the results on the coupling constant                
$G^{N\Delta}_{M1}$
in the single-baryon current
is shown.
The results refer to 
three values of $G^{N\Delta}_{M1}$: 
3.65 (dotted line), 5.16 (solid line) and 6.37 (long dashed one).
The experimental data are from Ref. \cite{kose} (squares) and Ref. \cite{aren}
(circles).
\end{figure}

\newpage
\begin{figure}
\mbox{\psfig{figure=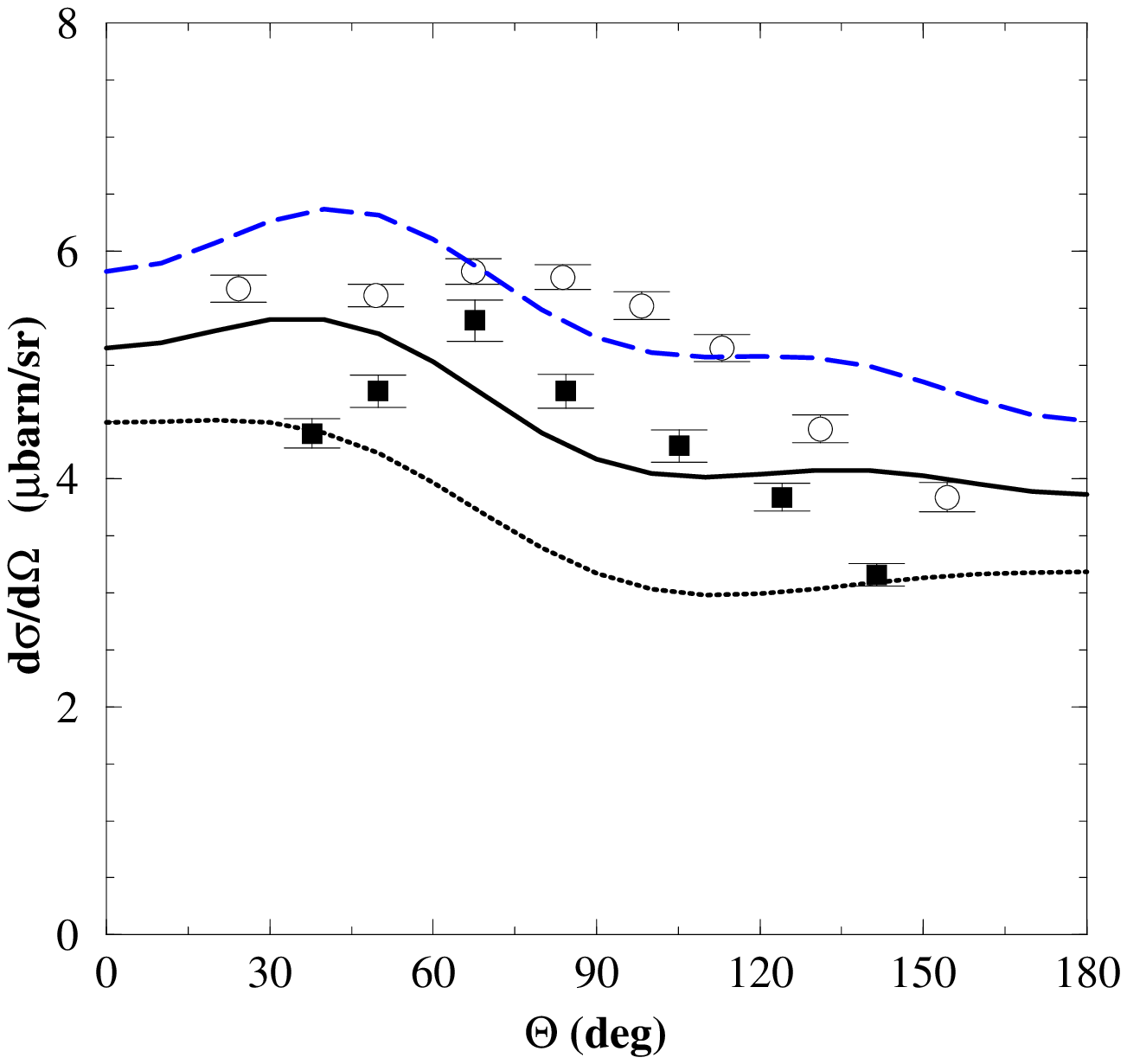,height=12cm,width=12cm}}

{\bf Fig. 4.2(a).} Differential cross-section of 
photo disintegration of the deuteron for the photon energy $E^{\gamma}_{lab} = 260$ MeV. 
The sensitivity of the results on the coupling constant 
$G^{N\Delta}_{M1}$ 
in the single-baryon current
is shown.
The results refer to 
three values of $G^{N\Delta}_{M1}$: 
3.65 (dotted line), 5.16 (solid line) and 6.37 (long dashed one).
The experimental data are from Ref. \cite{kose} (squares) and Ref. \cite{aren} 
(circles).
\end{figure}

\newpage
\begin{figure}
\mbox{\psfig{figure=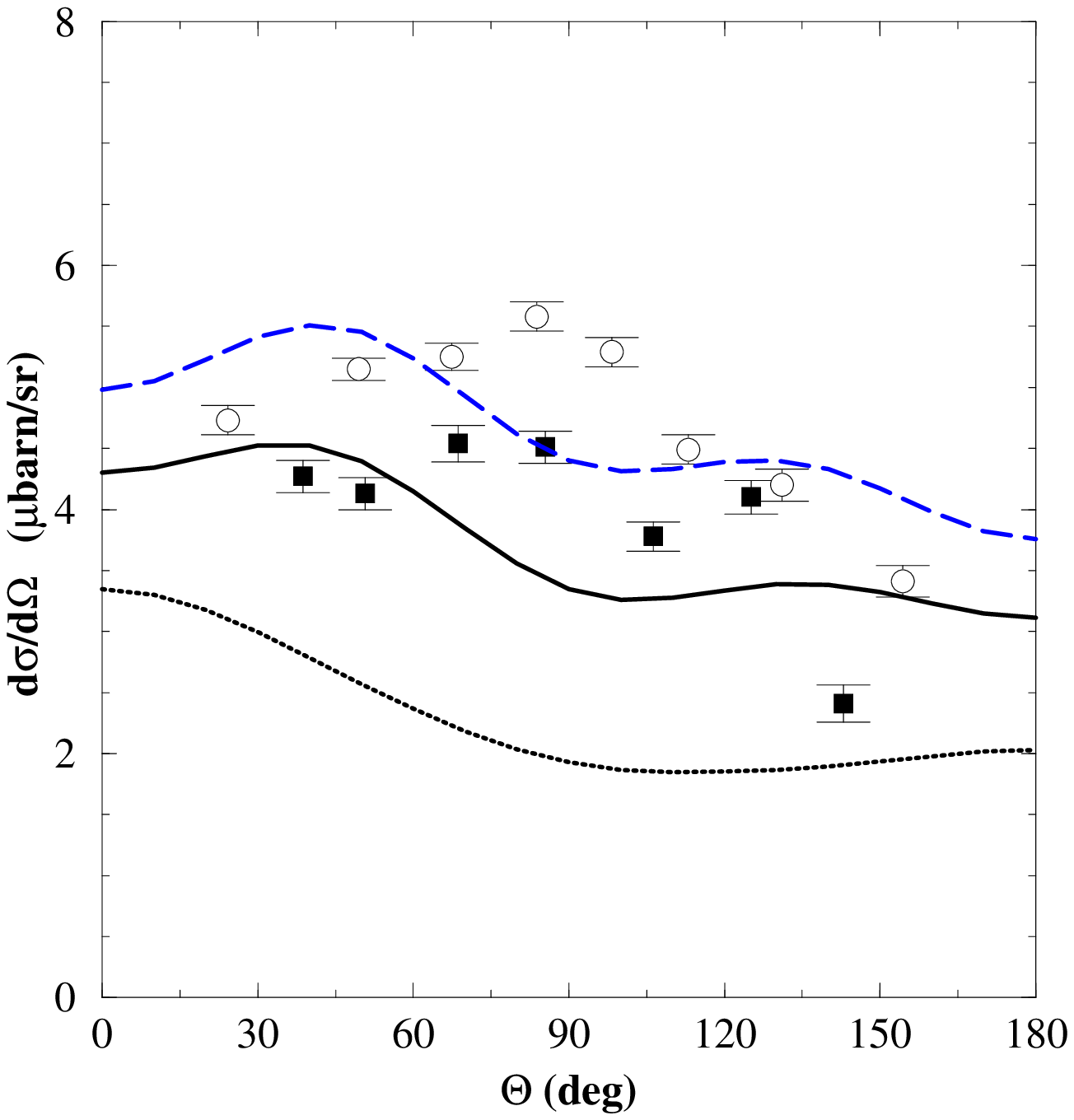,height=12cm,width=12cm}}

{\bf Fig. 4.2(b).} Differential cross-section of 
photo disintegration of the deuteron for the photon lab energy $E^{\gamma}_{lab} = 300$ MeV.
The legend corresponds to Fig. 4.2(a).
\end{figure}

\begin{figure}
\mbox{\psfig{figure=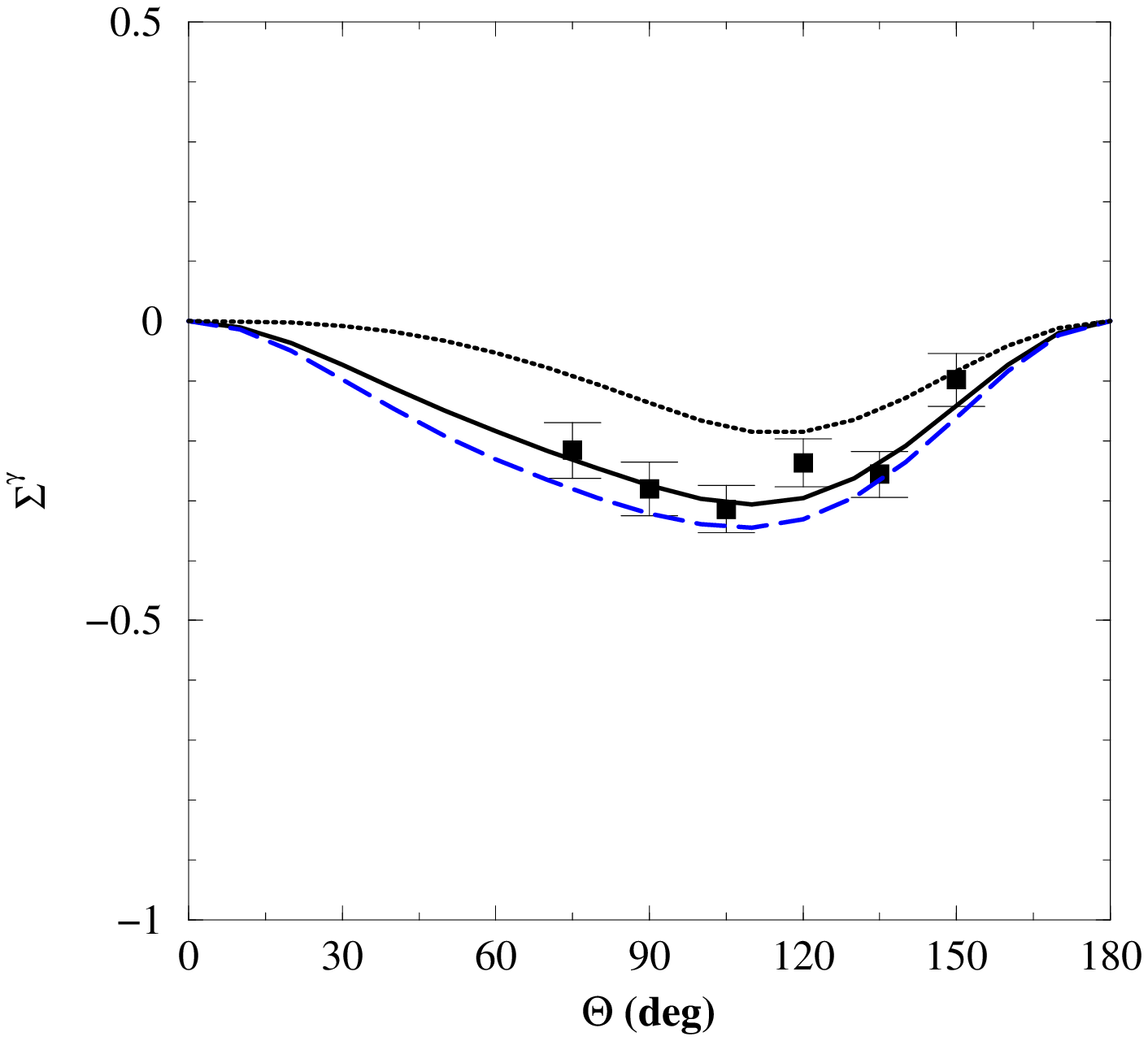,height=12cm,width=12cm}}

{\bf Fig. 4.3(a).}  Photon asymmetry of 
photo disintegration of the deuteron for the photon energy $E^{\gamma}_{lab} = 260$ MeV.
The sensitivity of the results on the coupling constant                
$G^{N\Delta}_{M1}$
in the single-baryon current
is shown.
The results refer to 
three values of $G^{N\Delta}_{M1}$: 
3.65 (dotted line), 5.16 (solid line) and 6.37 (long dashed one).
The experimental data are from Ref. \cite{gorb} (squares).
\end{figure}

\newpage

\begin{figure}
\mbox{\psfig{figure=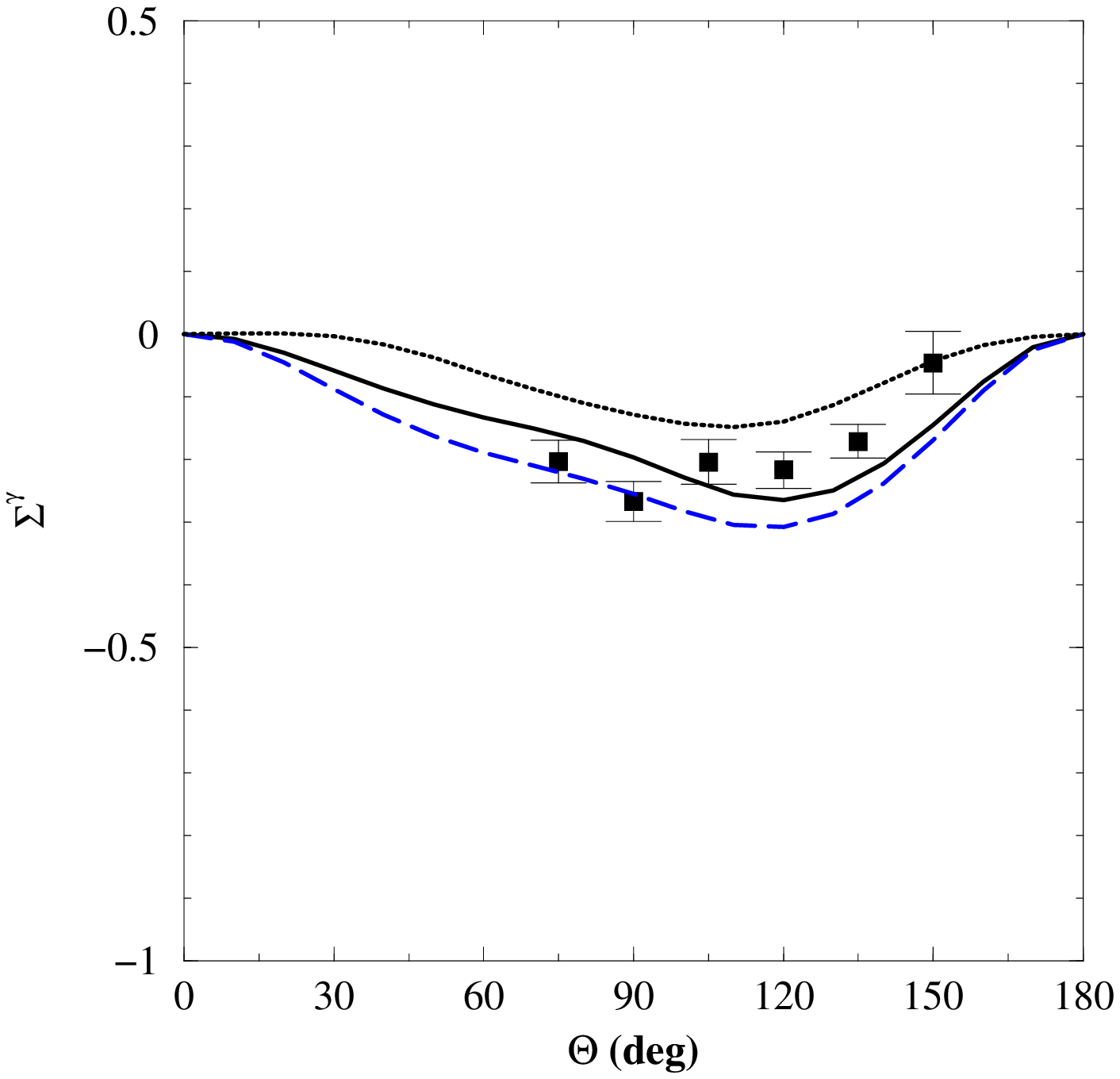,height=12cm,width=12cm}}

{\bf Fig. 4.3(b).}  Photon asymmetry of the
photo disintegration of the deuteron for the photon energy 
photo disintegration for the photon energy $E^{\gamma}_{lab} = 300$ MeV.
The legend corresponds to Fig. 4.3(a). 
\end{figure}

\newpage

\begin{figure}
\mbox{\psfig{figure=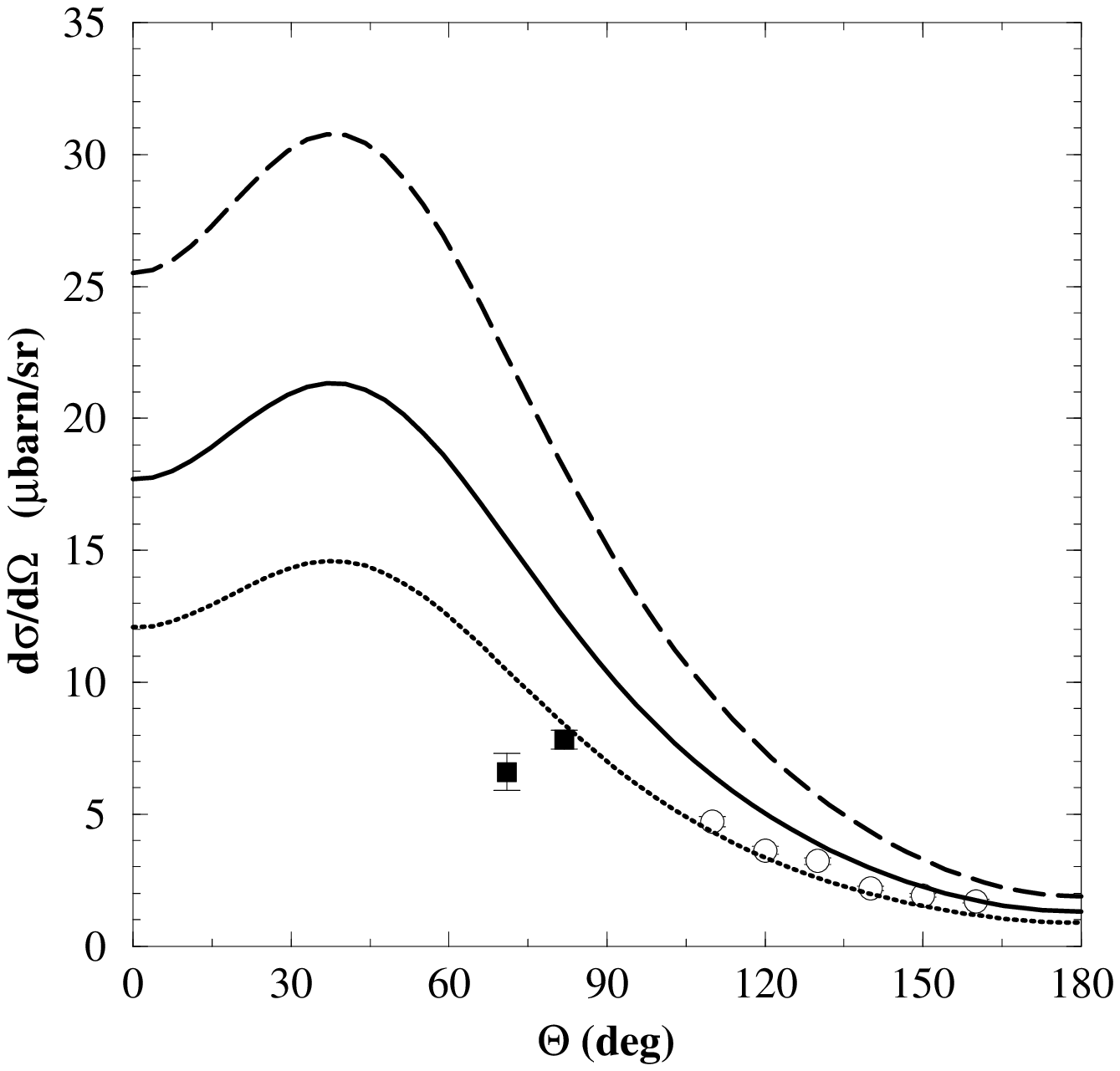,height=12cm,width=12cm}}

{\bf Fig. 4.4(a).}  Differential cross section of
photo pionproduction on the deuteron for the photon energy $E_{lab}^\g = 260$ MeV.
The sensitivity of the results on the coupling constant
$G^{N\Delta}_{M1}$
in the single-baryon current
is shown.
The results refer to 
three values of $G^{N\Delta}_{M1}$: 
3.65 (dotted line), 5.16 (solid line) and 6.37 (long dashed one).
The experimental data are from Ref. \cite{hol73} (circles) and Ref. \cite{bou74}
(squares).

\end{figure}

\newpage

\begin{figure}
\mbox{\psfig{figure=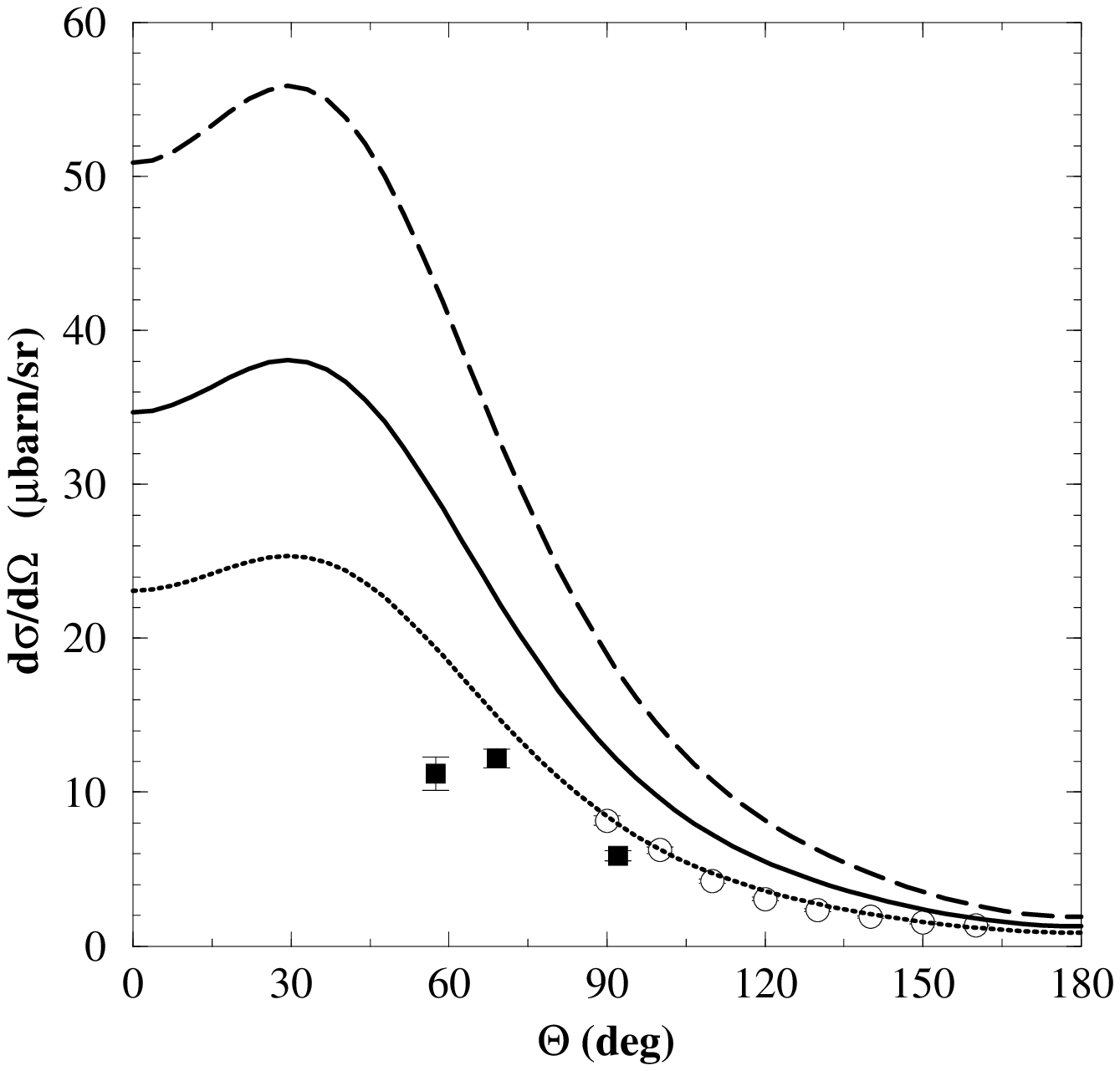,height=12cm,width=12cm}}

{\bf Fig. 4.4(b).}  Differential cross section of
photo pionproduction on the deuteron for the photon energy $E_{lab}^\g = 300$ MeV.
The legend corresponds to  Fig. 4.4(a). 
\end{figure}

\newpage

\begin{figure}
\mbox{\psfig{figure=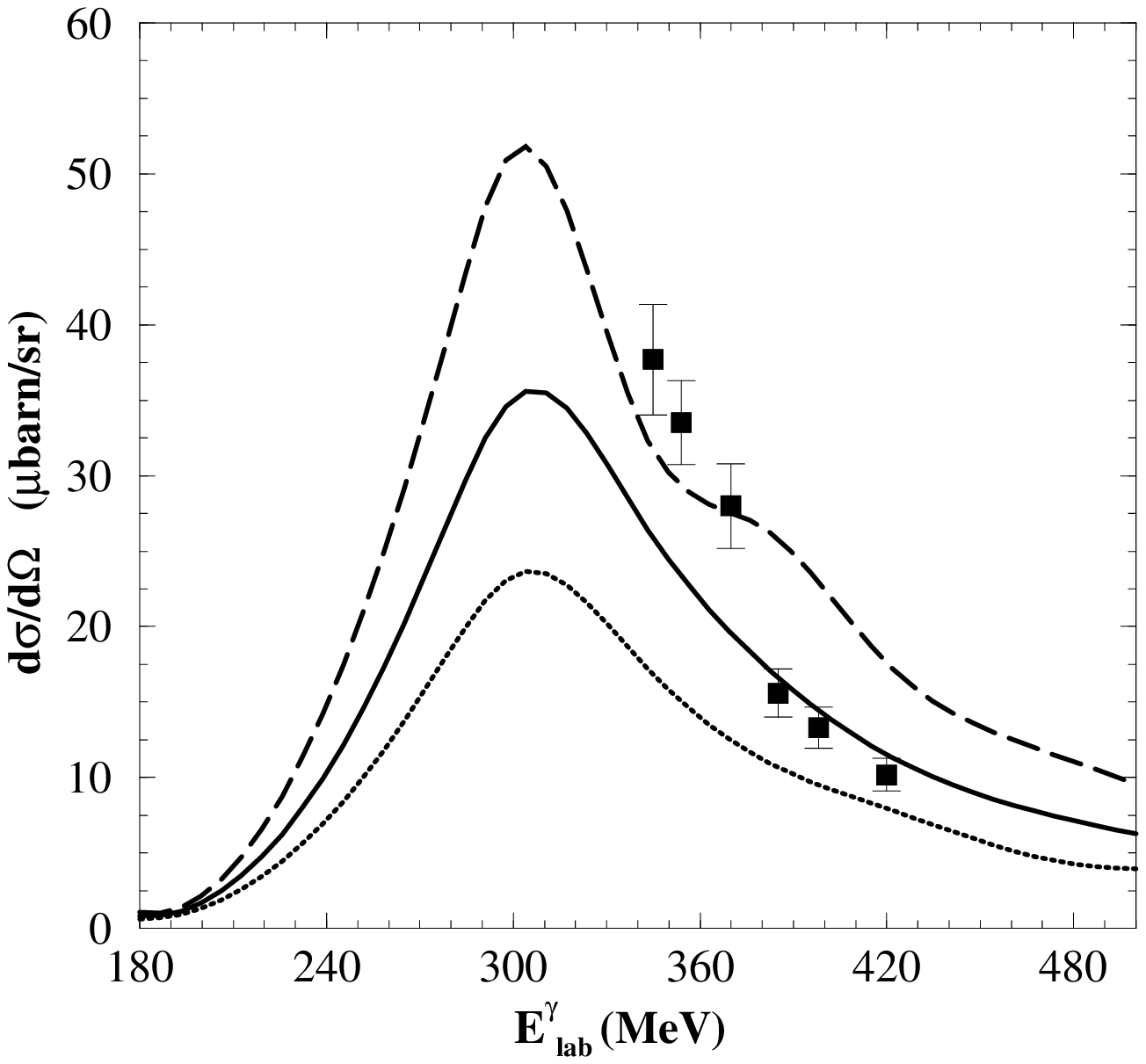,height=12cm,width=12cm}}

{\bf Fig. 4.5.}  Differential cross section of
photo pionproduction on the deuteron at the fixed pion angle $ 6^{\circ}$ in the c.m. system
as a function of the photon energy.
The sensitivity of the results on the coupling constant
$G^{N\Delta}_{M1}$
in the single-baryon current
is shown. 
The results refer to 
three values of $G^{N\Delta}_{M1}$: 
3.65 (dotted line), 5.16 (solid line) and 6.37 (long dashed one).
The experimental data are from Ref. \cite{hil75} (squares).

\end{figure}

\newpage

\begin{figure}
\mbox{\psfig{figure=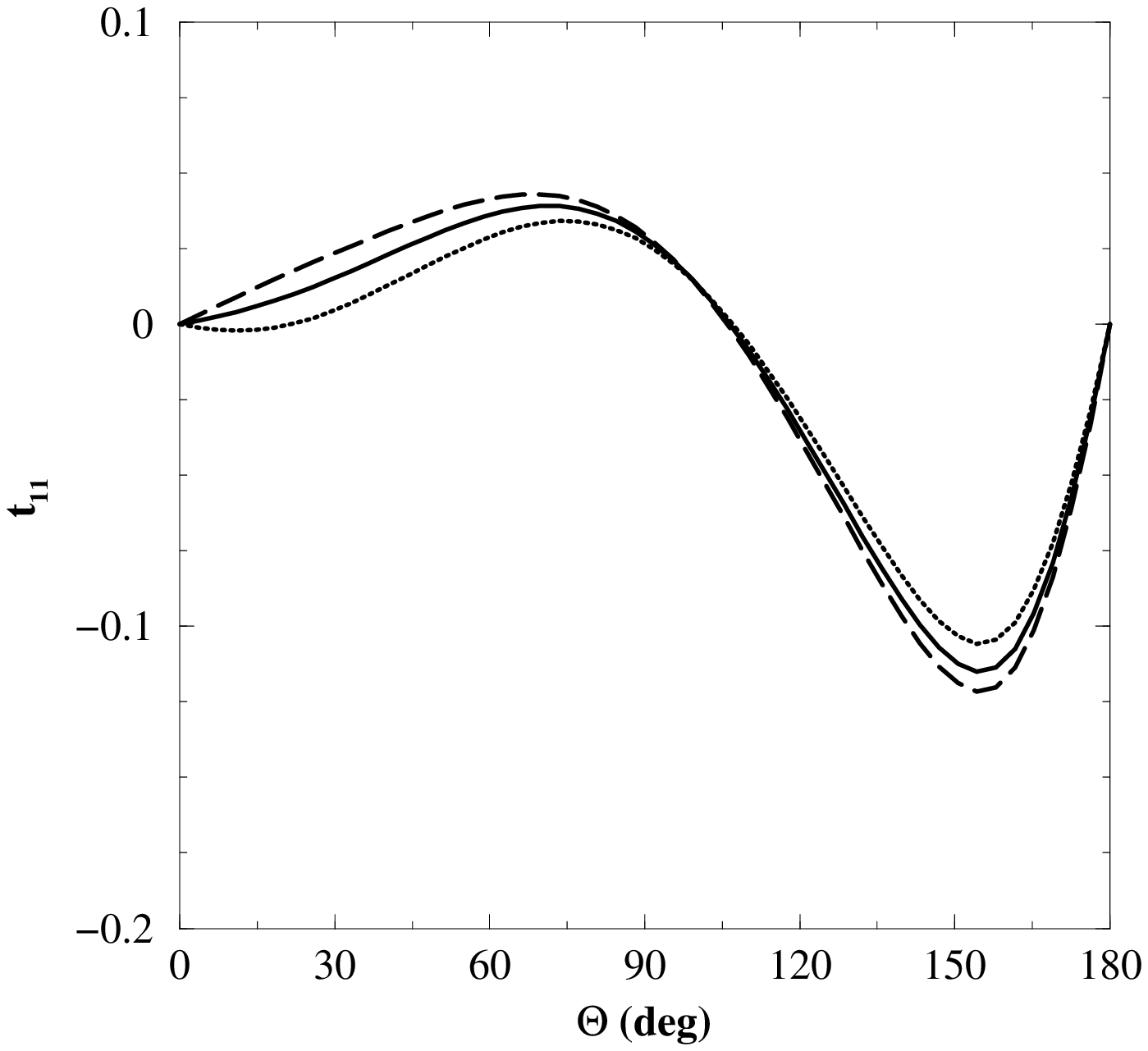,height=12cm,width=12cm}}

{\bf Fig. 4.6(a).}  Tensor polarization of photo pionproduction on the deuteron  for
the photon lab energy
$E_{lab}^\g = 260$ MeV.
The sensitivity of the results on the coupling constant
$G^{N\Delta}_{M1}$
in the single-baryon current
is shown.
The results refer to 
three values of $G^{N\Delta}_{M1}$: 
3.65 (dotted line), 5.16 (solid line) and 6.37 (long dashed one).
\end{figure}

\newpage

\begin{figure}
\mbox{\psfig{figure=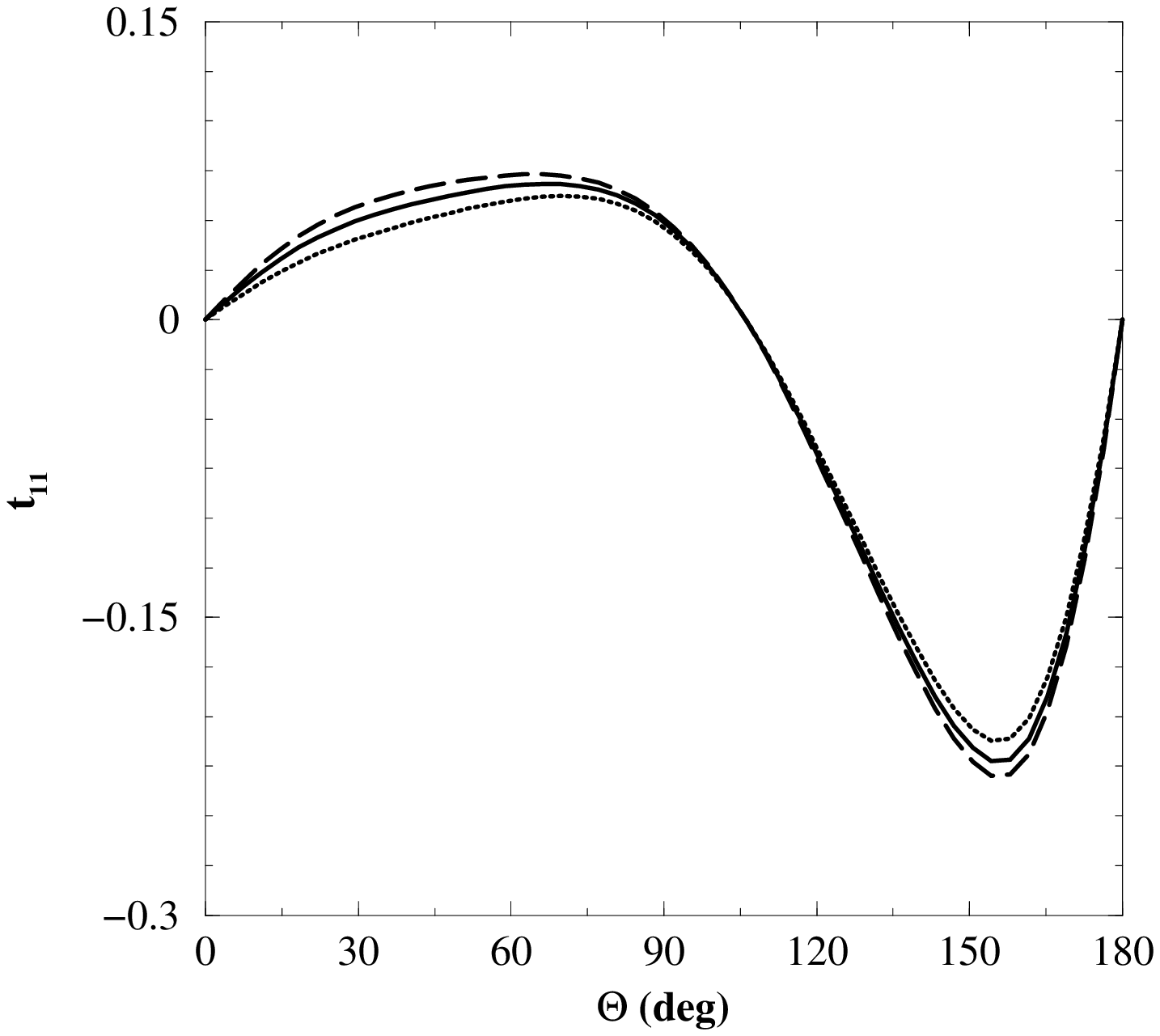,height=12cm,width=12cm}}

{\bf Fig. 4.6(b).}
Tensor polarization of photo pionproduction on the deuteron  for
the photon lab energy
$E_{lab}^\g = 300$ MeV.
The legend corresponds to Fig. 4.6(a).
\end{figure}

\newpage

\begin{figure}
\mbox{\psfig{figure=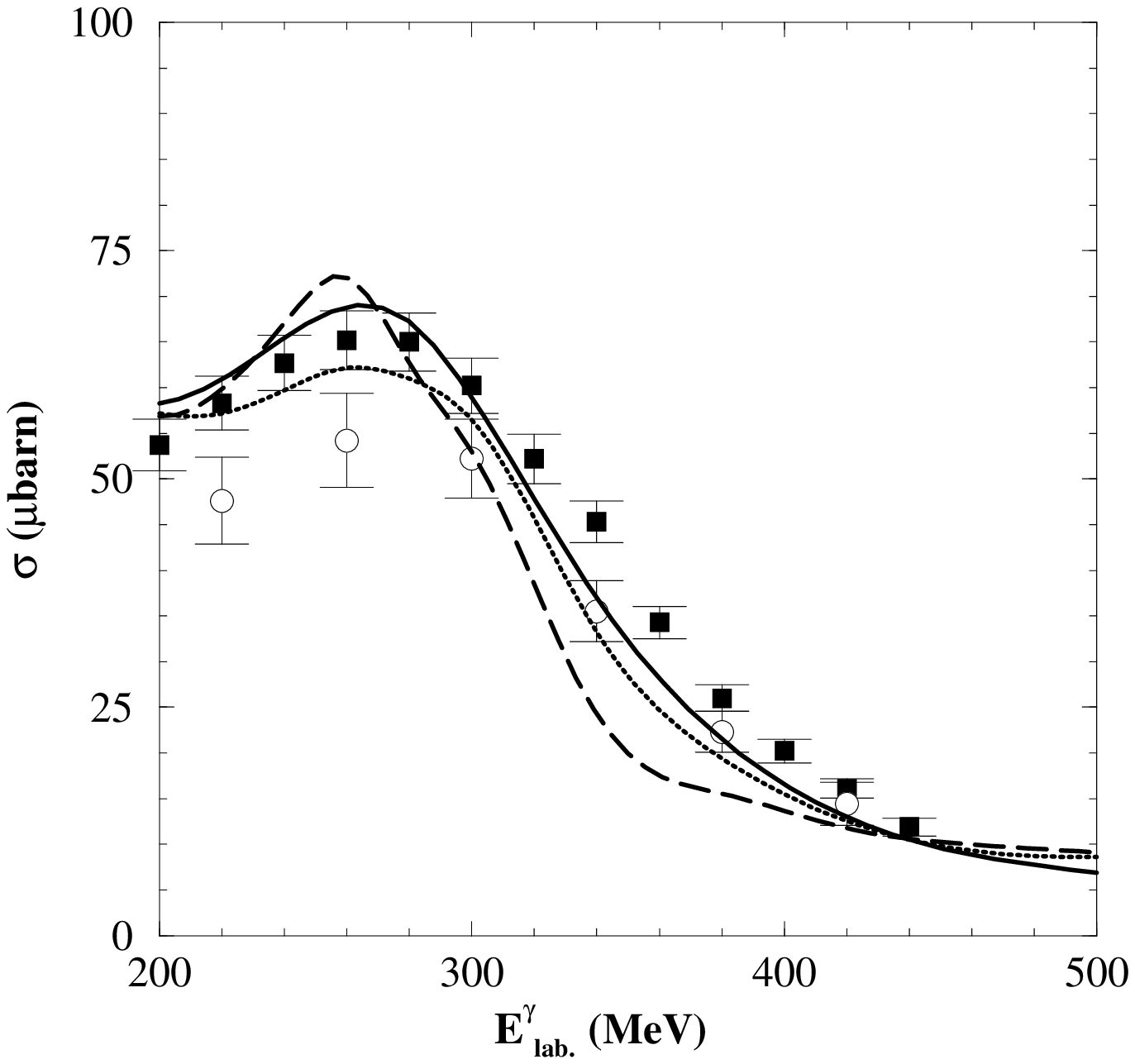,height=12cm,width=12cm}}

{\bf Fig. 4.7.}  Total cross section of 
photo disintegration of the deuteron as function of the photon lab energy. 
Sensitivity of results on three different choices of the 
nucleon-$\D$ potential in Figs. 1(c) and 1(d) is shown. 
They were chosen as follows:
The nucleon-$\D$ potential as described in Subsect. 3.1 is
the reference potential (solid line), the two others are  based on
meson exchange (long-dashed
line) and on a nonrelativistic quark model  (dotted line). 
The value 
of the  coupling strength  $G^{N\Delta}_{M1}$ is  6.37.
The experimental data are from Ref. \cite{kose} (squares) and Ref. \cite{aren}
(circles).
\end{figure}

\newpage

\begin{figure}
\mbox{\psfig{figure=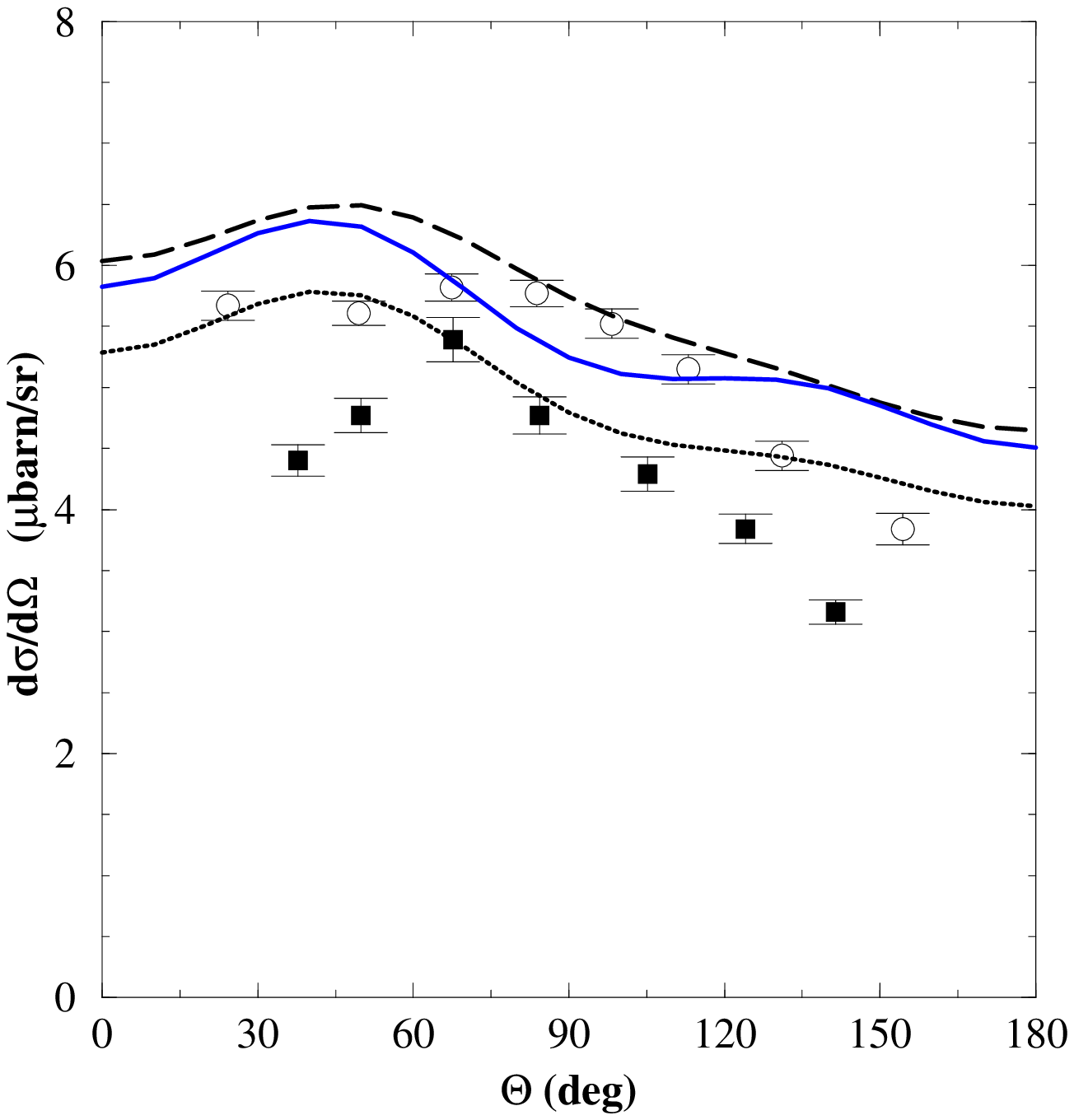,height=12cm,width=12cm}}

{\bf Fig. 4.8(a).}  Differential cross section of
photo disintegration of the deuteron for the photon lab energy
$E_{lab}^\g = 260$ MeV.
Sensitivity of results on three different choices of the
nucleon-$\D$ potential in Figs. 1(c) and 1(d) is shown.
They were chosen as follows:
The nucleon-$\D$ potential as described in Subsect. 3.1 is
the reference potential (solid line), the two others are  based on
meson exchange (long-dashed
line) and on a nonrelativistic quark model  (dotted line).                  
The value
of the  coupling strength  $G^{N\Delta}_{M1}$ is  6.37.
The experimental data are from Ref. \cite{kose} (squares) and Ref. \cite{aren}
(circles).
\end{figure}

\newpage

\begin{figure}
\mbox{\psfig{figure=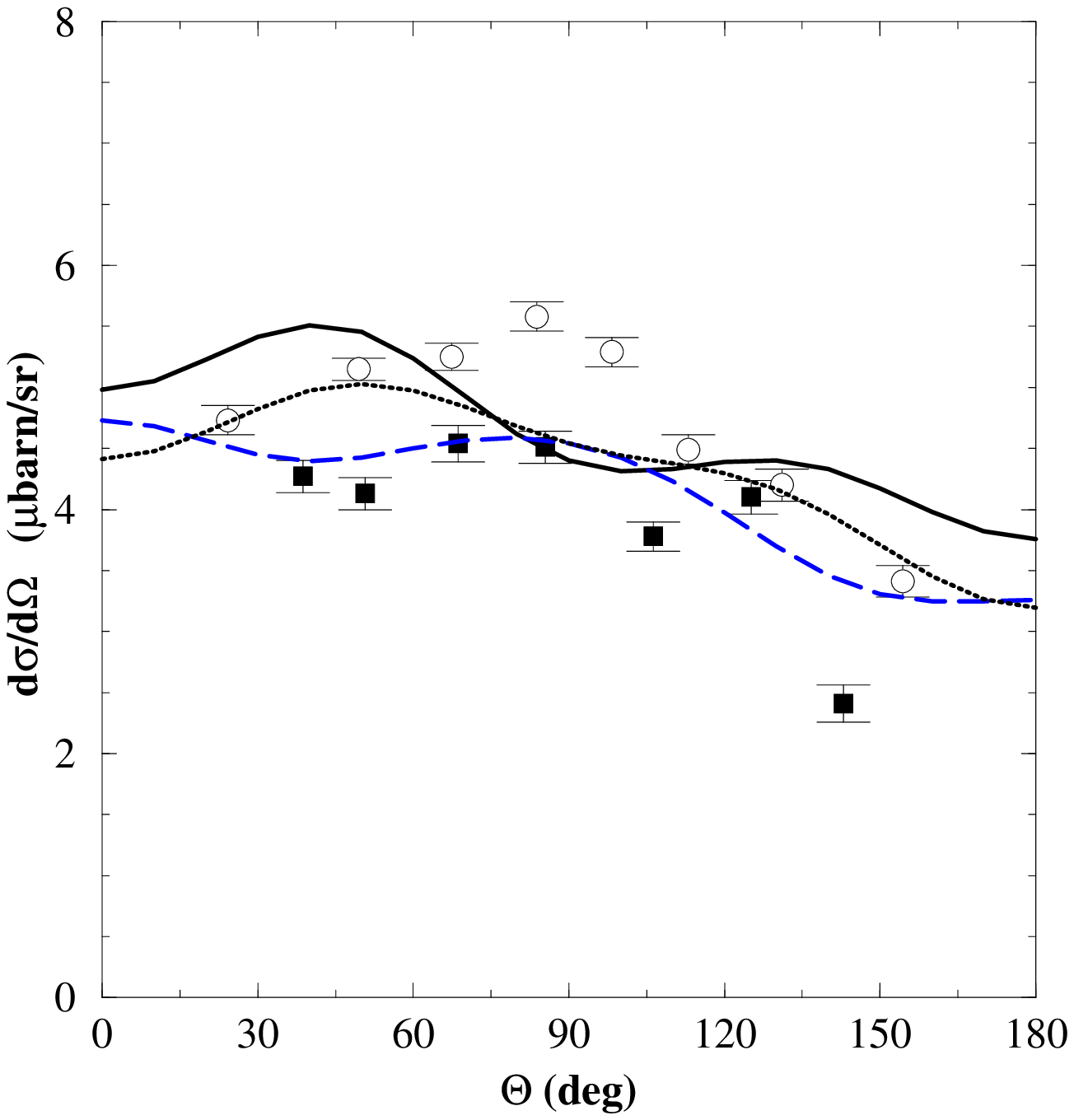,height=12cm,width=12cm}}

{\bf Fig. 4.8(b).}  Differential cross section of 
photo disintegration of the deuteron for the photon lab energy
$E_{lab}^\g = 300$ MeV.
The legend corresponds to  Fig. 4.8(a).
\end{figure}

\newpage

\begin{figure}
\mbox{\psfig{figure=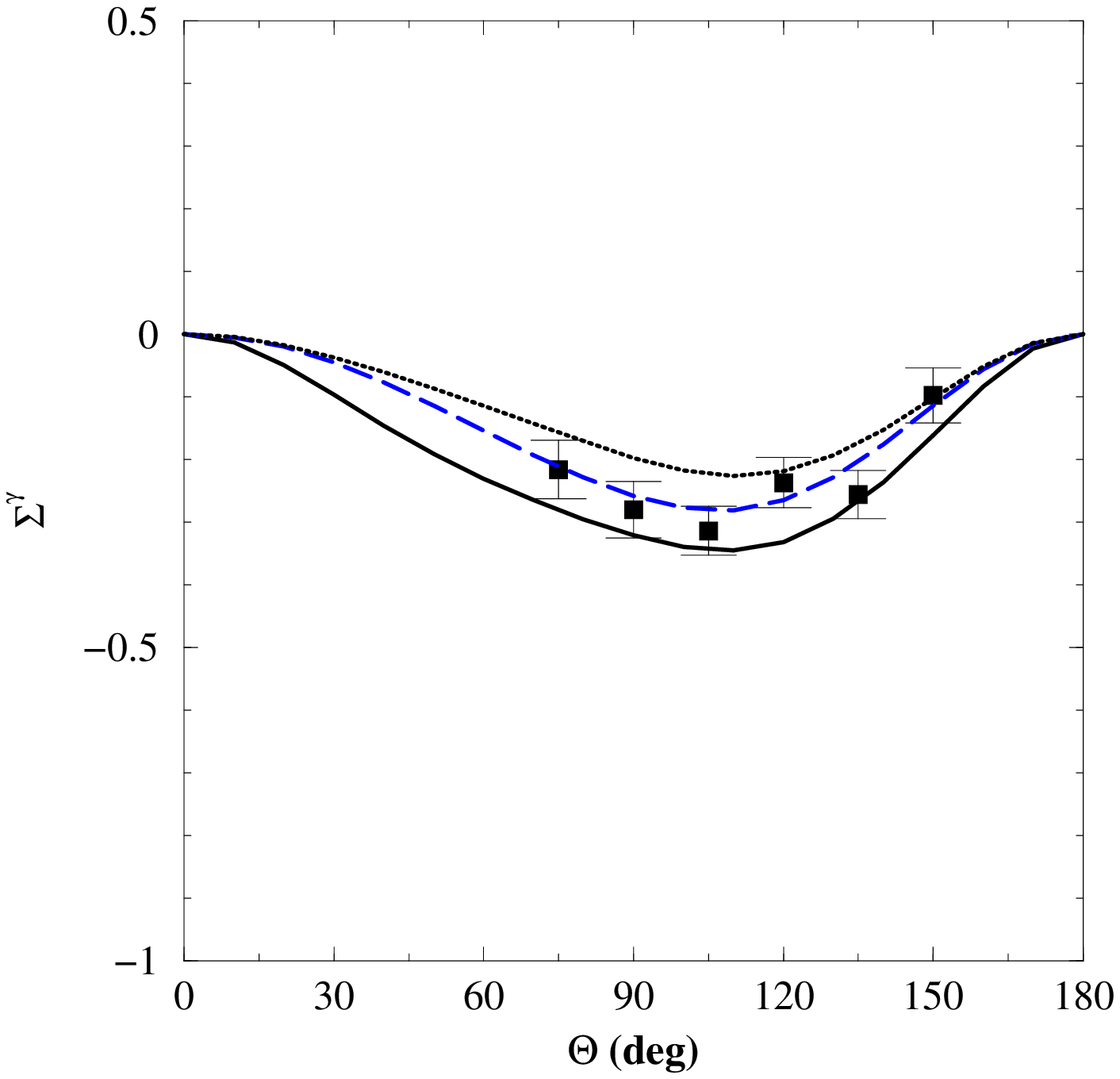,height=12cm,width=12cm}}

{\bf Fig. 4.9(a).}  Photon asymmetry of 
photo disintegration of the deuteron for the photon lab energy
$E_{lab}^\g = 260$ MeV.
Sensitivity of results on three different choices of the
nucleon-$\D$ potential in Figs. 1(c) and 1(d) is shown.
They were chosen as follows:
The nucleon-$\D$ potential as described in Subsect. 3.1 is
the reference potential (solid line), the two others are  based on
meson exchange (long-dashed
line) and on a nonrelativistic quark model  (dotted line).                  
The value
of the  coupling strength  $G^{N\Delta}_{M1}$ is  6.37.
The experimental data are from Ref. \cite{gorb}. 
\end{figure}

\newpage

\begin{figure}
\mbox{\psfig{figure=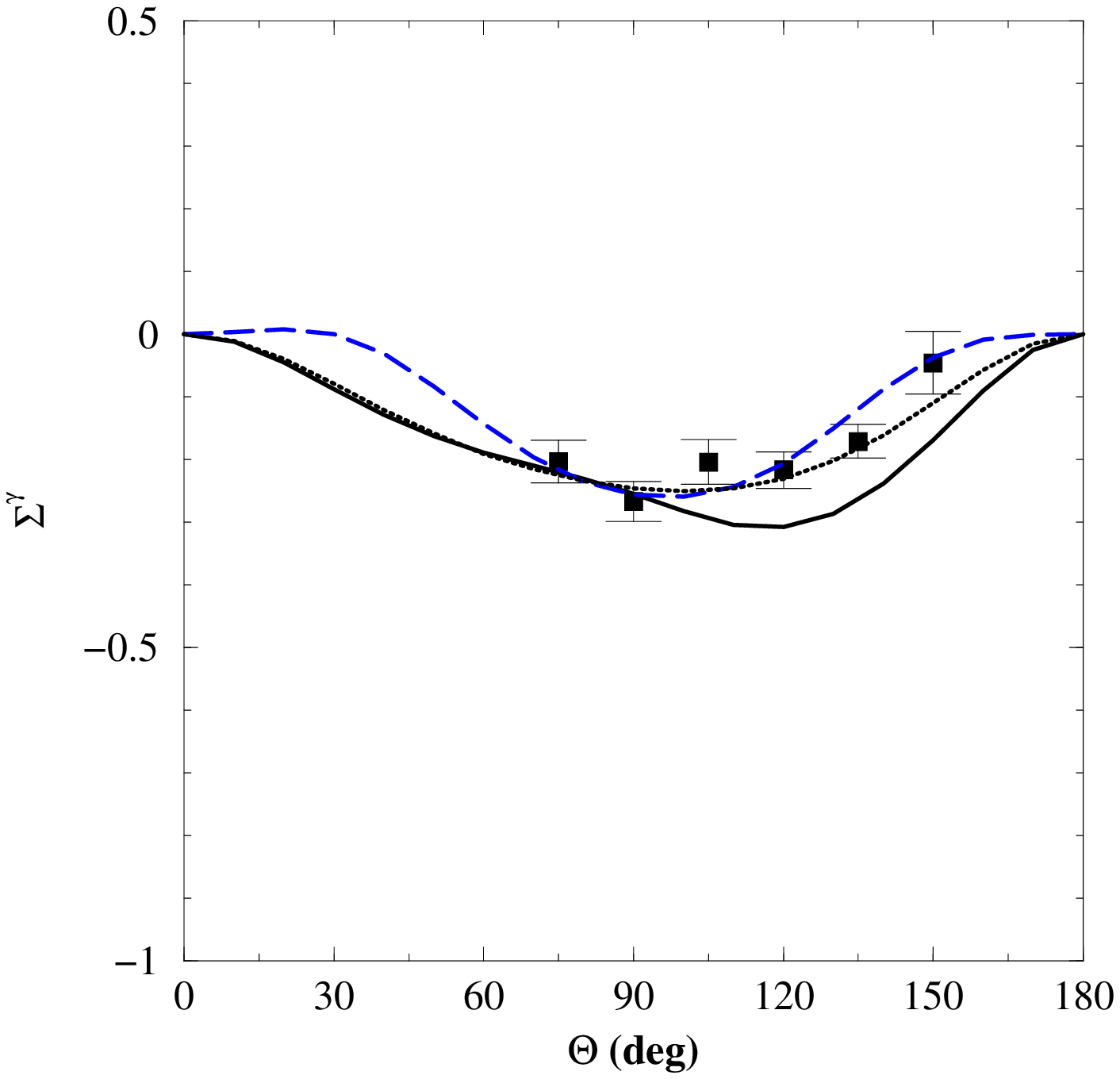,height=12cm,width=12cm}}

{\bf Fig. 4.9(b).} 
Photon asymmetry of
photo disintegration of the deuteron for the photon lab energy
$E_{lab}^\g = 300$ MeV.
The legend corresponds to  Fig. 4.9(a).
\end{figure}

\newpage

\begin{figure}
\mbox{\psfig{figure=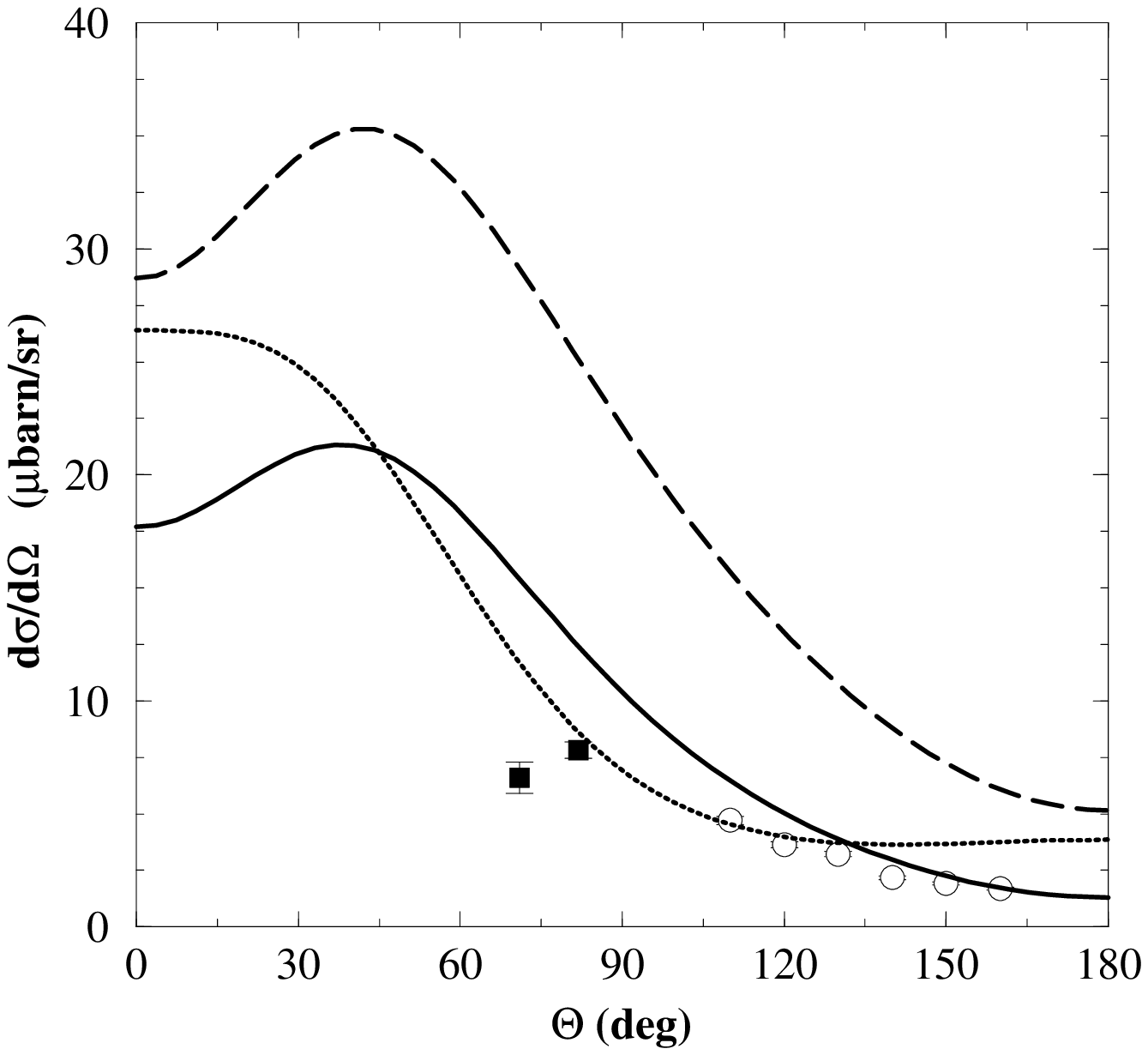,height=12cm,width=12cm}}

{\bf Fig. 4.10(a).}  Differential cross section of
photo pionproduction on the deuteron for the photon lab energy
$E_{lab}^\g = 260$ MeV.
Sensitivity of results on three different choices of the
nucleon-$\D$ potential in Figs. 1(c) and 1(d) is shown.
They were chosen as follows:
The nucleon-$\D$ potential as described in Subsect. 3.1 is
the reference potential (solid line), the two others are  based on
meson exchange (long-dashed
line) and on a nonrelativistic quark model  (dotted line).                  
The value
of the  coupling strength  $G^{N\Delta}_{M1}$ is  5.16.
The experimental data are from Ref. \cite{hol73} (circles) and Ref. \cite{bou74}
(squares).
\end{figure}

\newpage

\begin{figure}
\mbox{\psfig{figure=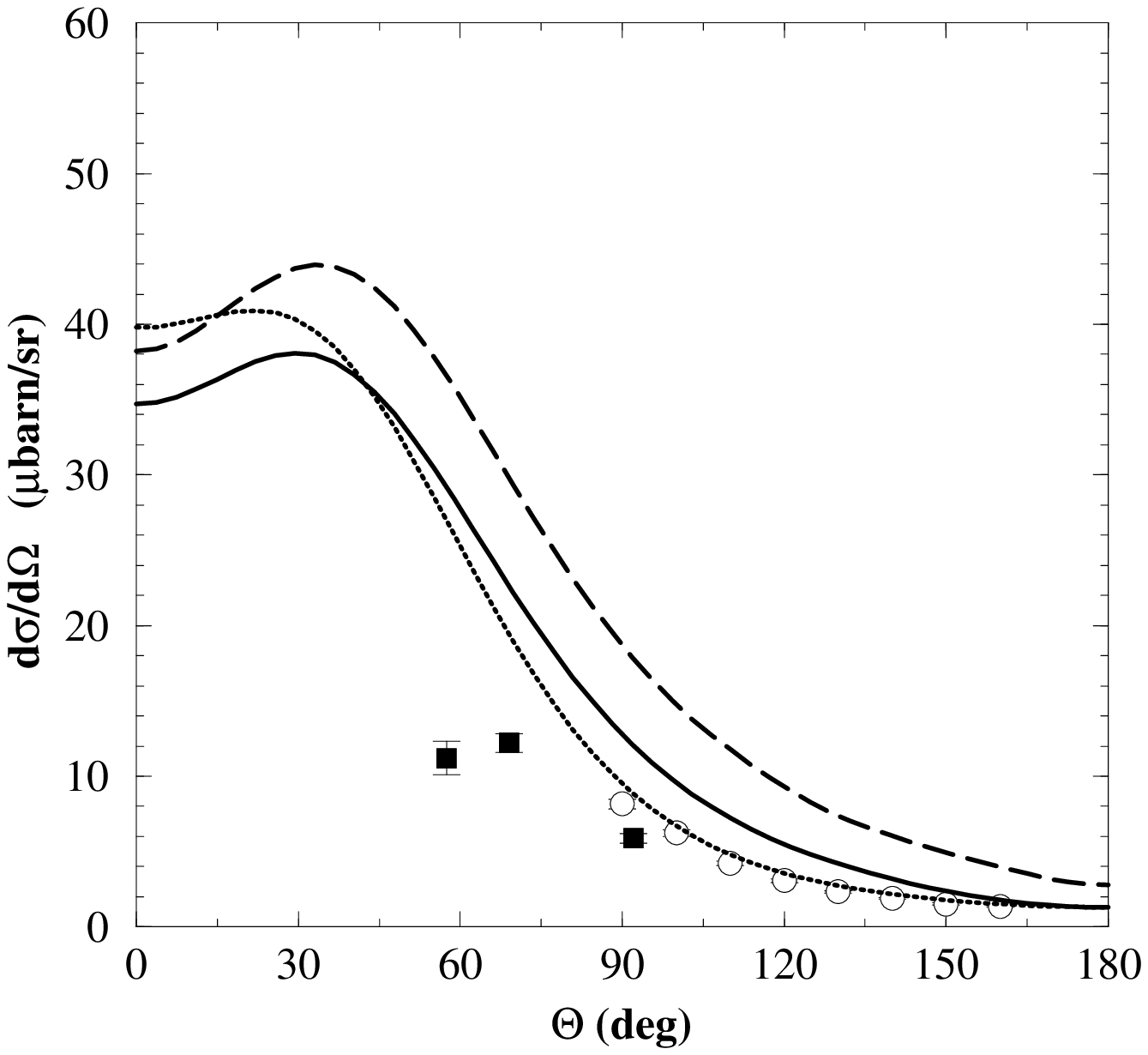,height=12cm,width=12cm}}

{\bf Fig. 4.10(b).}  Differential cross section of
photo pionproduction on the deuteron for the photon lab energy
$E_{lab}^\g = 300$ MeV.
The legend corresponds to  Fig. 4.10(a).
\end{figure}

\newpage
\clearpage

\begin{figure}
\mbox{\psfig{figure=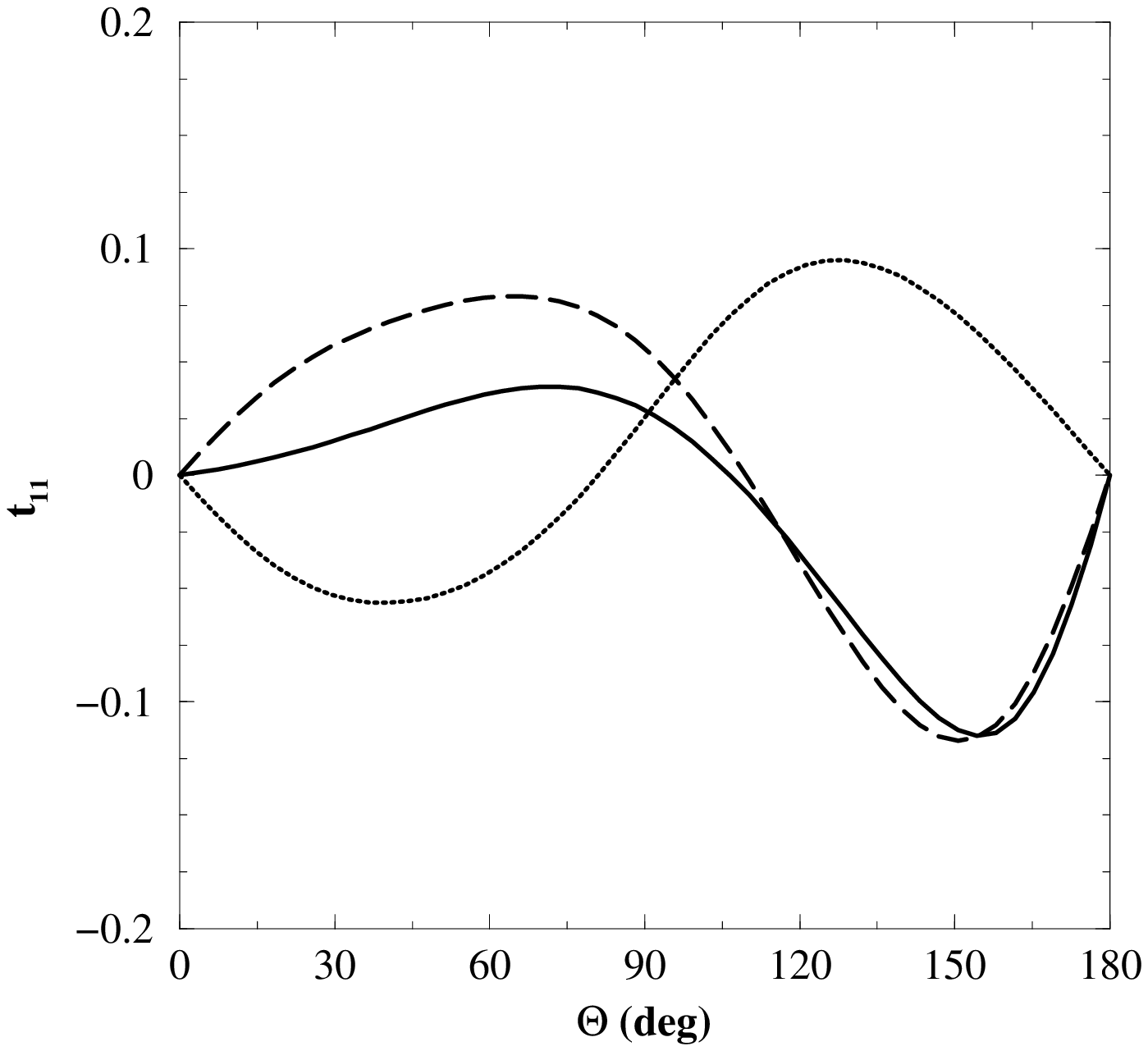,height=12cm,width=12cm}}

{\bf Fig. 4.11(a).}  Vector polarization of 
photo pionproduction on the deuteron for the photon lab energy
$E_{lab}^\g = 260$ MeV.
Sensitivity of results on three different choices of the
nucleon-$\D$ potential in Figs. 1(c) and 1(d) is shown.
They were chosen as follows:
The nucleon-$\D$ potential as described in Subsect. 3.1 is
the reference potential (solid line), the two others are  based on
meson exchange (long-dashed
line) and on a nonrelativistic quark model  (dotted line).                  
The value
of the  coupling strength  $G^{N\Delta}_{M1}$ is  5.16.
\end{figure}

\newpage
\clearpage

\begin{figure}
\mbox{\psfig{figure=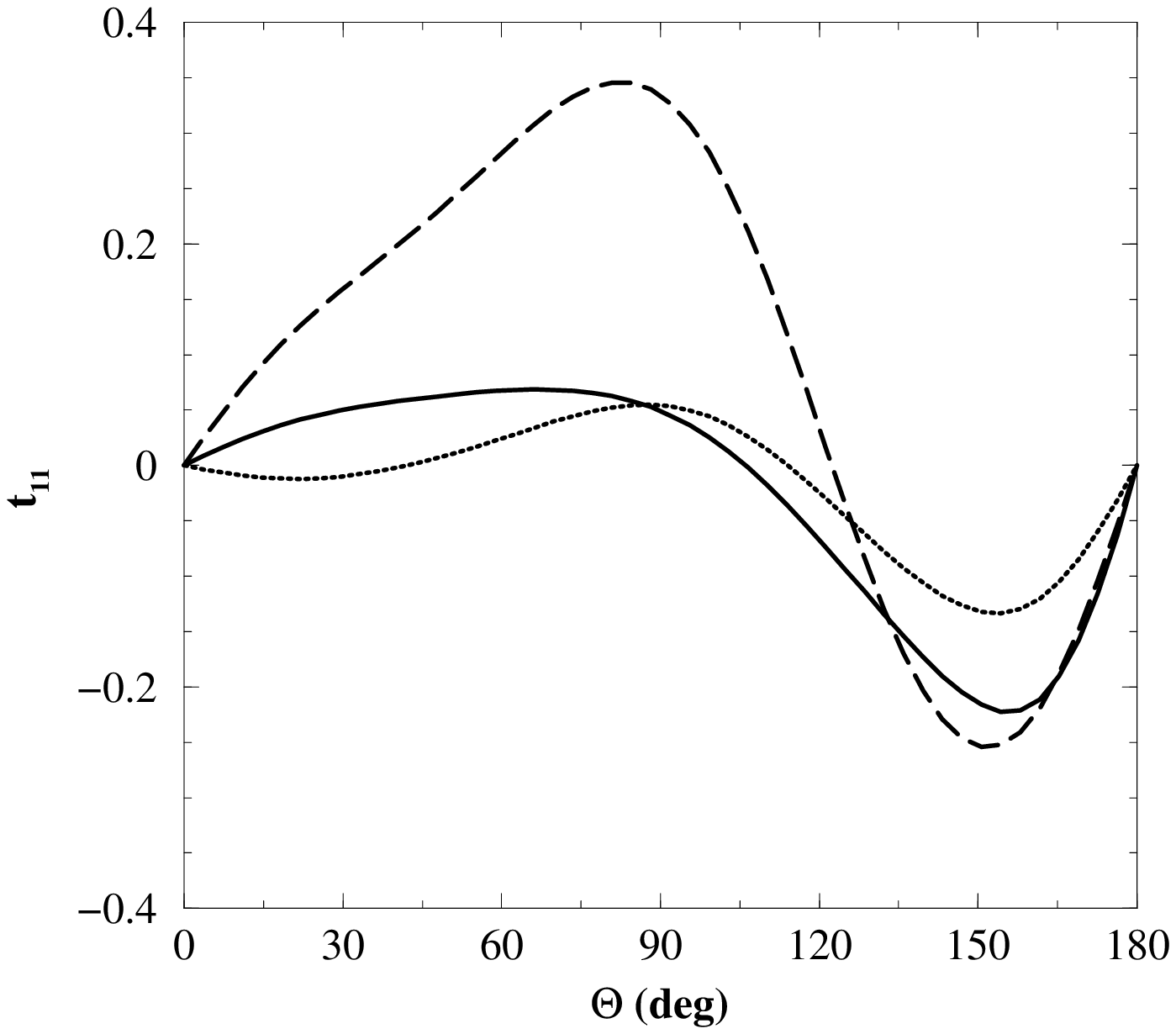,height=12cm,width=12cm}}

{\bf Fig. 4.11(b).}  
Vector polarization of
photo pionproduction on the deuteron for the photon lab energy
$E_{lab}^\g = 300$ MeV.
The legend  corresponds to  Fig. 4.11(a). 
\end{figure}

\newpage
\clearpage

\begin{figure}
\mbox{\psfig{figure=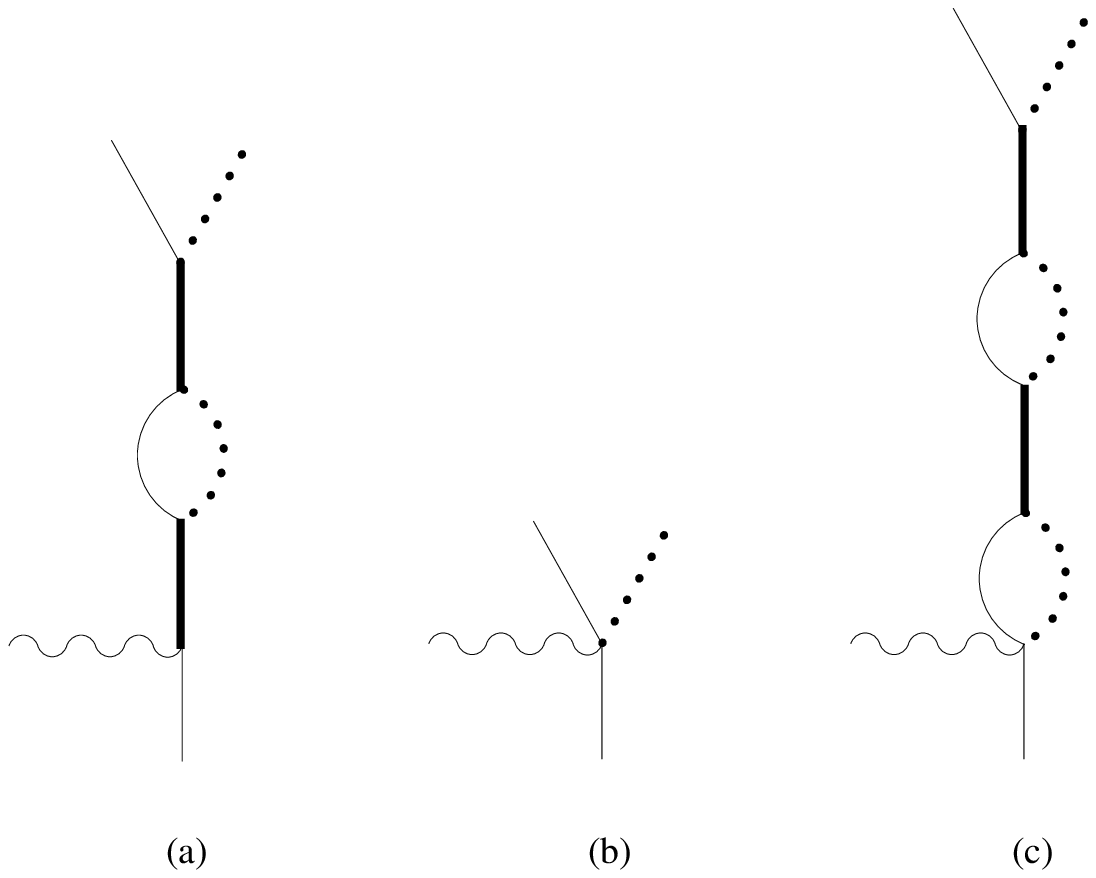,height=12cm,width=12cm}}

{\bf Fig. C.1.} Processes contributing to photo pionproduction  
on the single nucleon
in
the resonant multipole amplitudes $M_{1+}^{({3 \0 2})}$ and $E_{1+}^{({3 \0 2})}$
The processes correspond to the resonance (a), Born (b) and interference
contributions.
The hadronic rescattering occurring in the processes (a) and (c) up to 
infinite order is indicated only in a characteristic low oder.
\end{figure}

\newpage
\clearpage

\begin{figure}
\mbox{\psfig{figure=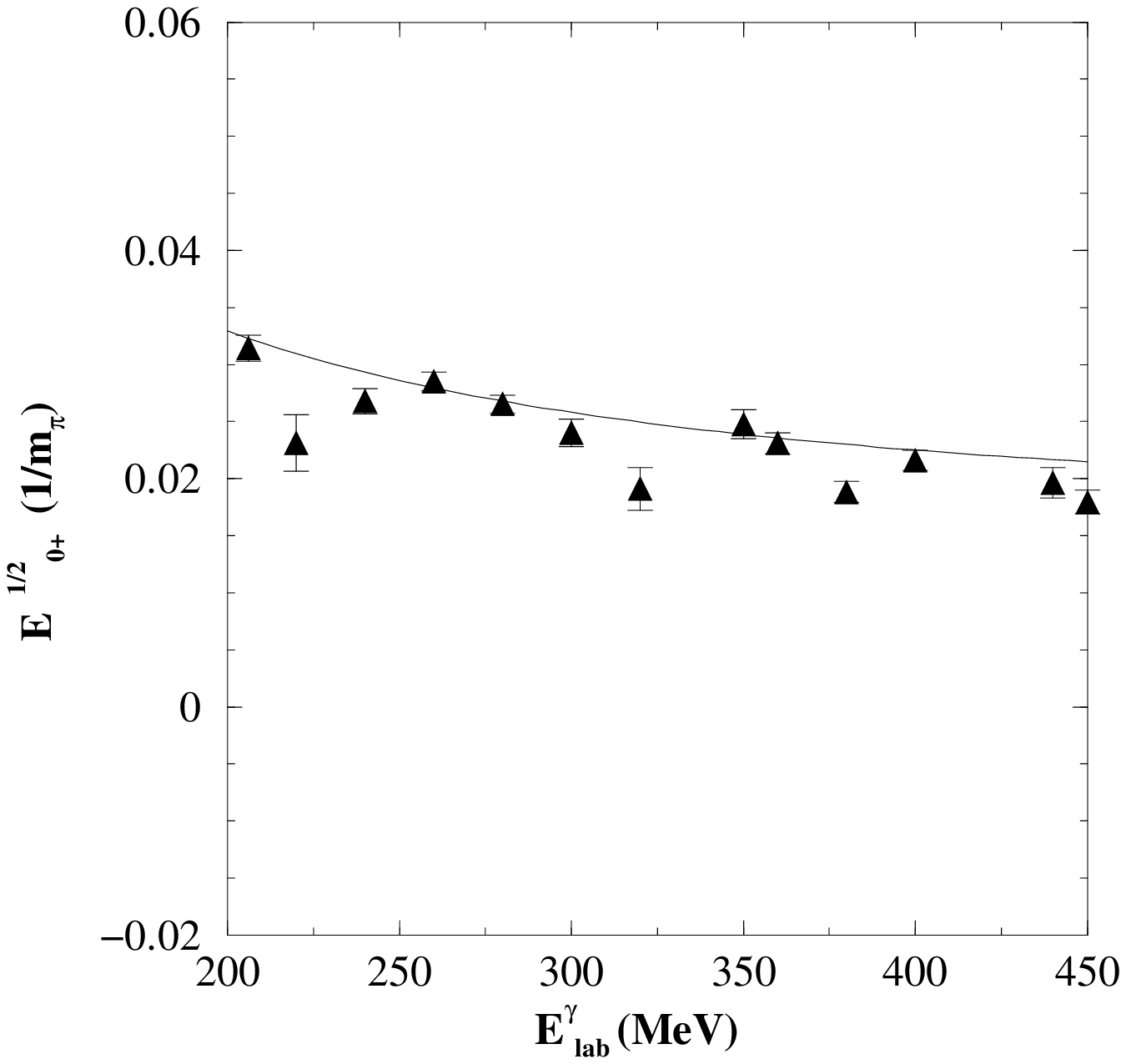,height=12cm,width=12cm}}

{\bf Fig. C.2.} $ E^{1/2}_{0+}$ multipole of
photo pionproduction on the single nucleon. 
The theoretical fit result (solid line) is compared with 
the experimental data of Ref. \cite{pfeil}.
\end{figure}

\newpage

\begin{figure}
\mbox{\psfig{figure=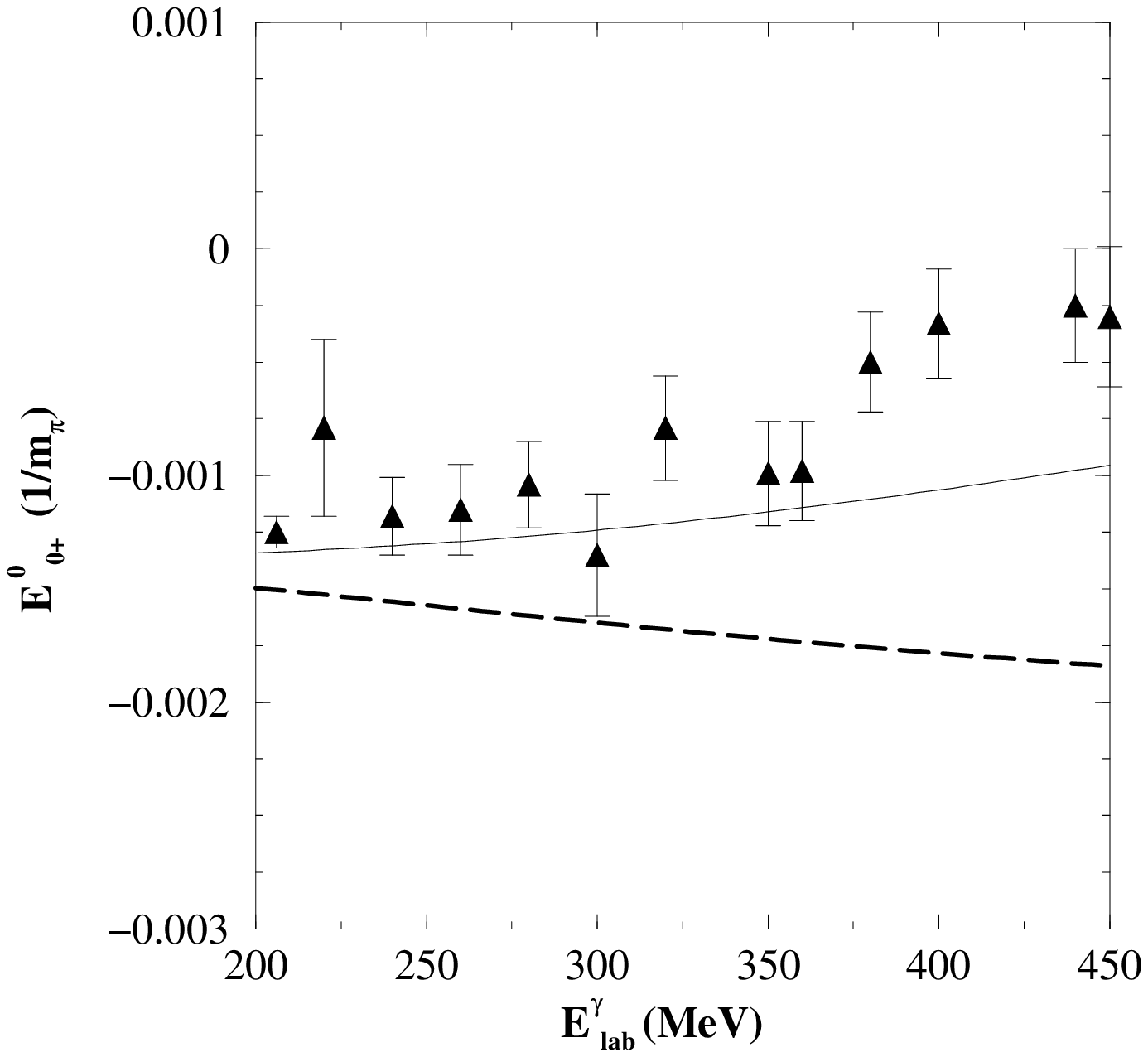,height=12cm,width=12cm}}

{\bf Fig. C.3.} $ E^{0}_{0+}$ multipole 
of
photo pionproduction on the single nucleon.
The theoretical fit result (solid line) is compared with
the experimental data of Ref. \cite{pfeil}.
The dashed line represents the result without $\rho$-meson
          contribution.
\end{figure}

\newpage
\clearpage

\begin{figure}
\mbox{\psfig{figure=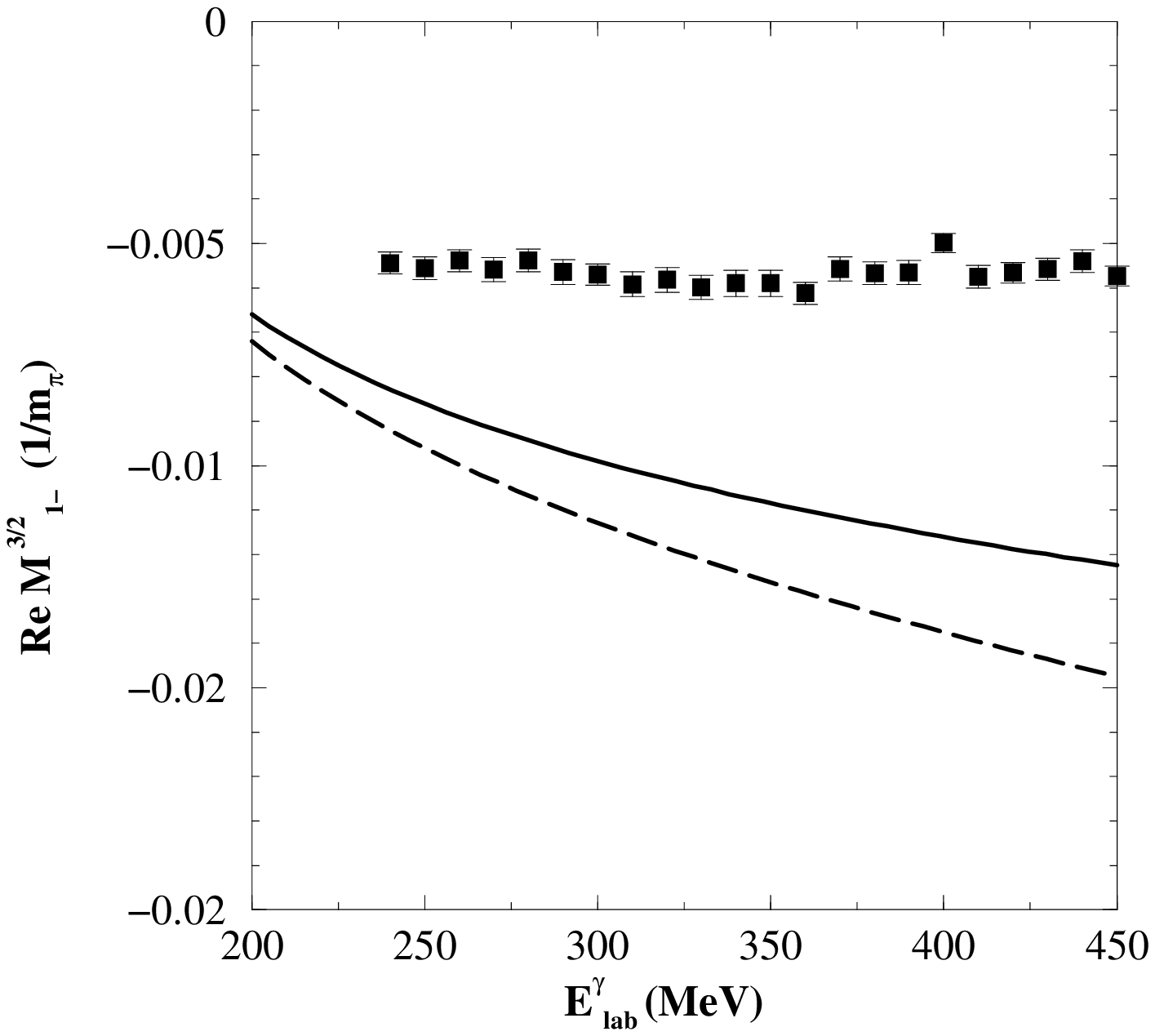,height=12cm,width=12cm}}

{\bf Fig. C.4.} $ M^{3/2}_{1-}$ multipole 
of
photo pionproduction on the single nucleon.
The theoretical fit result (solid line) is compared with
the experimental data of Ref.  \cite{don}.
          The dashed line represents the result without $\omega$-meson
          contribution. 
\end{figure}

\newpage

\begin{figure}
\mbox{\psfig{figure=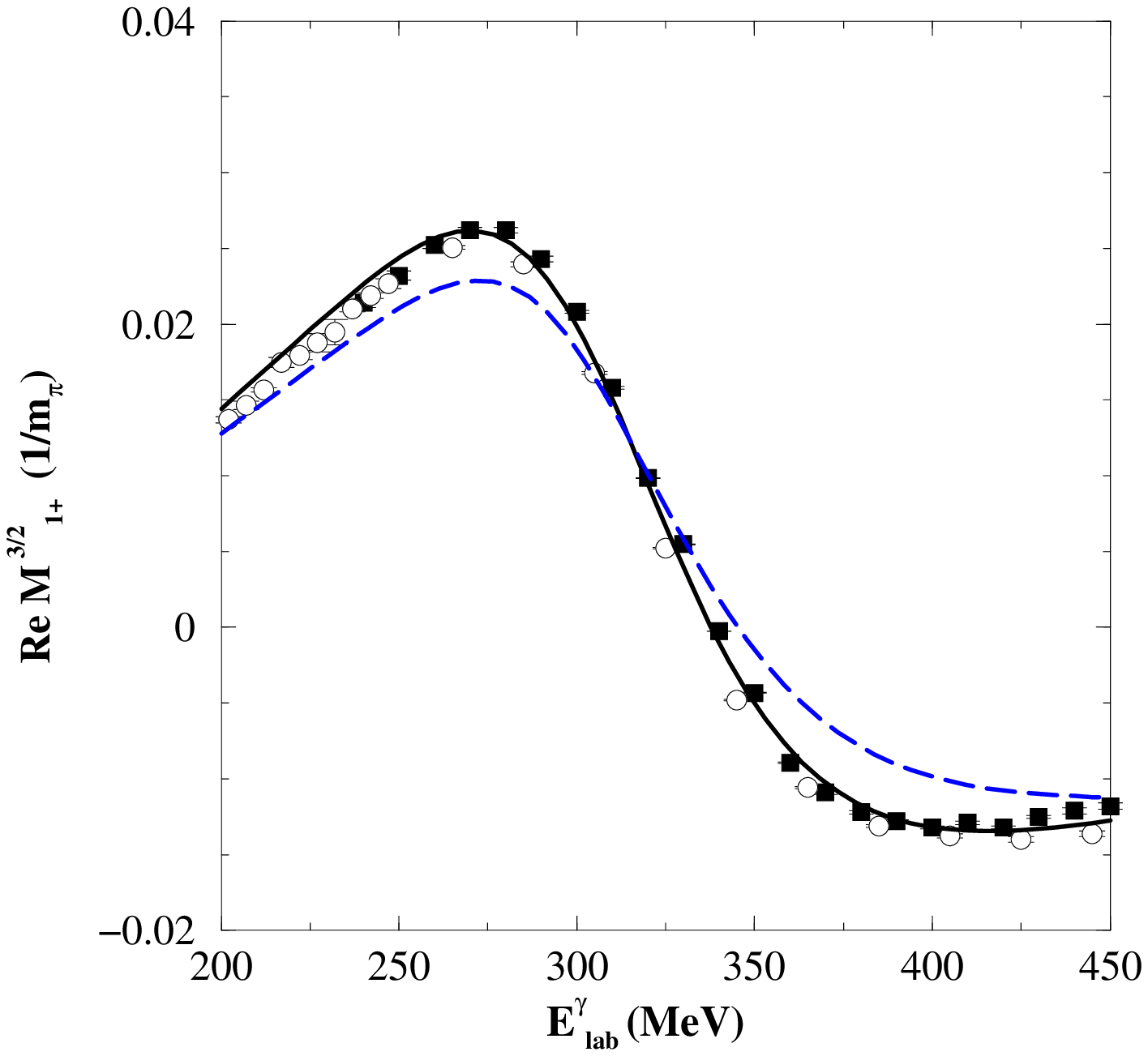,height=12cm,width=12cm}}

{\bf Fig. C.5.} Real part of $ M^{3/2}_{1+}$ multipole 
of
photo pionproduction on the single nucleon.
The theoretical fit result (solid line) is compared with
the experimental data of Ref. 
\cite{don} (filled-in squares) and of Ref. \cite{arnr2} (open circles).
The optimized fit parameters are $G_{M1}^{N \Delta} = 3.65$ and 
   $\Lambda_{B} = 245.5$ MeV.  The experimental data are from Ref. \cite{don} 
   (squares) and from Ref. \cite{arnr2}  (circles). The dashed line corresponds to
   the result without $\omega$-meson contribution.
\end{figure}

\newpage

\begin{figure}
\mbox{\psfig{figure=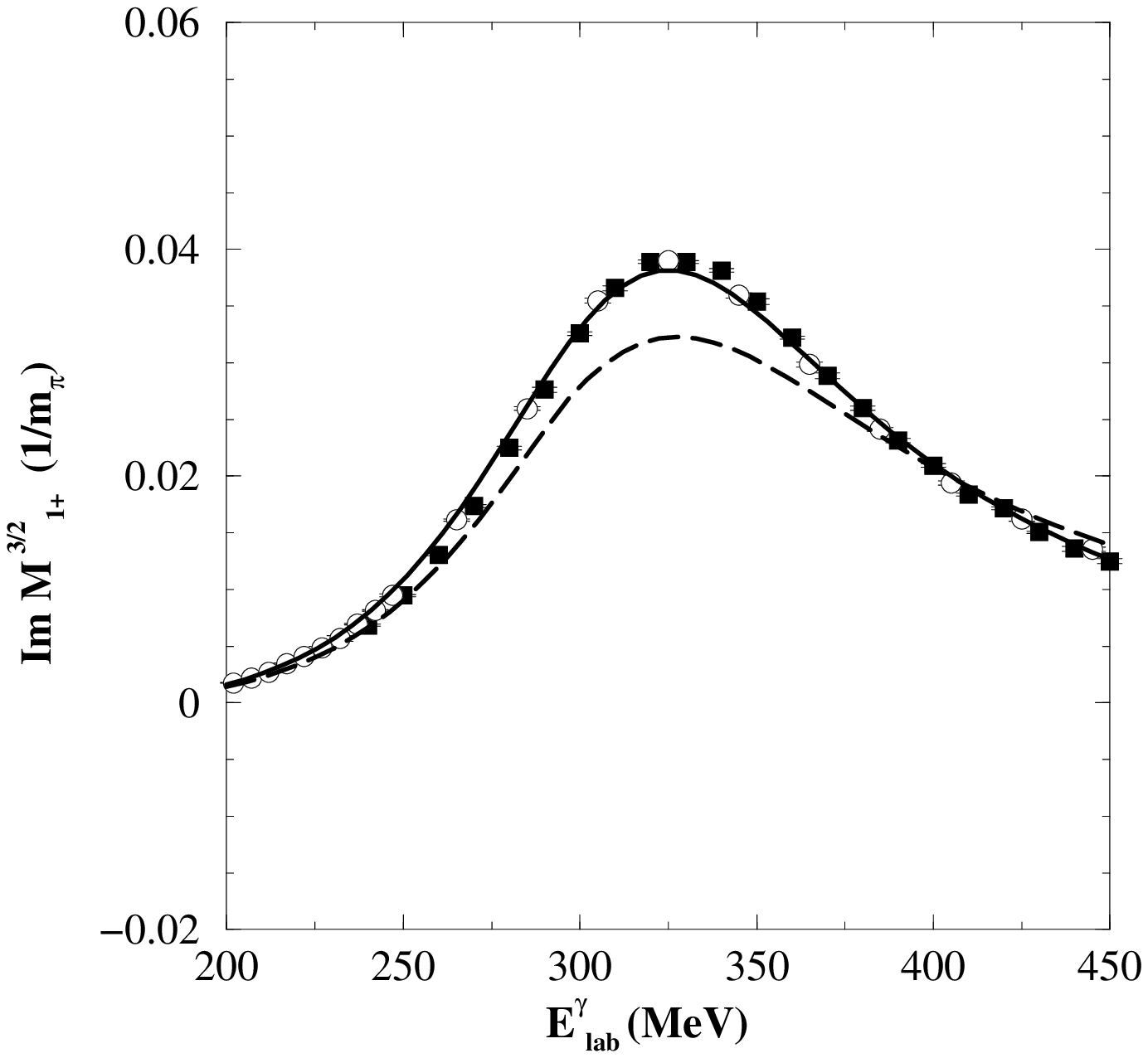,height=12cm,width=12cm}}

{\bf Fig. C.6.} Imaginary part of $M^{3/2}_{1+}$ multipole
of
photo pionproduction on the single nucleon.
The legend  corresponds to  Fig. C.5.
\end{figure}

\newpage

\begin{figure}
\mbox{\psfig{figure=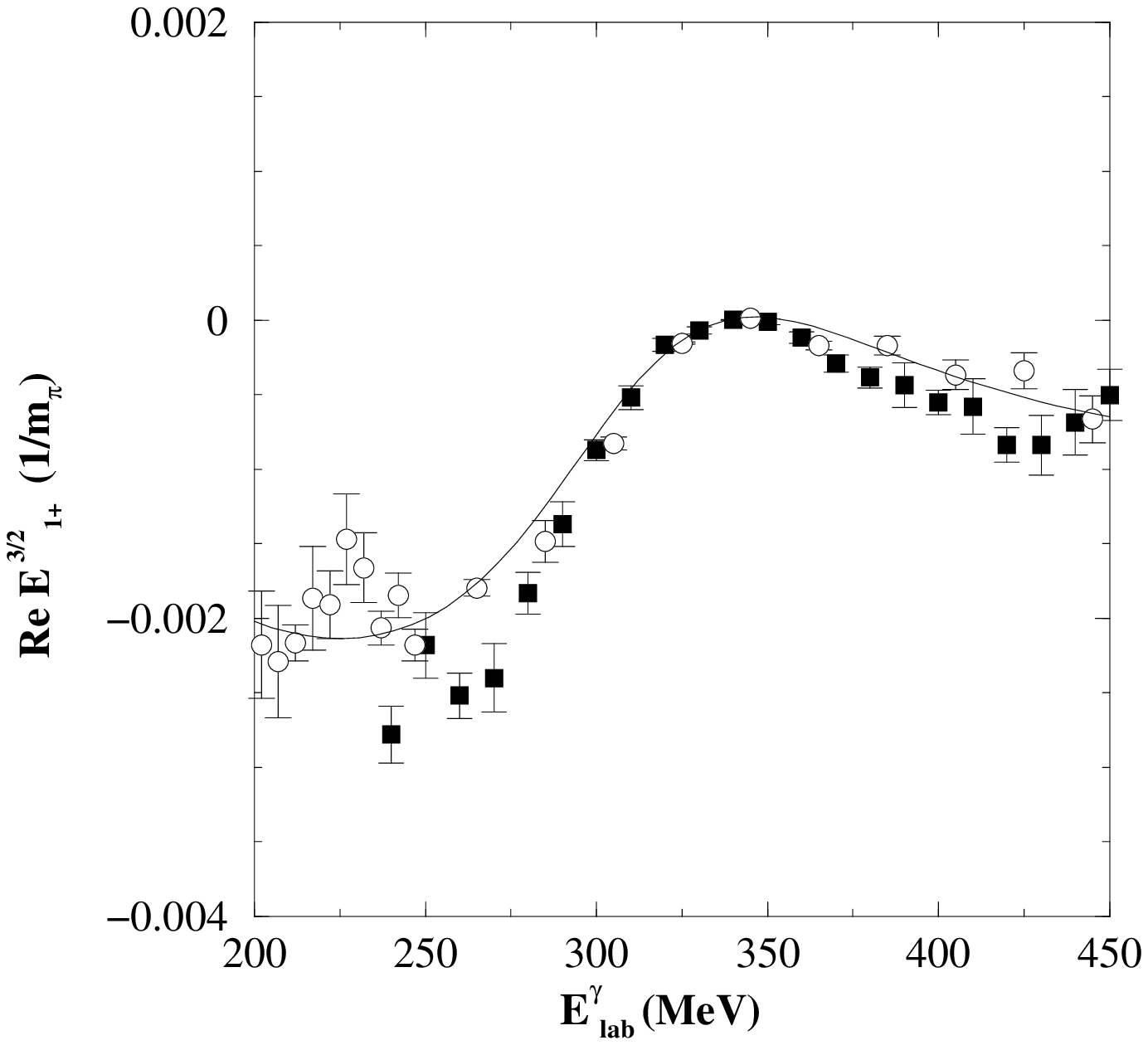,height=12cm,width=12cm}}

{\bf Fig. C.7.} Real part of $E^{3/2}_{1+}$ multipole
of
photo pionproduction on the single nucleon.
The theoretical fit result (solid line) is compared with
the experimental data of Ref.
\cite{don} (filled-in squares) and of Ref. \cite{arnr2} (open circles).
The optimized fit parameters are $G_{E2}^{N \Delta} = 0.1$ and
$\Lambda_{B} = 379.0$ MeV.
\end{figure}

\newpage

\begin{figure}
\mbox{\psfig{figure=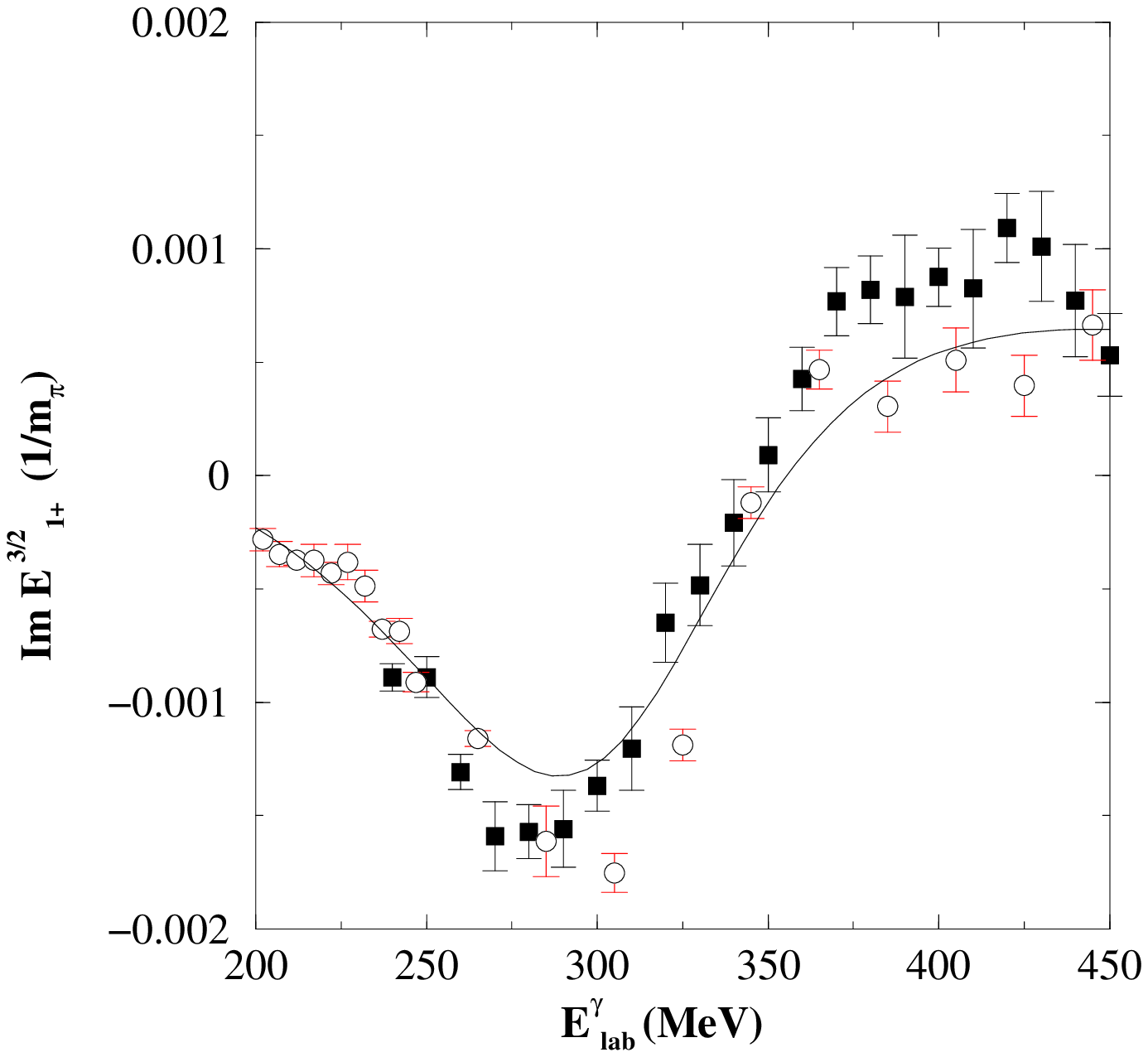,height=12cm,width=12cm}}

  {\bf Fig. C.8.} Imaginary part of $E^{3/2}_{1+}$ multipole
of
photo pionproduction on the single nucleon.
The legend  corresponds to  Fig. C.7.
\end{figure}